\documentclass[
 reprint,
superscriptaddress,
bibnotes,
 amsmath,amssymb,
 aps,
prb,
]{revtex4-2}
\usepackage[plainpages = false, pdfpagelabels,                 
                 bookmarks,
                 bookmarksopen = true,
                 bookmarksnumbered = true,
                 breaklinks = true,
                 linktocpage,
                 colorlinks = true, 
                 linkcolor = blue,
                 urlcolor  = blue,
                 citecolor = blue,
                 anchorcolor = green,
                 hyperindex = true,
                 hyperfigures
                 ]{hyperref} 
\usepackage{enumerate}
\usepackage{makecell, multirow}
\usepackage{tikz}
\usetikzlibrary{patterns}
\usepackage{slashed}
\usepackage{braket}
\usepackage{cancel}
\usepackage{bbold}
\usepackage{bm}
\usepackage{booktabs}
\usepackage{subfloat}
\usepackage[caption=false,singlelinecheck=off]{subfig}
\usepackage{blindtext}
\usepackage{bbm}
\usepackage{graphicx}
\usepackage{dcolumn}
\usepackage{bm}

\begin{document}
\title{Transport Signatures of Fractional Quantum Hall Binding Transitions}
\author{Christian Sp\r{a}nsl\"{a}tt}
 \email{christian.spanslatt@chalmers.se}
 \affiliation{Department of Microtechnology and Nanoscience (MC2),Chalmers University of Technology, S-412 96 G\"oteborg, Sweden}
\affiliation{Institute for Quantum Materials and Technologies, Karlsruhe Institute of Technology, 76021 Karlsruhe, Germany}
\affiliation{Institut f\"{u}r Theorie der Kondensierten Materie, Karlsruhe Institute of Technology, 76128 Karlsruhe, Germany}
\author{Ady Stern}
\affiliation{Department of Condensed Matter Physics, Weizmann Institute of Science, Rehovot 7610001, Israel}
\author{Alexander D. Mirlin}
\affiliation{Institute for Quantum Materials and Technologies, Karlsruhe Institute of Technology, 76021 Karlsruhe, Germany}
\affiliation{Institut f\"{u}r Theorie der Kondensierten Materie, Karlsruhe Institute of Technology, 76128 Karlsruhe, Germany}

\date{\today}
\begin{abstract}
Certain fractional quantum Hall edges have been predicted to undergo quantum phase transitions which reduce the number of edge channels and at the same time bind electrons together. However, detailed studies of experimental signatures of such a ``binding transition'' remain lacking. Here, we propose quantum transport signatures with focus on the edge at filling $\nu=9/5$. We demonstrate theoretically that in the regime of non-equilibrated edge transport, the bound and unbound edge phases have distinct conductance and noise characteristics. We also show that for a quantum point contact in the strong back-scattering regime, the bound phase produces a minimum Fano-factor $F_{SBS}=3$ corresponding to three-electron tunneling, whereas single electron tunneling is strongly suppressed at low energies. Together with recent experimental developments, our results will be useful for detecting binding transitions in the fractional quantum Hall regime.
\end{abstract}

\maketitle
\section{\label{sec:Introduction}Introduction}
Edges of fractional quantum Hall (FQH) states~\cite{Stormer1982,Laughlin1983} 
are outstanding platforms for strongly correlated electron physics. A FQH edge realizes the so-called chiral Luttinger liquid~\cite{Wen1990a,Wen1990b,Wen1992,Wen1994,Wen1995,Chang2003}, which is a set of one-dimensional conducting channels inheriting topological properties of the FQH bulk state. The chiral Luttinger liquid has been successfully used to investigate a wide variety of fundamental quantum phenomena, e.g. topological quantization, charge fractionalization~\cite{Saminadayar1997,DePicciotto1997,Lin2021}, anyonic statistics~\cite{Nakamura2020,Bartolomei2020}, topological quantum computation~\cite{Nayak2008}, or quantum phase transitions~\cite{Kane1994,Kane1995b,Moore1998}.

A particularly striking FQH edge quantum phase transition, called the binding transition, was proposed by Kao \textit{et.al.}~\cite{Kao1999}, based on earlier work by Haldane~\cite{Haldane1995}. In the binding transition, pairs of oppositely propagating edge channels localize due to an edge instability triggered by inter-channel particle tunneling and strong interactions. The remaining edge channels may then carry excitations with electrical charges different from those of the original edge; charges that can be viewed as bound composites of electrons. 

Binding transitions are possible only for so-called T-unstable FQH states, defined as those states permitting charge-neutral and bosonic quasiparticle excitations lacking topological content in their correlation function~\cite{Haldane1995,Moore1998}. On the edge, the creation and annihilation operators of such excitations describe charge tunneling between edge channels, and appear in the edge Hamiltonian without breaking any symmetries. Physically, it is only T-unstable edges that permit pairs of oppositely propagating channels to localize. Equivalently, T-unstable edges have low energy charge-neutral fixed points with an equal number of neutral modes (we use the terms ``mode'' and ``channel'' interchangeably in this work) propagating in each direction~\cite{Moore1998}. Importantly, the binding transition does not alter the topological order of the FQH bulk state, and is therefore a pure edge transition, amenable for detection in edge experiments. 

\begin{figure}[t!]
\begin{center}
\includegraphics[width=1.0\columnwidth]{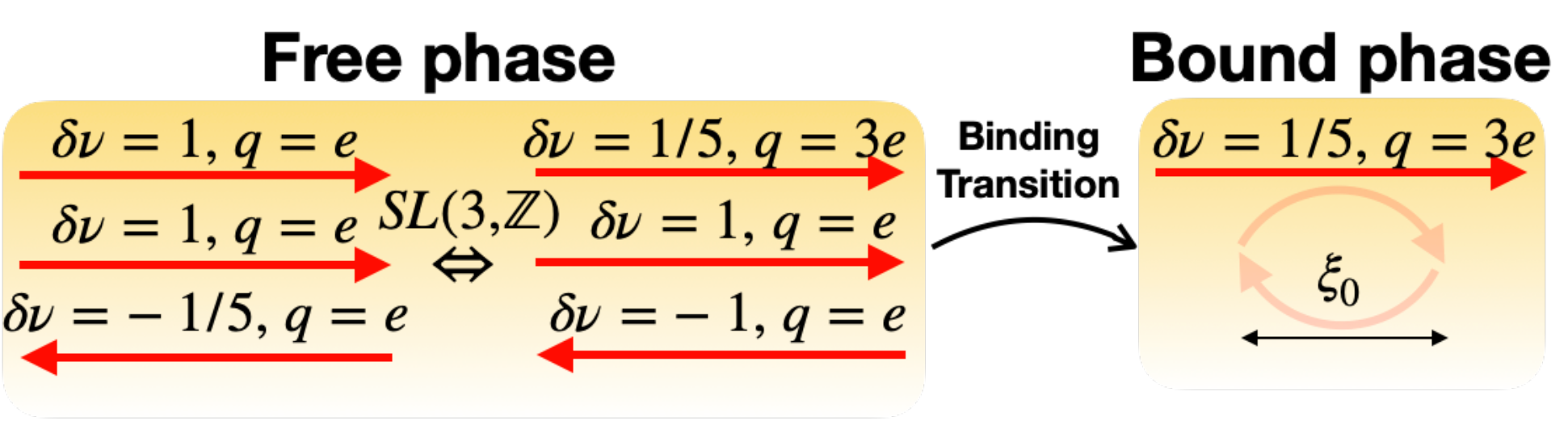}
\caption{\label{fig:95BindingTransition}The FQH binding transition at filling $\nu=9/5$. The free edge phase comprises two downstream propagating channels (right-pointing arrows), both with filling factor discontinuities $\delta \nu=1$ and charge $q=e$ (here, the charge $q$ is the minimum charge that can tunnel across vacuum into the channel). There is also one upstream channel filling factor discontinuity $\delta \nu =-1/5$ and charge $q=e$ (left-pointing arrow). This structure is equivalent (after a $SL(3,\mathbf{Z})$ basis change) to a Laughlin edge state with $\delta \nu =1/5$  composed of charge $q=3e$ composite particles, plus two counter-propagating channels with $\delta \nu=\pm 1$ and $q=e$. The binding transition is the phenomenon where inter-channel interactions and tunneling cause the counter-propagating channels to localize (with characteristic length scale $\xi_0$). In the bound phase, only the single composite channel remains.}
\end{center}
\vspace{-0.75cm}
\end{figure}

The simplest example of the binding transition was predicted for the edge at filling factor $\nu=9/5$ (see Fig.~\ref{fig:95BindingTransition}). In the free, or unbound, phase, the $9/5$ edge hosts three channels, which after the binding transition are reduced to a single channel. Most remarkably, in this so-called bound phase, single electron excitations become short ranged and do not participate in the low energy edge transport. By contrast, excitations with charges $3e$ remain long ranged and do contribute to the transport~\cite{Kao1999}. Despite such a striking re-organization of the edge structure, the prospects of experimentally observing a binding transition remain to large extent unexplored.

In this paper, we address this issue by proposing several experimental signatures of the binding transition. Our work is motivated by novel developments for probing FQH edges with quantum transport (a recent overview is given in Ref.~\cite{Heiblum2020}). More specifically, a growing body of experiments have demonstrated the existence of a wide range of different edge transport regimes. These range from complete charge and heat equilibration of the edge channels~\cite{Banerjee2017,Banerjee2018,Srivastav2019}, to intermediate regimes with full charge but no heat equilibration~\cite{Srivastav2021,Melcer2022,Kumar2022,Srivastav2022,LeBreton2022Sep}, to the extreme limit of non-equilibrated charge transport~\cite{Cohen2019,Lafont2019}. To detect a binding transition, access to non-equilibrated transport regimes is of particular interest, since in these regimes, the charge and heat conductances do not necessarily reflect the bulk topological order. Instead, these quantities can reveal the total number of edge channels and the charges they carry, quantities that both change across the binding transition. In addition, recent experiments~\cite{Dutta2022Sep,Dutta2022}, have demonstrated robust, engineered FQH edges formed between regions with different bulk fillings. For example, an ``artificial'' $9/5$ edge can be synthesized by proximitizing regions with fillings $\nu=2$ and $\nu=1/5$. Such structures, which might allow experimental control over Landau level spin polarizations and thereby tunneling rates between edge channels, facilitate the detection of binding transitions. 

As a key result, we find that the two edge phases have different charge and heat conductance characteristics (see Tab.~\ref{tab:TransportTab}).  With decreasing level of equilibration (i.e., with decreasing temperature $T$ and/or system size $L$), the free phase conductances monotonously increase from the values for equilibrated transport, which defines regime $\mathcal{III}$, to saturation at the non-equilibrated values, defining regime $\mathcal{II}$. By contrast, the bound phase is characterized by the existence of a localized regime, regime $\mathcal{I}$, with similar characteristics as regime $\mathcal{III}$. A transition between regimes $\mathcal{I}$ and $\mathcal{II}$ is possible for strong interactions and give rise to the unusual situation of increasing conductances with increasing temperature. Such an observation is a striking hallmark of the existence of edge localization. Complementing the conductances, we further argue that a current biased edge segment produces shot noise, $S$, when the charge transport is equilibrated but the heat transport is not. This feature occurs only for strong interactions, associated with the bound phase, in regime $\mathcal{II}$.

 We also demonstrate that in a quantum point contact (QPC) device in the strong back-scattering (SBS) regime, the bound phase yields a minimum shot noise Fano-factor $F_{SBS}=3$ corresponding to three-electron tunneling. No single electron tunneling is possible at low energies. This result stands in stark contrast to the free phase, where strong back-scattering favours single electron tunneling, i.e., $F_{SBS}=1$. By the same token, in the weak back-scattering (WBS) regime, the most relevant (in the renormalization group, RG, sense) quasiparticle tunneling yields in the bound phase $F_{WBS}=3/5$, in contrast to the free phase value $F_{WBS}=1/5$. Altogether, our set of derived transport signatures present several possibilities for experimentally detecting a FQH binding transition.

The remainder of this paper is organized as follows. In Sec.~\ref{sec:BindingTheory}, we review the basics of the FQH edge theory and the binding transition. We also perform a renormalization group (RG) treatment of the transition. In the main part of this paper, Sec.~\ref{sec:Signatures}, we derive several transport signatures of the bound phase and contrast them to those of the free phase. In Sec.~\ref{sec:Discussion}, we discuss possible experimental setups to detect the proposed signatures. We summarize in Sec.~\ref{sec:Summary} and also provide an outlook towards future studies.

Throughout this paper, we generally use units $e=\hbar =k_{\rm B}=1$, but we restore important units for transport observables. 

\begin{table}[t!]
\begin{ruledtabular}
\begin{tabular}{ccccc}  
\textbf{Regimes} & \textbf{Transport charact.} & \textbf{Free} & \textbf{Bound} \\
\hline
\multirow{3}{*}{$\mathcal{I}$}   &  $G / (e^2 / h)$ &-- & 9/5  \\ 
 & $G^Q / (\kappa T)$& -- & 1  \\
 & $S$& -- & $0$  \\
 \hline 
 \multirow{3}{*}{$\mathcal{II}$}   &  $G / (e^2 / h)$ &11/5 & 11/5  \\ 
 & $G^Q / (\kappa T)$& 3 & 3  \\
  & $S$ [$10^{-29}\rm{A}^2/(\rm {nA}\rm{Hz})$]& $\simeq 0$ & 0.25-0.7  \\
 \hline 
 \multirow{3}{*}{$\mathcal{III}$}   &  $G / (e^2 / h)$ &9/5 & 9/5  \\ 
 & $G^Q / (\kappa T)$& 1 & 1  \\
  & $S$& $\simeq 0$ & 0  \\
 \hline
  \multirow{2}{*}{QPC Fano factors}   &  $F_{SBS}$ &1& 3  \\ 
 & $F_{WBS}$& 1/5 & 3/5  \\
\end{tabular}
\end{ruledtabular}
\caption{\label{tab:TransportTab} Transport characteristics for the $\nu=9/5$ edge. For the free and bound edge phases, values for the two-terminal electrical ($G$) and heat ($G^{Q}$) conductances, the shot noise $S$ of a current biased edge segment, and QPC Fano actors for strong, $F_{SBS}$, and weak $F_{WBS}$ back scattering regimes are given. There are three relevant transport regimes. Regime $\mathcal{I}$: regime of localization, $\mathcal{II}$: no localization but vanishing equilibration, and $\mathcal{III}$ full edge equilibration. The emergence of regime $\mathcal{I}$ for the bound phase is a fundamental feature of the binding transition. The quantized $G^{Q}$ values in regime $\mathcal{II}$ are given under the assumption of vanishing interference in edge-contact plasmon scattering: $L\gg L_T\sim T^{-1}$. The noise in regime $\mathcal{II}$ is given under the condition~\eqref{eq:noisecondition} of efficient charge equilibration and poor thermal equilibration. This condition holds for strong interactions which is only the case in the bound phase.}
\end{table}

\section{\label{sec:BindingTheory}The Chiral Luttinger liquid and the binding transition}
For completeness, we review here key aspects of the chiral Luttinger theory and the FQH binding transition, closely following Refs.~\cite{Kane1995b,Moore1998,Kao1999}.
\subsection{The chiral Luttinger liquid}
\label{sec:ChiLL}
At low energies, an Abelian FQH edge is well described by the chiral Luttinger liquid ($\chi_{LL}$) model, specified by the pair ($K$, $\mathbf{t}$)~\cite{Wen1992}. Here, $K$ is an $n\times n$ integer valued symmetric matrix, and the charge vector $\mathbf{t}$ is an $n$-dimensional vector of integers. The generic $n$-channel action reads
\begin{align}
\label{eq:ChiLL}
	S_{\chi_{LL}} = -\frac{1}{4\pi}\int dt dx \big[ \partial_t\bm{\phi}^T K  \partial_x\bm{\phi} +  \partial_x\bm{\phi}^T V  \partial_x\bm{\phi}\big],
\end{align}
where $\bm{\phi}=(\phi_1,\hdots,\phi_n)$ is a set of bosonic fields obeying the commutation relations
\begin{align}
	[\phi_i(x),\phi_j(y)] = i\pi K^{-1}_{ij} \text{sgn}(x-y).
\end{align}
The symmetric matrix $V$ in~\eqref{eq:ChiLL} parametrizes the channel velocities, $V_{ii}$, and mutual, short-range Coulomb interactions, $V_{i\neq j}$. Generally, $V$ contains $n(n+1)/2$, independent, non-universal parameters, determined by microscopic details in the edge electrostatic confinement.

 The electrical charge enters the theory in the charge densities
 \begin{align}
 \rho_i \equiv t_i \frac{\partial_{x}\phi_i}{2\pi},	
 \end{align}
 which obey
\begin{align}
	[\rho_i(x),\rho_j(y)] =  \frac{i}{2\pi}t_i^2 K^{-1}_{ij} \partial_x \delta(x-y).	
\end{align}

Several quantities determined by the bulk topological order appear in the edge theory. The filling factor $\nu$ is given as
\begin{align}
\label{eq:nu}
	\nu = \mathbf{t}^T K^{-1} \mathbf{t},
\end{align}
and the ``thermal quantum number''
\begin{align}
\label{eq:nuQ}
	\nu_Q \equiv \text{Tr}(\eta) = n_d -n_u.
\end{align}
Here, $\eta$ is the signature matrix corresponding to $K$, so that $\nu_Q$ equals the difference in the number of positive and negative eigenvalues of $K$ (i.e., $\eta_{ij}=\pm 1\delta_{ij}$). In turn, this is equivalent to the number of ``downstream'' $n_d$ and ``upstream'' $n_u$ propagating channels (with respect to the chirality direction set by the magnetic field) respectively~\cite{Kane1997,Capelli2002}. For the present theory of  Abelian states, $\nu_Q \in \mathbb{Z}$, while for non-Abelian states, other values of $\nu_Q$ are possible. For example, Majorana edge channels allow half-integer values $\nu_Q\in \mathbb{Z}/2$~\cite{Banerjee2018}. As described below in Sec.~\ref{sec:Signatures}, $\nu_Q$ is closely connected to the edge heat transport characteristics.

Quasiparticles (including the special case of the electron) created or destroyed at position $(t,x)$ on the edge is described by vertex operators
\begin{align}
\label{eq:tunnelingOperator}
\mathcal{T}_\mathbf{l} (t,x) = \frac{e^{i\mathbf{l} \cdot \bm{\phi}(t,x)}}{(2\pi a)^{m/2}},
\end{align}
where $m$ is the number of involved bosons. A vertex operator is uniquely determined by specifying an integer valued vector $\mathbf{l}$ which describes how many of each quasiparticle species that are created or destroyed. The associated exchange statistics angle $\Theta_\mathbf{l}$ and electric charge $Q_\mathbf{l}$ of a vertex operator are given by
\begin{align}
	& \Theta_\mathbf{l} = \pi\mathbf{l}^T K^{-1} \mathbf{l} \quad \text{mod } 2\pi \label{eq:Theta},\\
	& Q_\mathbf{l} = \mathbf{t}^T K^{-1} \mathbf{l} \label{eq:Charge}.
\end{align}

With the use of vertex operators, tunneling of particles between edge channels is included in the theory by adding to~\eqref{eq:ChiLL} the term
\begin{align}
\label{eq:Spert}
	S_{\rm T} = \int dt dx \; \xi(x) \mathcal{T}_\mathbf{l} (t,x) + \text{H.c},
\end{align}
where $\xi(x)$ is the local tunneling strength at spatial location $x$. As discussed below in Eqs.~\eqref{eq:RandTunn}-\eqref{eq:UniTunn}, the function $\xi(x)$ determines the type of tunneling between the edge channels. 

To compute various observables in the theory, correlation functions involving $\mathcal{T}_\mathbf{l} (t,x)$ are needed. These are most easily obtained by diagonalizing $S_{\chi_{LL}}$ which is done by first taking $K$ to its signature matrix $\eta$ by a (non-unique) matrix $M_1$
\begin{align}
	M_1^T K M_1 = \eta.
\end{align}
Second, a matrix $M_2$ (also not unique) is sought which diagonalizes $V$ into $\tilde{V}$ but at the same time preserves $\eta$:
\begin{align}
	&\eta = M_2^TM_1^T K M_1 M_2,\\ 
	&\tilde{V} = M_2^TM_1^T V M_1 M_2.
\end{align}
In the diagonal basis, the action~\eqref{eq:ChiLL} becomes
\begin{align}
\label{eq:chiLLDiagonal}
	S_{\chi_{LL}} = -\frac{1}{4\pi}\int dt dx \big[ \partial_t\bm{\tilde{\phi}}^T \eta  \partial_x\bm{\tilde{\phi}} +  \partial_x\bm{\tilde{\phi}}^T \tilde{V}  \partial_x\bm{\tilde{\phi}}\big]
\end{align}
and the theory is now expressible in transformed quantities as
\begin{subequations}
\label{eq:rotations}
\begin{align}
	&\bm{\tilde{\phi}} = M^{-1}\bm{\phi},\\
	&\mathbf{\tilde{t}} = M^T\mathbf{t},\\
	&\mathbf{\tilde{l}} = M^T\mathbf{l},
\end{align}
\end{subequations}
with $M\equiv M_1M_2$. The free, ``diagonal'', bosons $\tilde{\phi}_i$ and their densities
 \begin{align}
 \tilde{\rho}_i \equiv \tilde{t}_i \frac{\partial_{x}\tilde{\phi}_i}{2\pi},	
 \end{align}
obey
\begin{align}
	&[\tilde{\phi}_i(x),\tilde{\phi}_j(y)] = i\pi  \eta_{ij} \text{sgn}(x-y),\\
	&[\tilde{\rho}_i(x),\tilde{\rho}_j(y)] =  \frac{i}{2\pi}\tilde{t}_i^2 \eta_{ij} \partial_x \delta(x-y).	
\end{align}
Note that all topological properties~\eqref{eq:nu},~\eqref{eq:nuQ},~\eqref{eq:Theta}, and \eqref{eq:Charge} are independent of the choice of basis. The total charge density is also preserved:
\begin{align}
	\sum_i \rho_i = \sum_i \tilde{\rho}_i.
\end{align}
In the diagonal basis, the action is quadratic and correlation functions of vertex operators follow from the identity
\begin{align}
\label{eq:CorrSimp}
	&\langle e^{i \tilde{\phi}_i(t,x)} e^{-i \tilde{\phi}_i(0,0)}  \rangle = e^{\langle \tilde{\phi}_i(t,x) \tilde{\phi}_i(0,0) - \langle \tilde{\phi}^2_i(0,0) \rangle } \notag\\
	&= \frac{a}{a+x-i\eta_i\tilde{v}_i t},
\end{align}
upon use of the zero temperature correlation function
\begin{align}
\label{eq:DiagGF}
	\langle\tilde{\phi}_i(t,x) \tilde{\phi}_i(0,0) - \langle \tilde{\phi}^2_i(0,0) \rangle =-\log\left[ \frac{a+x-i\eta_i\tilde{v}_i t}{a} \right].
\end{align}
In Eqs.~\eqref{eq:CorrSimp} and~\eqref{eq:DiagGF}, $\eta_i\equiv\eta_{ii}$ and $\tilde{v}_i\equiv\tilde{V}_{ii}$ is the chirality and the speed of mode $\tilde{\phi}_i$, respectively. We also introduced a short distance (ultraviolet, UV) cutoff $a$ on the order of the characteristic magnetic length. At finite temperature $T$, the correlation function~\eqref{eq:DiagGF} changes to
\begin{align}
\label{eq:DiagGF2}
	&\langle\tilde{\phi}_i(t,x) \tilde{\phi}_i(0,0) - \langle \tilde{\phi}^2_i(0,0) \rangle \notag \\
	&=-\log\left[ \frac{\sin \left(\frac{\pi T}{\tilde{v}_i} (a+x-i\eta_i \tilde{v}_i t) \right)}{\pi aT/\tilde{v}_i}\right].
\end{align}

By combining the single mode correlation~\eqref{eq:CorrSimp} with the transformation rules~\eqref{eq:rotations}, the (zero $T$) correlation function of $\mathcal{T}_\mathbf{l}(t,x)$ is obtained as
\begin{align}
	&\langle \mathcal{T}_\mathbf{l}(t,x) \mathcal{T}_\mathbf{l}^\dagger(0,0) \rangle \sim \langle e^{i\mathbf{l} \cdot \bm{\phi}(t,x)} e^{-i\mathbf{l} \cdot \bm{\phi}(0,0)} \rangle\notag \\
	& = \langle e^{i\mathbf{\tilde{l}} \cdot \bm{\tilde{\phi}}(t,x)} e^{-i\mathbf{\tilde{l}} \cdot \bm{\tilde{\phi}}(0,0)} \rangle = \prod_i \left(\frac{a}{a+x-i\eta_i\tilde{v}_i t}\right)^{((\mathbf{\tilde{l}})_i)^2}.
\end{align}
The long time behaviour
\begin{align}
\label{eq:gencorr}
\langle \mathcal{T}_\mathbf{l}(t,0) \mathcal{T}_\mathbf{l}^\dagger(0,0) \rangle\propto \frac{1}{t^{\kappa(\mathbf{l})}}\frac{1}{|t|^{2\Delta(\mathbf{l})-\kappa(\mathbf{l})}},
\end{align}
where
\begin{align}
\label{eq:GenScalingDim}
	&\Delta(\mathbf{l}) \equiv  \frac{1}{2} \mathbf{\tilde{l}}^T \mathbf{\tilde{l}} = \frac{1}{2}\mathbf{l}^T M M^T \mathbf{l},\\
		\label{eq:Kform} &\kappa(\mathbf{l}) \equiv \mathbf{\tilde{l}}^T \eta^{-1}\mathbf{\tilde{l}} = \mathbf{l}^T K^{-1}\mathbf{l}\end{align}
defines the scaling dimension and the topological part of the correlation function, respectively. The generic scaling dimension~\eqref{eq:GenScalingDim} is non-universal, as it depends not only on the topological matrix $K$ but also on the components of $V$. An exception occurs for so-called maximally chiral edges, in which all channels propagate in the same direction, i.e., either $n_d$ or $n_u$ equals zero. Then, the scaling dimensions of tunneling operators are fully specified by $K$ alone.

 Importantly, $\kappa(\mathbf{l})$ and $\Delta(\mathbf{l})$ obey the following inequality~\cite{Moore1998}
\begin{align}
\label{eq:KvsDelta}
	|\kappa(\mathbf{l})|\leq 2\Delta(\mathbf{l}),
\end{align}
with equality for vanishing interactions (diagonal $V$). For maximally chiral edges, Eq.~\eqref{eq:KvsDelta} becomes an equality independently of interactions. 
 
We now consider tunnel coupled edge channels, so that the total action $S=S_{\chi_{LL}}+S_{\rm T}$.  Depending on the nature of the tunneling events, $S_{\rm T}$ becomes an RG relevant perturbation when
\begin{align}
	&3-2\Delta(\mathbf{l})>0, \;\; \langle \xi (x)\xi^{*}(y)\rangle = D\delta (x-y), \label{eq:RandTunn}  \\
	&1-\Delta(\mathbf{l})>0, \quad \xi(x) = \Gamma_0 \delta (x), \label{eq:PointTunn} \\
	&2-\Delta(\mathbf{l})>0, \quad \xi(x) = \Gamma_0, \label{eq:UniTunn}
\end{align}
corresponding to Gaussian random (characterized by the strength $D$), single point, and uniform tunneling, respectively. In this paper, we mainly focus on the common situation of random tunneling due to edge disorder (point tunneling as realized in a QPC is considered in Sec.~\ref{sec:QPC}). Then, to guarantee the relevancy of $\mathcal{T}_\mathbf{l}$,  Eq.~\eqref{eq:RandTunn} implies that $\Delta(\mathbf{l})<3/2$. By virtue of Eq.~\eqref{eq:KvsDelta}, this implies further
\begin{align}
\label{eq:Krestrict}
|\kappa(\mathbf{l})|\leq 2\Delta(\mathbf{l})<3.
\end{align}
In turn, any tunneling operator must be bosonic, which by use of Eq.~\eqref{eq:Theta} implies that $\kappa(\mathbf{l})$ must be an even integer. This fact, combined with Eq.~\eqref{eq:Krestrict}, implies that
\begin{align}
	\kappa(\mathbf{l}) = -2,0,2,
\end{align}
for random, relevant tunneling operators. As we shall see next, when a very particular class of such tunneling operators exist and are relevant, a FQH edge will undergo a binding transition. 
 
\subsection{Review of the binding transition}
The binding transition is only possible for so-called T-unstable edges. These are defined as those edges permitting a special kind of quasiparticles (which we parametrize for convenience by $\mathbf{m}$ rather than $\mathbf{l}$), satisfying the two constraints~\cite{Kao1999,Haldane1995}
\begin{align}
	& \label{eq:condNeutral}\mathbf{m}^T K^{-1}\mathbf{m} = 0, \\
	& \label{eq:condBoson}\mathbf{t}^T K^{-1}\mathbf{m} = 0.
\end{align}
A non-zero string $\mathbf{m}$ obeying these constraints is called a null-vector, and Eqs.~\eqref{eq:condNeutral} and~\eqref{eq:condBoson} are called the null conditions. They are invariant under basis transformations.

The possibility to satisfy the null conditions can be traced to the existence of counter-propagating neutral modes in the charge-neutral basis~\cite{Moore1998}. Physically, the null operators create charge-neutral and bosonic particles without any topological part in their correlation function~\eqref{eq:gencorr}. Then, and only then, is it possible for pairs of edge channels to undergo localization. We may view this feature as the edge structure containing a non-topological part which can be removed by the non-topological disorder and interactions.

We can readily check that no null-vectors exist for edges (those with non-zero Hall conductance) with one or two channels, i.e., for $n=1,2$. For $n=1$, we have $\mathbf{m}=m$, $\mathbf{t}=t$ and $K^{-1}$ is an odd integer. Eqs.~\eqref{eq:condNeutral} and~\eqref{eq:condBoson} then read
\begin{align}
	K^{-1} m^2 = 0, \\
	tK^{-1} m = 0,
\end{align}
which is only trivially satisfied by $m=0$. 

For $n=2$, we may choose without loss of generality
\begin{align}
	&K = \begin{pmatrix}
		1/\delta \nu_1 & 0 \\
		0 &  1/\delta \nu_2.
	\end{pmatrix},\notag\\ &\mathbf{m}^T = (m_1,m_2),  \quad \mathbf{t}^T = (1,1).
\end{align}
Here, $\delta \nu_{1,2}$ (the eigenvalues of $K^{-1}$) are known as ``filling factor discontinuities'', and specify jumps in the Hall fluid density close to the edge. For example, the $\nu=2/3$ edge is specified by $\delta \nu_1=1$ and $\delta \nu_2=-1/3$~\cite{Kane1994}.
The null conditions~\eqref{eq:condNeutral}-\eqref{eq:condBoson} now read
\begin{align}
\label{eq:NullEqs21}
	m^2_1 \delta \nu_1 + m_2^2\delta \nu_2 = 0, \\
	\label{eq:NullEqs22}
	m_1\delta \nu_1 + m_2\delta \nu_2 = 0.
\end{align}
These equations have only the trivial $m_1=m_2=0$ solution, under the condition $\delta \nu_1 \neq \delta \nu_2$, which must be satisfied for the $n=2$ chiral Luttinger liquid~\cite{Wen1992}. With the present formalism, we can however readily see that for a standard, spinless Luttinger liquid, i.e., for $\delta \nu_1 =-\delta \nu_2 =1$, all charge conserving ($m_1=m_2$) operators satisfy~\eqref{eq:NullEqs21}-\eqref{eq:NullEqs22} and may cause instabilities localizing the edge channels~\cite{Giamarchi1988,Gornyi2007} (see also Ref.~\cite{Murthy2020} for a recent discussion).

For $n=3$, there exists several T-unstable FQH states. Specifically, we focus as follows on the state at filling $\nu=9/5$. The corresponding edge theory is defined by
\begin{align}
\label{eq:KFree}
	K = \begin{pmatrix}
		1 & 0 & 0 \\
		0 & 1 & 0 \\
		0 & 0 & -5
	\end{pmatrix},\quad \mathbf{t}^T = (1,1,1).
\end{align}
Using the null conditions~\eqref{eq:condNeutral} and~\eqref{eq:condBoson}, it can readily be checked that
\begin{align}
\label{eq:95NullVectors}
	\mathbf{m}^T_1 = (-1,2,5), \quad 	\mathbf{m}^T_2 = (2,-1,5),
\end{align}
are the two possible null vectors (changing an overall sign does not count as a new null-vector). We denote the corresponding null-operators by
\begin{align}
\label{eq:Null_Operator}
	\mathcal{V}_{\mathbf{m}_j}(t,x) \sim  e^{i\mathbf{m}_j \cdot \bm{\phi}(t,x)}, \quad j=1,2.
\end{align} 
With the scaling dimension formula~\eqref{eq:GenScalingDim}, one can check that in the absence of interactions, i.e., when $V$ is diagonal in basis~\eqref{eq:KFree}, the scaling dimensions of these operators are $\Delta(\mathbf{m}_1)=\Delta(\mathbf{m}_2)=5$, whereas interactions reduce these values. 

The binding transition becomes possible when at least one of the null-operators~\eqref{eq:Null_Operator} is RG relevant~\cite{Kao1999}, i.e., $\Delta(\mathbf{m}_j)<3/2$, according to Eq.~\eqref{eq:RandTunn}. This requires sufficiently strong edge interactions. As follows, we now assume, without loss of generality~\cite{Kao1999}, that $\mathcal{V}_{\mathbf{m}_1}$ is a relevant operator and ignore all effects from $\mathcal{V}_{\mathbf{m}_2}$. 

To understand the binding transition, it is very useful to perform a basis transformation with a matrix $W\in SL(3,\mathbb{Z})$. With the appropriate choice of $W$ (the exact details of $W$ are not important, but its existence follows from the existence of nullvectors), we can equally represent the $9/5$ edge as
\begin{align}
\label{eq:Kfree2}
	K' = \begin{pmatrix}
		5 & 0 & 0 \\
		0 & 1 & 0 \\
		0 & 0 & -1
	\end{pmatrix},\quad \mathbf{t'}^T = (3,1,1).
\end{align}
In this basis, the nullvectors read
\begin{align}
\label{eq:95NullVectors2}
	\mathbf{m'}^T_1 = (0,1,1), \quad 	\mathbf{m'}^T_2 = (15,-2,7).
\end{align}
We focus only on the null vector $\mathbf{m'}_1$ and its associated null operator $\mathcal{V}_{\mathbf{m'}_1}$. Note that this operator only couples two of the three modes. In the absence of interactions in the basis~\eqref{eq:Kfree2}, the scaling dimension of $\mathcal{V}_{\mathbf{m'}_1}$ is $\Delta(\mathbf{m'}_1)=1$. 

Eq.~\eqref{eq:Kfree2} suggests that the $9/5$ edge can be viewed as two counter-propagating channels with filling factor discontinuities $\pm 1$ carrying unit charge, and one downstream channel with filling factor discontinuity $1/5$ carrying particles with charge $3e$ (see Fig.~\ref{fig:95BindingTransition}). These composite particles can be thought of ``Cooper-triplons'': bound states of three electrons. Unlike conventional Cooper-pairs, however, the composite particles here are fermionic and therefore cannot condense.

The binding transition is the manifestation of localization of the two equal and counter-propagating channels due to interactions and disorder~\cite{Giamarchi1988,Gornyi2007}. After localization, only one channel remains. In some sense, this can be thought of as an ``inverted edge reconstruction'', where non-topological pairs of channels vanish rather than appear. On length scales much larger than a characteristic localization length (to be discussed in more detail below in Sec.~\ref{sec:scaling}), the edge structure is then effectively given as
\begin{align}
\label{eq:Kbound}
	K = (5),\quad \mathbf{t}^T = (3),
\end{align}
which is nothing but a $\nu=1/5$ Laughlin state made out of charge $3e$ composites. The peculiarity of the binding transition is that when two counter-propagating but otherwise equal channels localize, the remaining channel carries a changed charge as to preserve the topological quantum numbers from the bulk. This picture clarifies also why $n=3$ is the minimum number of edge channels required for a binding transition.

From Eqs.~\eqref{eq:Kfree2} and~\eqref{eq:95NullVectors2}, we see that excitations~\eqref{eq:tunnelingOperator} with vectors
\begin{align}
	\mathbf{l'}^T=(0,q,q), \quad q\in \mathbb{Z},
\end{align}
will become localized on the edge, i.e., those $\mathbf{l'}$ satisfying
\begin{align}
\label{eq:gaplessExc}
	\mathbf{l'}^TK'^{-1}\mathbf{m'}_1=0.
\end{align}
Due to the transformation rules~\eqref{eq:rotations}, the condition~\eqref{eq:gaplessExc} holds in any basis. In contrast, an excitation on the form 
\begin{align}
\label{eq:prop_ells}
	\mathbf{l'}^T=(p,0,0), \quad p\in \mathbb{Z}
\end{align}
will propagate freely along the edge. In more technical terms, in the original three-dimensional vector-space of excitations, only excitations in the one-dimensional subspace of vectors $\mathbf{l'}$ satisfying
\begin{align}
	\mathbf{l'}^TK'^{-1}\mathbf{m'}_1=0 \text{ and } \mathbf{l'} \text{ not proportional to } \mathbf{m'}_1
\end{align}
will propagate freely on the bound edge. From Eqs.~\eqref{eq:Theta} and~\eqref{eq:Charge}, we see that the statistics and the charge of the propagating excitations~\eqref{eq:prop_ells} are
\begin{align}
	& \Theta_{\mathbf{l'}} = \pi \frac{p^2}{5},\\
	& Q_{\mathbf{l'}} = \frac{3p}{5}.
\end{align}
Since only $p=5N$, with $N$ an odd integer produces quantum numbers consistent with the electron, it follows that only electron excitations in bunches of $3$ can propagate. In contrast, single electron excitations become localized on the edge. The consequence of this effect is explored in more detail in Sec.~\ref{sec:QPC}. In passing, we note that for bosonic FQH states, $N$ must be even to preserve Bose statistics. This implies that binding transitions for such states always generate even multiples of bound composites. 

The goal of this paper can now be formulated as investigating the transport properties for the edge in the two different edge phases. To this end, we next perform an RG analysis to find the phase diagram of the edge.

\subsection{Scaling analysis of the binding transition}
\label{sec:scaling}
Our simple scaling analysis of the binding transition follows the approach in Refs.~\cite{Giamarchi1988,Gornyi2007,Protopopov2017}. Throughout this section, we assume that $\mathcal{V}_{\mathbf{m'}_1}$ [see Eq.~\eqref{eq:Null_Operator}] is more relevant than $\mathcal{V}_{\mathbf{m'}_2}$ and analyze here only the influence of $\mathcal{V}_{\mathbf{m'}_1}$. The opposite situation can be treated in a perfectly analogous manner.

\begin{figure*}[t!]
  \centering
    \includegraphics[width=1\textwidth]{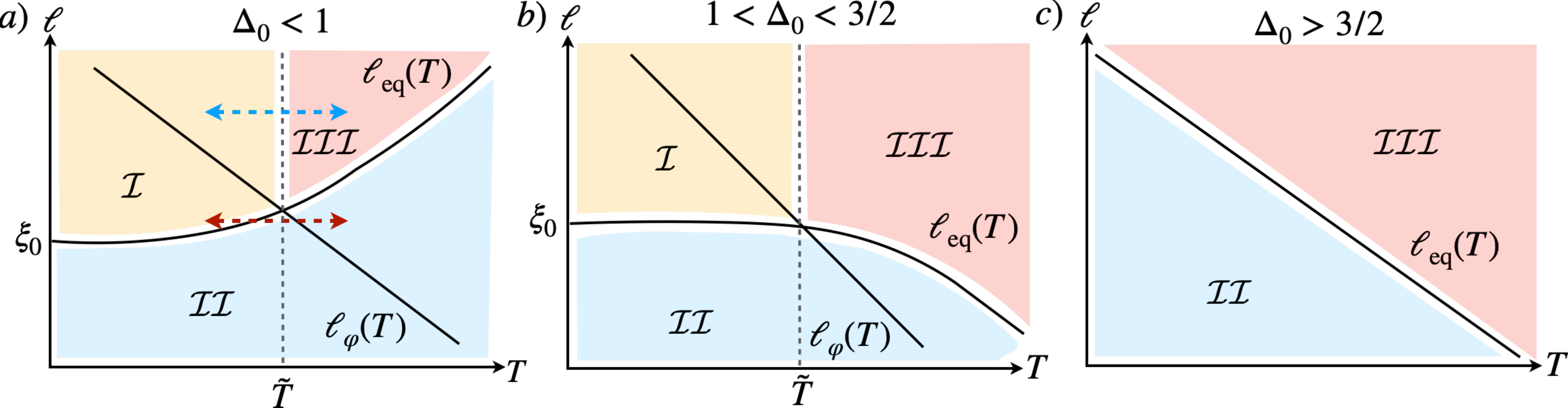}
    \caption{Schematic log-log phase diagrams of the temperature ($T$) dependence of equilibration lengths $\ell_{\rm eq}\sim T^{2-2\Delta_0}$. Here, $\ell$ is the length scale at which the system is observed, $\Delta_0$ is the bare scaling dimension of a nulloperator, $\ell_\varphi\sim 1/T$ is the thermal length in the presence of strong interactions, and $\xi_0$ is the zero temperature localization length. The gray, dashed line is the characteristic temperature scale for onset of localization: $\ell_\varphi(\tilde{T})=\ell_{\rm eq}(\tilde{T})$. In $(a)$ and $(b)$, localization occurs in regime $\mathcal{I}$ (yellow region), whereas regimes $\mathcal{II}$ (blue region) and $\mathcal{III}$ (red region) correspond to non equilibrated and equilibrated transport, respectively. In $(c)$, localization is absent.}
    \label{fig:phasediagram}
\end{figure*}

For convenience, we next redefine the disorder strength in Eq.~\eqref{eq:RandTunn} so that it becomes dimensionless (see Appendix~\ref{sec:DimAppendix}). We thus set
\begin{align}
	D \rightarrow D\times \frac{a^{3-m}}{v^2}.
\end{align}
Here, $m$ is the number of bosonic fields involved in the null-operator $\mathcal{V}_{\mathbf{m}_1}$, i.e., $m=3$, and $v$ is a characteristic velocity (i.e., some combination of the $v_i$; the exact details are not important here).
The renormalization equation for $D$ then reads~\cite{Giamarchi1988}
\begin{align}
\label{eq:disorder_renorm}
\frac{\partial D}{\partial \ell} = (3-2\Delta_0)D,	
\end{align}
where $\Delta_0$ is the initial scaling dimension of $\mathcal{V}_{\mathbf{m'}_1}$ and $\ell = \log(L/a)$, with $L$ the system size and $a$ is the UV length cutoff (e.g., the magnetic length). As follows, we ignore effects of renormalization of $\Delta_0$, as they are assumed weak~\cite{Gornyi2007,Protopopov2017}. In any case, the renormalization of $\Delta_0$ is non-essential for the qualitative analysis presented here. Hence, as follows we take
\begin{align}
\Delta\approx \Delta_0
\end{align}
during the RG flow. We now study the consequences of Eq.~\eqref{eq:disorder_renorm} in the two regions $\Delta_0>3/2$ and $\Delta_0<3/2$. 

\subsubsection{Weak interactions and disorder: $\Delta_0>3/2$}
From Eq.~\eqref{eq:disorder_renorm}, it follows that 
\begin{align}
\label{eq:disorder_renorm2}
D(L)=D_0\left(\frac{L}{a}\right)^{3-2\Delta_0},
\end{align}
where $D_0\equiv D(L=a)$ is the bare (non-renormalized) disorder strength, assumed to be weak: $D_0\ll1$. For $\Delta_0>3/2$, $D$ decreases upon renormalization towards lower energies (i.e., with increasing $\ell$) and the edge is in the free phase, described by Eq.~\eqref{eq:KFree} or~\eqref{eq:Kfree2}. Tunneling between the edge modes remains weak at all energies. This tunneling exchanges charge and heat among the edge channels which causes equilibration. We denote the characteristic length scales for charge and heat equilibration as $\ell_{\rm C}$ respectively $\ell_{\rm Q}$. They are both non-universal as they depend on microscopic details such as disorder, interactions, and the temperature. Generically, a FQH edge is characterized by two such length scales for each pair of edge modes~\cite{Spanslatt2019,Asasi2020}. However, only tunneling between counter-propagating channels can partition charge and energy flows and influence transport coefficients. The quantities $\ell_{\rm C}$ and $\ell_{\rm Q}$ should then be understood as the dominating ones in the full set.

The temperature scalings of $\ell_{\rm C}$ and $\ell_{\rm Q}$ are expected to be the same~\cite{Protopopov2017}. In principle, strong interactions can cause parametrically different equilibration lengths~\cite{Srivastav2021}, but that case is excluded for the case $\Delta_0>3/2$. In the remainder of this subsection, as we are interested in only the temperature scaling, we treat the length scales on the same footing by setting
\begin{align}
	\ell_{\rm C}\sim \ell_{\rm Q}\sim \ell_{\rm eq},
\end{align}
and $\ell_{\rm eq}$ is to be understood as meaning both $\ell_{\rm C}$ and $\ell_{\rm Q}$.

To find the temperature scaling of $\ell_{\rm eq}$, we first note that with weakening disorder, the RG flow is cut at the thermal length
\begin{align}
	L_T\sim T^{-1}.
\end{align}
At this scale, the disorder has, according to Eq.~\eqref{eq:disorder_renorm2}, the strength
\begin{align}
\label{eq:WeakScaling1}
D(L_T) = D_0\left(\frac{L_T}{a}\right)^{3-2\Delta_0}\sim T^{2\Delta_0-3}.	
\end{align}
At larger length scales $L>L_T$, $D(L)$ scales linearly (trivially) in $L$ as
\begin{align}
\label{eq:WeakScaling2}
	D(L)\sim \frac{L}{L_T}D(L_T)\sim \frac{L}{L_T}D_0 \left(\frac{L_T}{a} \right)^{3-2\Delta_0},
\end{align}
where used Eq.~\eqref{eq:WeakScaling1} in the final step. The equilibration length is defined as
\begin{align}
	D(\ell_{\rm eq})\sim 1,
\end{align}
which by use of Eqs.~\eqref{eq:WeakScaling1} and~\eqref{eq:WeakScaling2} becomes
\begin{align}
\label{eq:equilibration_length}
	\ell_{\rm eq} \sim D_0^{-1}a\left(\frac{L_T}{a}\right)^{2\Delta_0-2}\sim T^{2-2\Delta_0}.
\end{align} 
Note that $2-2\Delta_0<0$ for weak interactions $\Delta_0>3/2$ so that both $\ell_{\rm C}$ and $\ell_{\rm Q}$ increase with decreasing temperature~\cite{Kane1995b,Srivastav2021}. 

We can then identify two possible regimes in the free phase:
\begin{itemize}
	\item At lowest temperatures, such that $L\ll \ell_{\rm C},\ell_{\rm Q}\sim T^{2-2\Delta_0}$, the three modes in the free phase are neither charge nor heat equilibrated. The temperature scaling exponent depends non-universally on inter-channel interactions and velocities.  We call this regime $\mathcal{II}$: the regime of absent localization and poor equilibration.   
	 
	\item At higher temperatures: $\ell_{\rm C},\ell_{\rm Q}\ll L$ the three edge modes become fully equilibrated, which we denote as regime $\mathcal{III}$. Here, the edge is expected to be governed by hydrodynamic behaviour due to inter-channel scattering~\cite{Kane1995}. 
\end{itemize}

We depict regimes $\mathcal{II}-\mathcal{III}$ for the free phase in Fig.~\ref{fig:phasediagram}\textcolor{blue}{(c)}.

\subsubsection{Strong interactions and disorder: $\Delta_0<3/2$}
For $\Delta_0<3/2$, Eq.~\eqref{eq:disorder_renorm} shows that $D$ grows during renormalization. The characteristic length scale where the disorder becomes strong is defined as~\cite{Giamarchi1988,Gornyi2007} 
\begin{align}
	D(\xi_0)\sim 1.
\end{align}
Using this condition in Eq.~\eqref{eq:disorder_renorm} gives the characteristic zero temperature localization length
\begin{align}
\xi_0 \propto aD_{0}^{-1/(3-2\Delta_0)}.
\end{align}
Approaching the critical value $\Delta_0=3/2$ from below, $\xi_0$ diverges, and no localization is possible for $\Delta_0>3/2$ as expected. 

Remarkably, we find that the strongly interacting regime, $\Delta_0<3/2$, is split into two parameter regions with very distinct behaviour:  Eq.~\eqref{eq:equilibration_length} implies that $\ell_{\rm eq}$ increases with increasing $T$ if $\Delta_0<1$. In contrast, when $\Delta_0>1$ $\ell_{\rm eq}$ decreases with increasing $T$. These two distinct cases are depicted in Figs.~\ref{fig:phasediagram}\textcolor{blue}{(a),(b)}, respectively. The existence of the regime $\Delta_0<1$ is a striking manifestation of the fact that the $\nu=9/5$ edge is T-unstable and thus susceptible to the binding transition. This is a key result of this paper.

Let us now analyze the onset of edge localization.  At finite temperature, the disorder is dressed by temperature dependent Friedel oscillations and $\xi_0$ is replaced by $\ell_{\rm eq}$ in Eq.~\eqref{eq:equilibration_length}. The parameter $\ell_{\rm eq}$ thus plays the role as a finite temperature localization length. The temperature scale $\tilde{T}$ where the localization becomes strong is determined by the condition
\begin{align}
\ell_{\rm eq}(\tilde{T}) \sim \ell_\varphi(\tilde{T})	
\end{align}
where $\ell_\varphi$ is a characteristic phase coherence length. For this length, we use the estimate~\cite{Gornyi2007}
\begin{align}
\label{eq:localization_onset}
	\ell_\varphi \sim L_T \sim T^{-1}.
\end{align}
It is important to note that this estimate can only be true for strong interactions (which is assumed here), as it is clear that non-interacting electrons are subject to localization independently of the temperature: $\ell_\varphi = \infty$. Hence, we expect also that~\cite{Gornyi2007} 
\begin{align}
	\tilde{T}\propto (\Delta_0-1)^{-2},
\end{align}
since $\Delta_0=1$ at vanishing interaction. Here, the exponent $-2$ follows from a more detailed analysis~\cite{Gornyi2007}. The scale $\tilde{T}$ marks the onset where weak localization corrections to the conductivity start to dominate over the classical Drude contribution and, as mentioned above, signals Anderson (i.e., strong) localization.

We now analyze the edge under the assumption that $\Delta_0<3/2$. Three possibles regimes appear:
\begin{itemize}
	\item At lowest $T\ll \tilde{T}$ and $L>\ell_{\rm eq},\xi_0$, the edge is in the strongly localized or bound phase~\eqref{eq:Kbound}. This is depicted as regime $\mathcal{I}$ in Fig.~\ref{fig:phasediagram}\textcolor{blue}{(a)-(b)}. 
	
	\item For $L<\ell_{\rm eq},\xi_0$ for all $T$, disorder remains weak and no localization or equilibration occurs. We call this regime $\mathcal{II}$.
	
	\item For $L>\ell_{\rm eq},\xi_0$ and $T>\tilde{T}$, dephasing suppresses the localization and at the same time, the three edge modes are equilibrated. This is regime $\mathcal{III}$. 
\end{itemize}

The conclusion of the present analysis is that the bound phase is characterized by three transport regimes: the localization regime $\mathcal{I}$, and non-equilibrated and equilibrated regimes $\mathcal{II}$ and $\mathcal{III}$, respectively. This stands in stark contrast to the free phase which exhibit only regimes $\mathcal{II}-\mathcal{III}$. The localized regime $\mathcal{I}$ in the bound phase is depicted as the yellow regions in Fig.~\ref{fig:phasediagram}. With this phase diagram in mind, we now move on to a transport analysis for the regimes $\mathcal{I}-\mathcal{III}$.

\section{\label{sec:Signatures}Transport signatures of the binding transition}
Electrical and thermal transport on FQH edges are usually quantized, reflecting a non-trivial bulk topological order~\cite{Wen1990a}. Specifically, the electrical Hall and two-terminal conductances, $G_H$ and $G$, are commonly proportional to the bulk filling factor [see Eq.~\eqref{eq:nu}]
 \begin{equation}
G_H = \nu \frac{e^2}{h}, \qquad G = |G_H|.
\label{eq:G}
\end{equation}
By contrast, thermal Hall and two-terminal conductances, $G^Q_H$ and $G^Q$ are determined by $\nu_Q$ [see Eq.~\eqref{eq:nuQ}] as
\begin{equation}
G^Q_H = \nu_Q\kappa T, \qquad G^Q = |G^Q_H|,
\label{eq:GQ}
\end{equation}
where the heat conductance quantum $\kappa T = \pi^2 k_{\rm B}^2 T/(3h)$, in which $T$ is the temperature, and $k_{\rm B}$ and $h$ are the Boltzmann and Planck constants, respectively. 

 For edges with counter-propagating channels, the quantizations~\eqref{eq:G} and~\eqref{eq:GQ} hold only in the transport regime of full edge channel equilibration \cite{Protopopov2017,Nosiglia2018}. To observe deviations from~\eqref{eq:G} and~\eqref{eq:GQ} requires poor equilibration, which can be achieved by very low temperatures, strong inter-channel interactions~\cite{Srivastav2021,Melcer2022}, very small inter-contact distance, or a detailed control over the inter-channel tunneling strength~\cite{Cohen2019}. Since $\nu_Q$ takes negative values on some edges (e.g., at filling $\nu=3/5$), it has further been predicted that heat may flow in the opposite direction of the charge~\cite{Kane1997}. However, most experiments measure $|\nu_Q|$. It was therefore proposed that the direction of heat flow (i.e., the sign of $\nu_Q$) is in direct correspondence with the scaling behaviour of the electrical shot noise with the edge length $L$~\cite{Park2019,Spanslatt2019,Spanslatt2020,Park2020NAB,Hein2022Nov}. These insights have led to a deeper understanding of the FQH edge structure. In particular, recent measurements of the heat conductance~\cite{Banerjee2018,Dutta2022Sep} and noise~\cite{Dutta2022}  now strongly point towards GaAs/AlGaAs hosting the non-Abelian particle-hole-Pfaffian edge structure~\cite{Fidkowski2013,Son2015,Zucker2016,Antonic2018} at filling  $\nu=5/2$~\cite{Willet1987,Moore1991}.

As described in Sec.~\ref{sec:BindingTheory}, the binding transition preserves the topological transport coefficients $\nu$ and $\nu_Q$. It is therefore clear that charge and heat conductances in the fully equilibrated regime cannot distinguish between the free~\eqref{eq:KFree} and bound~\eqref{eq:Kbound} edge phases. Indeed, in both phases, the equilibrated transport coefficients are given as
\begin{align}
	& G_H = G = \frac{9}{5}\frac{e^2}{h},\\
	& G^Q_H = G^Q = 1\kappa T.
\end{align}
However, by building upon the results in Sec.~\ref{sec:scaling} we will next argue  that signatures of the binding transition can be deduced from edge transport experiments by accessing regimes with absent equilibration. 

\subsection{Two-terminal charge conductance}
\label{sec:G}

\begin{figure}[t!]
\begin{center}
\includegraphics[width=1.0\columnwidth]{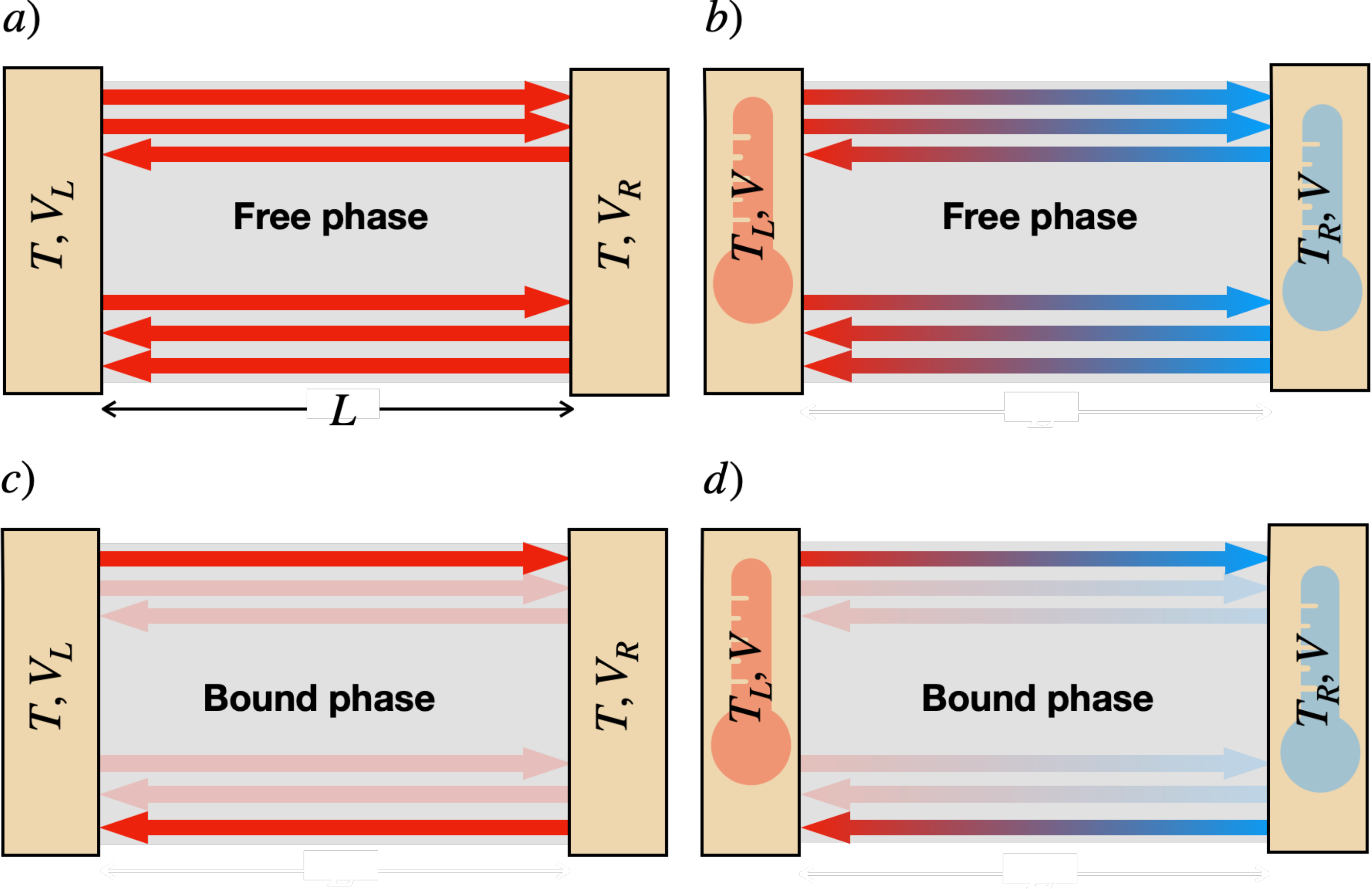}
\caption{\label{fig:TwoTerminalSetup}Two-terminal setup for depicting edge transport at filling $\nu=9/5$ in the free (a-b) and bound (c-d) edge phases. The device length is $L$. Two contacts at different potentials $V_L\neq V_R$ but the same temperature $T$ allows extraction of the two-terminal conductance $G$ (left figures). For contacts at the same potential $V_L=V_R=V$ but different temperatures $T_L\neq T_R$ (right figures) one can determine the two-terminal heat conductance $G^Q$.}
\end{center}
\end{figure}

We first consider the two terminal charge conductance $G$ (see Figs.~\ref{fig:TwoTerminalSetup}\textcolor{blue}{a}~and~\ref{fig:TwoTerminalSetup}\textcolor{blue}{c}). As follows, we assume for simplicity a sharp enough edge potential such that no edge reconstruction occurs. We also assume that in contact regions, screening causes the inter-channel interactions to vanish. 
 
\subsubsection{The free phase}
When the edge is in the free phase, $\Delta_0>3/2$, we expect with decreasing system size $L$ and/or decreasing temperature $T$, a crossover in $G$. When $L\gg \ell_{\rm C}$, i.e., for equilibration (i.e., regime $\mathcal{III}$, cf. Sec.~\ref{sec:scaling}) we expect $G/(e^2/h)=\nu=9/5$ in accordance with Eq.~\eqref{eq:G}. Since $\Delta_0>3/2$, we have from Eq.~\eqref{eq:equilibration_length} that $\ell_{\rm C}$ increases with decreasing $T$. When $L\sim \ell_{\rm C}$, charge begins to propagate upstream, which increases $G$. In the limit of $L\ll \ell_{\rm C}$, i.e., no equilibration (regime $\mathcal{II}$), the upstream channels transport charge upstream and also remain in equilibrium with the contact they were emitted from. The individual channel conductance contributions then add~\cite{Protopopov2017,Spanslatt2021Contacts}, i.e., $G/(e^2/h)=1+1+1/5=11/5$. Hence, we expect the conductance characteristics
\begin{align}
\label{eq:GFree}
	G/(e^2/h)= 11/5 \rightarrow 9/5,
\end{align}
with increasing $L$ and/or $T$ in the free edge phase. This corresponds to moving from the blue to the red region in Fig.~\ref{fig:phasediagram}\textcolor{blue}{(c)}.

\subsubsection{The bound phase}
In the localized regime $\mathcal{I}$, only a single channel transports charge over distances $L>\ell_{\rm C},\xi_0$. Hence, Eq.~\eqref{eq:G} gives the charge conductance $G/(e^2/h)=9/5$. The analysis of regimes $\mathcal{II}$ and $\mathcal{III}$ proceed just as for the free phase and give charge conductances $G/(e^2/h)=11/5$ and $G/(e^2/h)=9/5$, respectively.

We next analyze two important situations. For $1<\Delta_0<3/2$, Fig.~\ref{fig:phasediagram}\textcolor{blue}{(b)} indicates that, for fixed $L>\ell_{\rm C},\xi_0$ the conductance remains at $G/(e^2/h)=9/5$ with decreasing $T$, i.e., transitioning from regime $\mathcal{III}$ to regime $\mathcal{I}$. However, for $L<\ell_{\rm C},\xi_0$, the conductance increases from $G/(e^2/h)=9/5$ to $11/5$ since one channel begins to conduct more and more charge upstream. The other case is $\Delta_0<1$, see  Fig.~\ref{fig:phasediagram}\textcolor{blue}{(a)}. Just as in the previous case, a transition between regimes $\mathcal{I}$ and $\mathcal{III}$ is not visible in the charge conductance which remains at $G/(e^2/h)=9/5$. This is the blue, dashed line in Fig.~\ref{fig:phasediagram}\textcolor{blue}{(a)} However, we see that it is possible to crossover directly between regimes $\mathcal{I}$ and $\mathcal{II}$. This is depicted as the red, dashed line. Such a transition would give rise to a change in the conductance
\begin{align}
G/(e^2/h)=11/5 \rightarrow 9/5,
\end{align}
with \textit{decreasing} temperature, which is a quite unusual situation, only possible due to the existence of the localized regime. Observing this crossover would be a strong hallmark of edge localization and the binding transition. 

\subsection{Two-terminal heat conductance}
\label{sec:GQ}
Here, we consider the setup in Figs.~\ref{fig:TwoTerminalSetup}\textcolor{blue}{b}~and~\ref{fig:TwoTerminalSetup}\textcolor{blue}{d} and analyze the two-terminal heat conductance $G^Q$. We do note that state-of-the-art measurements of $G^Q$ use a different geometry (see, e.g., Ref.~\cite{Jezouin2013}). This does not however change the validity of the results in this section. 

In contrast to the topological quantization \eqref{eq:GQ}, in the regime of vanishing heat equilibration, $G^Q$ becomes proportional to the total number of edge channels
\begin{align}
\label{eq:GQNEQ}
G^Q = (n_d + n_u)\kappa T.
\end{align} 
This holds under the condition $L<L_T$~\cite{Krive1998}, which we assume is fulfilled as follows. Deviations from this assumption is commented upon below.

\subsubsection{The free phase}
Abelian FQH edge channels carry the same heat conductance regardless of their filling factor discontinuity $\delta \nu_{i}$ and charge $t_i$. Simple channel counting gives, for the free phase, the crossover 
\begin{align}
G^Q/(\kappa T)=3\rightarrow 1.	
\end{align}
with decreasing $L$ and/or $T$.  This corresponds to  the two limiting values~\eqref{eq:GQ} and~\eqref{eq:GQNEQ} when moving from regime $\mathcal{III}$ to $\mathcal{II}$. Analogously to the charge transport, the crossover is governed by a characteristic, non-universal, heat equilibration length $\ell_{\rm Q}$ [$\sim \ell_{\rm eq}$ in Fig.~\ref{fig:phasediagram}\textcolor{blue}{(c)}]. 

\subsubsection{The bound phase}
Also in the bound phase, we have in regime $\mathcal{III}$ a heat conductance $G^Q/(\kappa T)=1$ according to Eq.~\eqref{eq:GQ}. In regime $\mathcal{II}$, Eq.~\eqref{eq:GQNEQ} gives $G^Q/(\kappa T)=3$. For $1<\Delta_0<3/2$, Fig.~\ref{fig:phasediagram}\textcolor{blue}{(b)} indicates that, for fixed $L>\ell_{\rm Q},\xi_0$ the heat conductance remains at $G^Q/(\kappa T)=1$ with decreasing $T$. For $L<\ell_{\rm Q},\xi_0$, the heat conductance instead increases from $G^Q/(\kappa T)=1$ to $G^Q/(\kappa T)=3$. For strong interactions, $\Delta_0<1$, the transition between regimes $\mathcal{I}$ and $\mathcal{III}$ is not visible as $G^Q/(\kappa T)=1$ across the transition; the blue, dashed line in Fig.~\ref{fig:phasediagram}\textcolor{blue}{(a)}. Crossing over from $\mathcal{II}$ to $\mathcal{I}$ directly (the red, dashed line) yields the crossover
\begin{align}
G^Q/(\kappa T) =3 \rightarrow 1,
\end{align}
with \textit{decreasing} temperature, due localization. Similarly to the charge conductance, such an unusual crossover is a hallmark of localization on the edge. 

Finally, we comment on the situation $L>L_T$. Then, plasmon scattering on interfaces between edges and contacts lead to quantum interference effects~\cite{Krive1998}. This interference reduces the conductance from the value in Eq.~\eqref{eq:GQNEQ}. For example, the strongly interacting $\nu=2/3$ edge with $L_T<L<\ell_{\rm eq}$ produces a heat conductance $G^Q/(\kappa T)=1$~\cite{Protopopov2017,Melcer2022}. A prerequisite for this interference effect is the presence of counter-propagating channels which do not thermally equilibrate. The only regime where the interference effect therefore could potentially have an impact is regime $\mathcal{II}$, where it would reduce $G^Q/(\kappa T)$ from $3$ to slightly lower values. 

\subsection{Shot noise on a voltage biased edge segment}
\label{sec:2TNoise}

\begin{figure}[t!]
\begin{center}
\includegraphics[width=1.0\columnwidth]{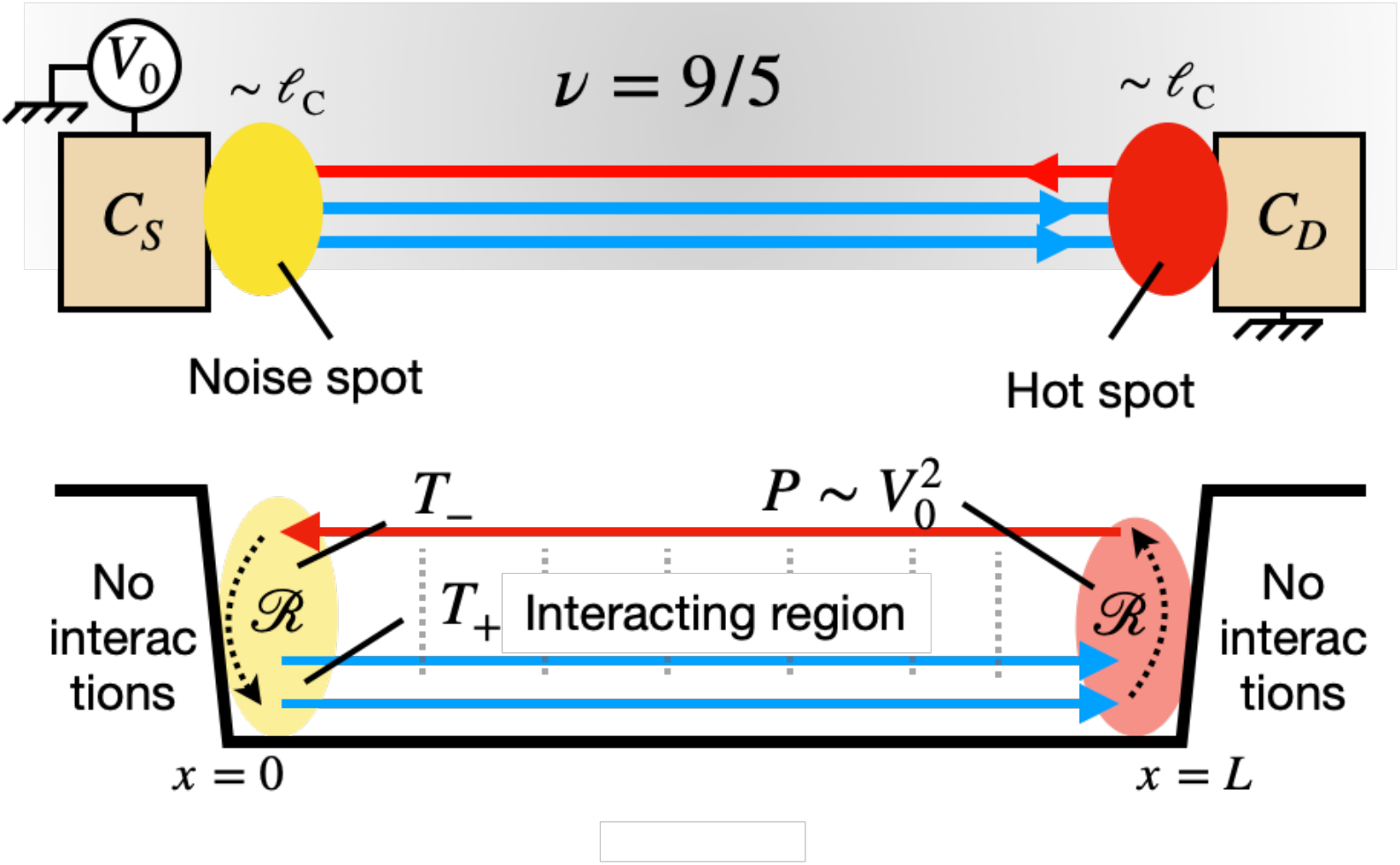}
\caption{\label{fig:NoiseSegment} Top panel: noise generation in regime of full charge equilibration but no thermal equilibration [condition~\eqref{eq:noisecondition}] in regime $\mathcal{II}$ (cf. Tab.~\ref{tab:TransportTab}). With voltage bias $V_0$, heat is generated by the voltage dropping in a region of size $\sim \ell_{\rm C}$ to the \textit{right} contact: the hot spot. The dissipated power $P\sim V_0^2$ [see Eq.~\eqref{eq:PowerDiss}]. Dc noise by partitioning of the downstream (left to right direction) charge current is only generated in a region of size $\sim \ell_{\rm C}$ close to the \textit{left} contact. This happens if heat is transported upstream (right to left) via the upstream channel, which is here only possible in the absence of thermal equilibration.  Bottom panel: By dissipation in the hot spot and reflection of plasmons at contacts (reflection coefficient $\mathcal{R}$), the downstream and upstream modes acquire steady state temperatures $T_+$ respectively $T_-$ in the noise spot.}
\end{center}
\vspace{-0.75cm}
\end{figure}

Here, we analyze the shot (or excess dc) noise, $S$, generated on a single edge segment, of length $L$, bridging two contacts (see Fig.~\ref{fig:NoiseSegment}). This setup was studied theoretically in Refs.~\cite{Park2019,Spanslatt2019,Spanslatt2020,Park2020NAB} and has been realized experimentally for both conventional~\cite{Kumar2022} and interfaced~\cite{Dutta2022} edge structures. 

Under conditions of strongly equilibrated charge transport, $\ell_{\rm C}\ll L$, noise in this setup is generated by an interplay between the charge and heat transport characteristics: When a charge current is driven between the two contacts, heat is generated only near the drain contact (in a region called the hot spot; see right hand side in Fig.~\ref{fig:NoiseSegment}). This is a consequence of downstream ballistic charge transport due to the efficient charge equilibration. By contrast, excess noise can only be generated near the source contact (in a region called the noise spot, see left hand side in Fig.~\ref{fig:NoiseSegment}) due to thermal enhancement of current partitioning. Partitioning beyond this spot leads, due to repeated charge scattering and the chiral nature of the edge, to scattered particles ending up in the same contact and no noise is generated. Non-zero shot noise in the contacts is therefore possible only if (i) the edge hosts counter-propagating modes and (ii) there is an upstream heat flow from the hot spot to the noise spot. 

Let us apply this reasoning to the $\nu=9/5$ edge. In the localized regime $\mathcal{I}$ there is no upstream heat flow and therefore no noise; $S=0$. In regime $\mathcal{III}$, the edge is fully thermally equilibrated and the upstream heat flow as well. The noise is then exponentially suppressed in $L$: $S\propto\exp[-L/\ell_{\rm C}]\simeq 0$. These two regimes stand in stark contrast to regime $\mathcal{II}$, under the additional condition
\begin{align}
\label{eq:noisecondition}
\ell_{\rm C}\ll L\ll \ell_{\rm Q}.
\end{align}
In this case, the edge has three charge equilibrated but not thermally equilibrated channels, conditions which have been experimentally observed~\cite{Melcer2022,Kumar2022}. Upstream heat transport is then possible. This heat reaches the noise spot through ballistic upstream flow and noise is generated. The condition~\eqref{eq:noisecondition} can arise due to strong interactions~\cite{Srivastav2021}, so we expect that it holds only for $\Delta_0\approx 0$, i.e., deep in the bound phase. 

Our goal in the remainder of this subsection is to estimate the magnitude of $S$ in regime $\mathcal{II}$, under the condition~\eqref{eq:noisecondition}. In our analysis, we consider a large voltage bias $eV_0\gg T$, which allows us to effectively set $T\approx 0$ in the following calculations. 

The shot noise in any of the two contacts [see Fig.~\ref{fig:NoiseSegment}] (equal due to current conservation) can be written as~\cite{Melcer2022}  
 \begin{align}
 	S &=\frac{2e^2}{h \ell_{\rm C}}\frac{ \nu_-}{ \nu_+} (\nu_+-\nu_-) \int_0^L dx\;\Lambda(x)e^{-\frac{2x}{\ell_{\rm C}}}. 	\label{eq:MicroNoise}
  \end{align}
  Here, $\nu_+$ and $\nu_-$ are the combined filling factor discontinuities of the downstream $(+)$ and upstream $(-)$ edge modes respectively. They satisfy the relation $\nu=\nu_+-\nu_-$ where $\nu$ is the bulk filling factor. For the $\nu=9/5$ edge in regime $\mathcal{II}$, we have $\nu_+=2$ and $\nu_-=1/5$. The exponential factor in the integral is a result of chiral, equilibrated charge transport as described above. It indicates that noise is dominantly generated in a region of size $\sim l^C_{\rm eq}$ close to the upstream contact, i.e.,  the noise spot.   

The key quantity to compute in Eq.~\eqref{eq:MicroNoise} is the local noise kernel
\begin{align}
\label{eq:Lambda}
    \Lambda(x) \equiv \frac{S_{\rm loc}[\delta V(x),T_+(x),T_-(x),\Delta]}{2g_{\rm loc}[\delta V(x),T_+(x),T_-(x),\Delta]}.
\end{align}
It is composed of $S_{\rm loc}$ and $g_{\rm loc}$ which is the local electron tunneling dc noise and the (dimensionless) tunneling conductance, respectively. Importantly, the conductance $g_{\rm loc}\propto\ell^{-1}_{\rm C}$, where the proportionality factor is the typical distance between scattering points~\cite{Park2019,Asasi2020}. Both $S_{\rm loc}$ and $g_{\rm loc}$ depend on microscopic details of the edge: Inter-channel interactions, the edge disorder strength, the local voltage difference between the modes $\delta V(x)$, and the effective temperatures $T_\pm(x)$ of downstream and upstream edge modes. Importantly, the interactions enter via the scaling dimension $\Delta$ of the most relevant inter-channel charge tunneling operator~\eqref{eq:tunnelingOperator}. 

Under the condition~\eqref{eq:noisecondition}, both the voltage difference and the temperatures are to excellent approximation constant across the noise spot: 
\begin{align}
\label{eq:noiseApprox}
	\delta V(x)\approx 0 \quad \text{and} \quad T_\pm(x)\approx T_\pm.
\end{align}
The first approximation holds because the channels equilibrate to the same voltage along the edge (except at the hot spot), whereas the second holds because of assumed poor thermal equilibration. With the approximations~\eqref{eq:noiseApprox} we can write 
\begin{align}
\label{eq:LambdaConst}
	\Lambda(x)\approx \Lambda(T_\pm, \Delta).
\end{align}

By using Eq.~\eqref{eq:LambdaConst} in Eq.~\eqref{eq:MicroNoise}, the integral can be trivially performed to give
\begin{align}
\label{eq:Final}
	S = \frac{9e^2}{50h}\Lambda(T_\pm,\Delta).
\end{align}	
We now want to find $\Lambda(T_\pm ,\Delta)$. The first step is to specify the edge tunneling operator, which we take as the null operator~\eqref{eq:Null_Operator} with $\mathbf{m}_1$ from  Eq.~\eqref{eq:95NullVectors}. We use the basis~\eqref{eq:KFree}. 

For this tunneling process, we next compute $\Lambda(T_\pm,\Delta)$ with the approach described in Ref.~\cite{Kumar2022} (see also Sec.~\ref{sec:QPC} below). For weak tunneling, a perturbative approach for the local noise and conductance gives
\begin{align}
	&S_{\rm loc} \propto 2\int dt\; \cos \left(\delta Vt\right)\langle \mathcal{V}^{}_{\mathbf{m}_1}(t,0)\mathcal{V}^\dagger_{\mathbf{m}_1}(0,0)  \rangle \\
	&g_{\rm loc} \propto i \partial_{\delta V}\left[ \int dt\; \sin \left(\delta Vt\right)\langle \mathcal{V}^{}_{\mathbf{m}_1}(t,0)\mathcal{V}^\dagger_{\mathbf{m}_1}(0,0)\rangle \right]_{\delta V\rightarrow 0}.
\end{align}
Here, the non-universal proportionality constant depends on a short distance cutoff, but importantly, the constant is the same for $S_{\rm loc}$ and $g_{\rm loc}$ and therefore cancels in $\Lambda$. Inserting the above expressions into Eq.~\eqref{eq:Lambda} and using $\delta V=0$ gives
\begin{align}
\label{eq:lambdaGF}
	\Lambda(T_\pm,\Delta) = \frac{\int dt\; \langle \mathcal{V}^{}_{\mathbf{m}_1}(t,0)\mathcal{V}^\dagger_{\mathbf{m}_1}(0,0)  \rangle}{i\int dt \; t\langle \mathcal{V}^{}_{\mathbf{m}_1}(t,0)\mathcal{V}^\dagger_{\mathbf{m}_1}(0,0)  \rangle}.
\end{align}
In Eq.~\eqref{eq:lambdaGF} the temperature dependence enters via the correlation functions
\begin{align}
\label{eq:VVCorr}
	\langle\mathcal{V}^{}_{\mathbf{m}_1}(t,0)\mathcal{V}^\dagger_{\mathbf{m}_1}(0,0)  \rangle \propto \prod_{j=1,2,3}\tilde{G}_j(t,0)^{\left((\mathbf{\tilde{m}}_1)_j\right)^2},
\end{align}
where the finite temperature Green's functions
\begin{align}
\label{eq:GFs}
	\tilde{G}_j(t,0) = \frac{\pi a T_j /\tilde{v}_j}{\sin \left[\frac{\pi T_j}{\tilde{v}_j}(a-i\tilde{v}_jt) \right] }, \quad j\in \lbrace 1,2,3\rbrace,
\end{align}
Note that the Green's functions are those of of the modes in the diagonal basis~\eqref{eq:chiLLDiagonal}. From Eq.~\eqref{eq:GenScalingDim}, we have that the exponents in~\eqref{eq:VVCorr}  satisfy
\begin{align}
	\left((\mathbf{\tilde{m}}_1)_1\right)^2+\left((\mathbf{\tilde{m}}_1)_2\right)^2+\left((\mathbf{\tilde{m}}_1)_3\right)^2=2\Delta.
\end{align}
The formula for the noise~\eqref{eq:MicroNoise} assumes that all downstream modes have the temperature $T_+$, and the upstream modes are at $T_-$. We thus set
\begin{subequations}
\begin{align}
	&T_1 = T_2= T_+,\\ 
	&T_3=T_-.
\end{align}
\end{subequations}
By plugging these temperatures into the Green's functions~\eqref{eq:GFs}, and expanding to leading order in $a$, we can cast Eq.~\eqref{eq:lambdaGF} on the form
\begin{align}
	&\Lambda(T_\pm,\Delta) = \notag\\
&\frac{\int dz \sin \left[\frac{T_+}{T_-}(\frac{\pi}{2}+iz) \right]^{-\delta_+}\cosh \left[z\right]^{-\delta_-}}{
	\int dz \left(\frac{1}{2T_-}+\frac{iz}{\pi T_-}\right)\sin \left[\frac{T_+}{T_-}(\frac{\pi}{2}+iz) \right]^{-\delta_+}\cosh \left[z \right]^{-\delta_-}},
\end{align}
where the exponents
\begin{subequations}	
\label{eq:deltapm}
\begin{align}
	&\delta_+\equiv\left((\mathbf{\tilde{m}}_1)_1\right)^2+\left((\mathbf{\tilde{m}}_1)_2\right)^2,\\
	&\delta_-\equiv\left((\mathbf{\tilde{m}}_1)_3\right)^2.
\end{align}
\end{subequations}

Our next step is to find $T_\pm$ in terms of the bias voltage $V_0$. When $n_+$ downstream and $n_-$ upstream edge modes are not thermally equilibrated, the local downstream and upstream temperatures at the noise spot, $T_\pm$, were computed in Ref.~\cite{Kumar2022}. They are given as
\begin{subequations}
\label{eq:Tpm}
\begin{align}
	& k_{\rm B} T_{+} = \left(\frac{6hP}{\pi^2}\times \frac{\mathcal{R}(n_-+\mathcal{R})}{(n_++n_-)(n_+n_--\mathcal{R}^2)}\right)^{1/2},\\
	& k_{\rm B} T_{-} = \left(\frac{6hP}{\pi^2}\times \frac{n_+ (n_-+\mathcal{R})}{(n_++n_-)(n_+n_--\mathcal{R}^2)}\right)^{1/2}.
\end{align}
\end{subequations}
Here $\mathcal{R}\in[0,1]$ is the reflection coefficient between the contacts and the edge. This reflection depends explicitly on the sharp change in interaction strength between the contact region and the edge~\cite{Protopopov2017,Spanslatt2021Contacts,Kumar2022} (see Fig.~\ref{fig:NoiseSegment}). We have assumed that both contacts have the same $\mathcal{R}$. Eq.~\eqref{eq:Tpm} also includes $P$, the power dissipated in the hot spot, which was computed in Ref.~\cite{Spanslatt2020} as
\begin{align}
\label{eq:PowerDiss}
	P = \frac{e^2 V_0^2}{2h}\times \frac{(\nu_+-\nu_-)\nu_-}{\nu_+},
\end{align}	
under the assumption~\eqref{eq:noisecondition}. For the $\nu=9/5$ edge in regime $\mathcal{II}$, we have $\nu_+=2$, $\nu_-=1/5$, $n_+=2$, and $n_-=1$. Plugging these numbers into Eqs.~\eqref{eq:Tpm} and~\eqref{eq:PowerDiss}, we find
\begin{align}
P=\frac{9e^2V_0^2}{100h},
\end{align}
and
\begin{subequations}
\label{eq:eigentemps}
\begin{align}
& k_{\rm B} T_+	=\frac{3eV_0}{5\pi}\sqrt{\frac{\mathcal{R}(1+\mathcal{R})}{4-2\mathcal{R}^2}},\\
& k_{\rm B} T_-	=\frac{3eV_0}{5\pi}\sqrt{\frac{1+\mathcal{R}}{2-\mathcal{R}^2}}.
\end{align}
\end{subequations}
For $\mathcal{R}=0$, which corresponds to vanishing edge interactions, we have $k_{\rm B}T_+=0$ and $k_{\rm B}T_-=3eV_0/(5\pi \sqrt{2})$. This is the situation where the downstream channels remain in equilibrium with the left contact (in our approximation, at $T=0$) and only the upstream channel is heated by dissipation at the hot spot. Even in the absence of thermal equilibration by edge impurities, a finite reflection probability at the two contacts distributes the hot spot power to all edge channels. 

\begin{figure}[t!]
\begin{center}
\includegraphics[width=1.0\columnwidth]{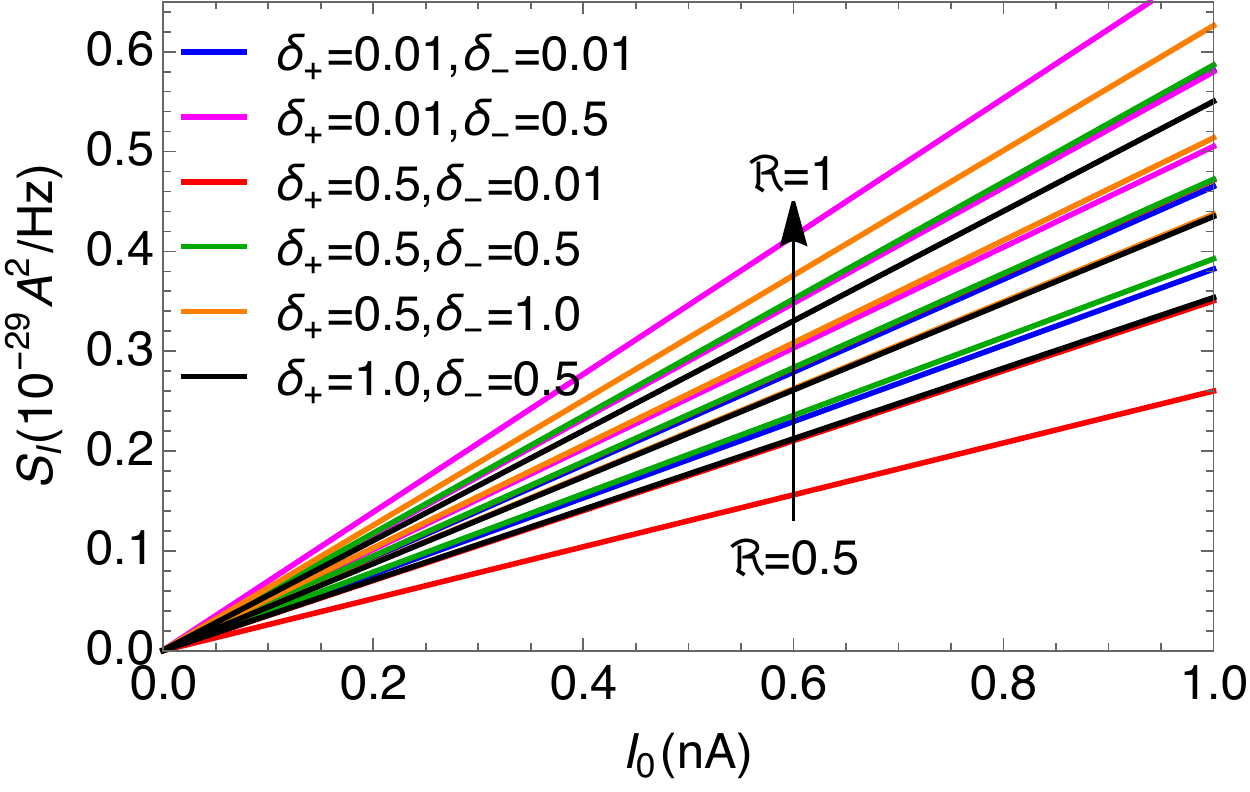}
\caption{\label{fig:Noise_vs_I0} Noise $S$ vs the bias current $I_0$ for different values of the scaling dimensions $\delta_\pm$ [see Eq.~\eqref{eq:deltapm}] and the edge contact reflection coefficient $\mathcal{R}$. The calculation is done for the $\nu=9/5$ edge under the condition~\eqref{eq:noisecondition} for the edge equilibration length scales.}
\end{center}
\vspace{-0.75cm}
\end{figure}
 
For simpler comparison with the experimental convention of plotting $S$ versus the source current $I_0$, we next convert the bias voltage $V_0$ to $I_0$ via
\begin{align}
	I_0=\frac{9}{5}\frac{e^2}{h}V_0,
\end{align}
which is valid under the condition~\eqref{eq:noisecondition}. 

We now have all ingredients to compute the noise in Eq.~\eqref{eq:Final} for a given bias current $I_0$. We evaluate the integrals in Eq.~\eqref{eq:lambdaGF} numerically for various values of $\mathcal{R}$ and $\delta_\pm$ and plot the result in Fig.~\ref{fig:Noise_vs_I0}. Since we expect the condition~\eqref{eq:noisecondition} to be obtained for strong interactions which also tend to increase $\mathcal{R}$, we limit the range of $\mathcal{R}$ to the range $\mathcal{R}\in[0.5,1]$, for consistency.  For experimental comparison, we have reinstated experimentally convenient units where noise is measured in $10^{-29}\rm{A}^2/Hz$ and currents in $\rm{nA}$. Qualitatively, we see that for fixed $\mathcal{R}$, the dependencies on $\delta_\pm$ are quite weak, whereas the dependence on $\mathcal{R}$ is pronounced. This can be understood on physical grounds since $\mathcal{R}$ strongly affects the temperatures $T_\pm$ at the noise spot. By inspection, we give the rough estimate
\begin{align}
	S \approx 0.25-0.7\times 10^{-29}\frac{\rm{A}^2}{\rm {nA}\rm{Hz}}.
\end{align}
This magnitude is around half of that detected for the $\nu=2/3$ edge in Ref.~\cite{Kumar2022}. The noise in regime $\mathcal{II}$ should therefore be detectable with present technology. 

\subsection{Shot noise in a QPC device}
\label{sec:QPC}
\begin{figure}[t]
\captionsetup[subfigure]{position=top,justification=raggedright}
\subfloat[]{
\includegraphics[width=0.5\columnwidth]{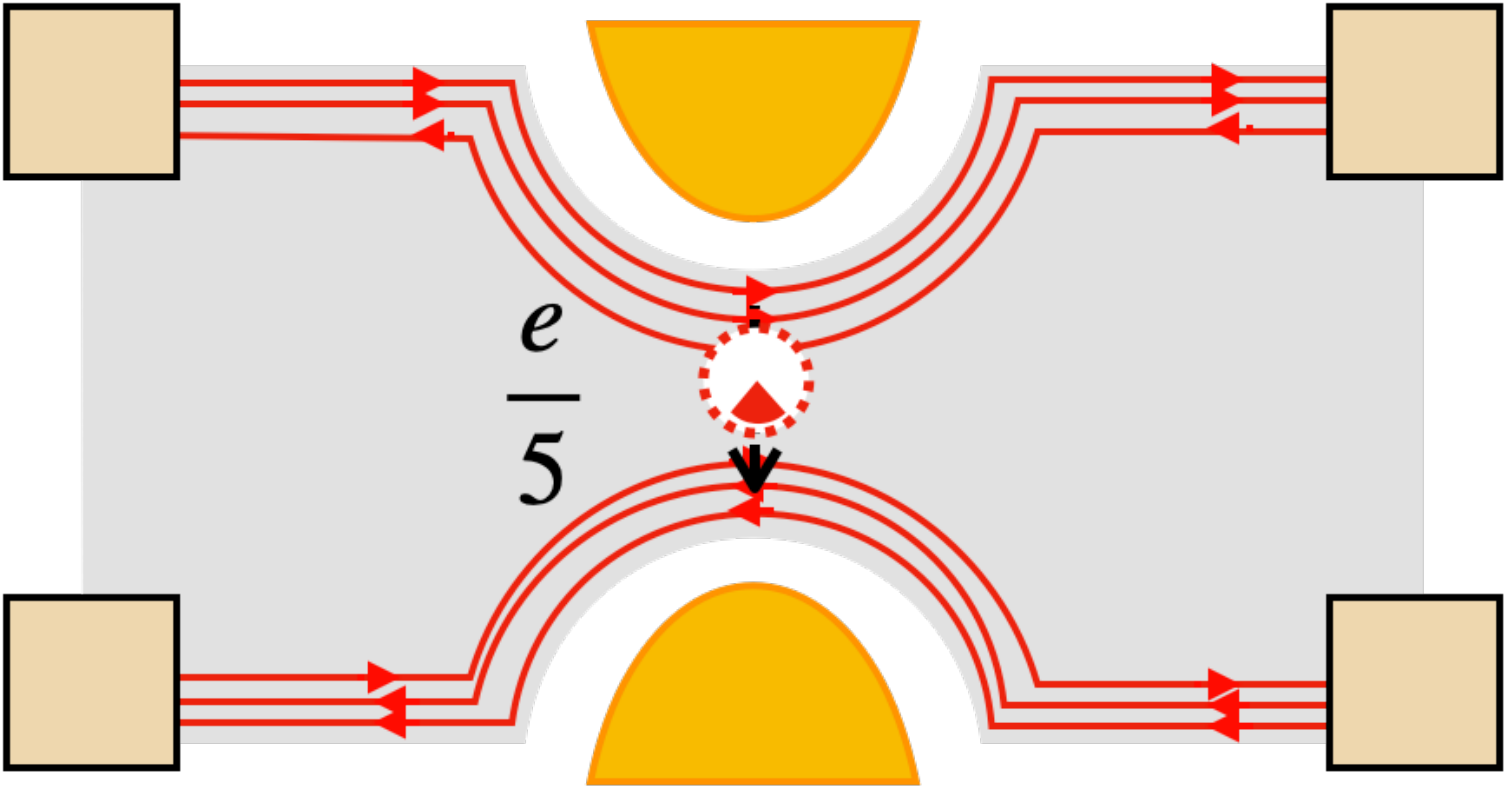}
\label{fig:QPC15}}
\subfloat[]{
\includegraphics[width =0.5\columnwidth]{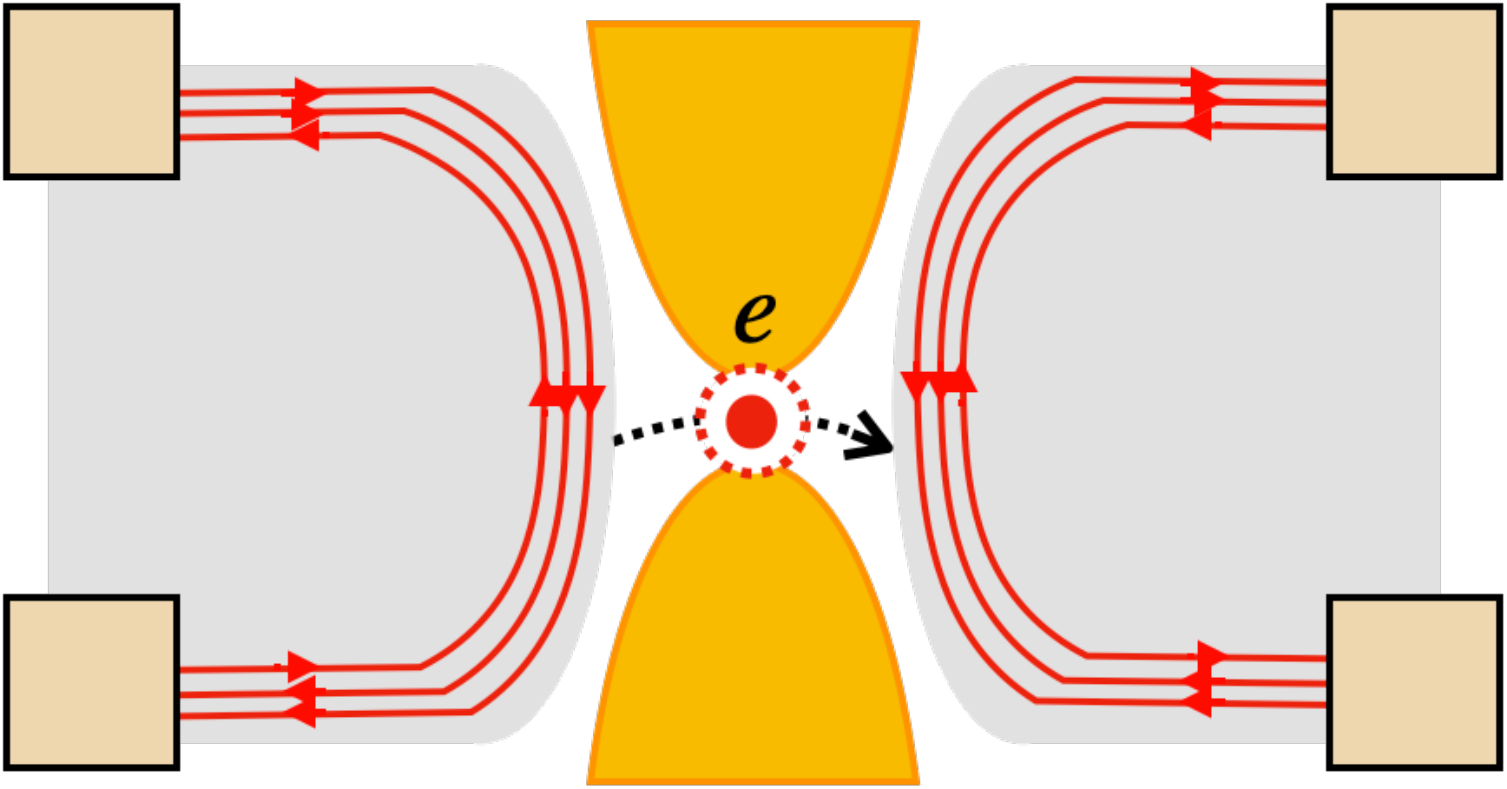}
\label{fig:QPC1}}
\\
\subfloat[]{
\includegraphics[width=0.5\columnwidth]{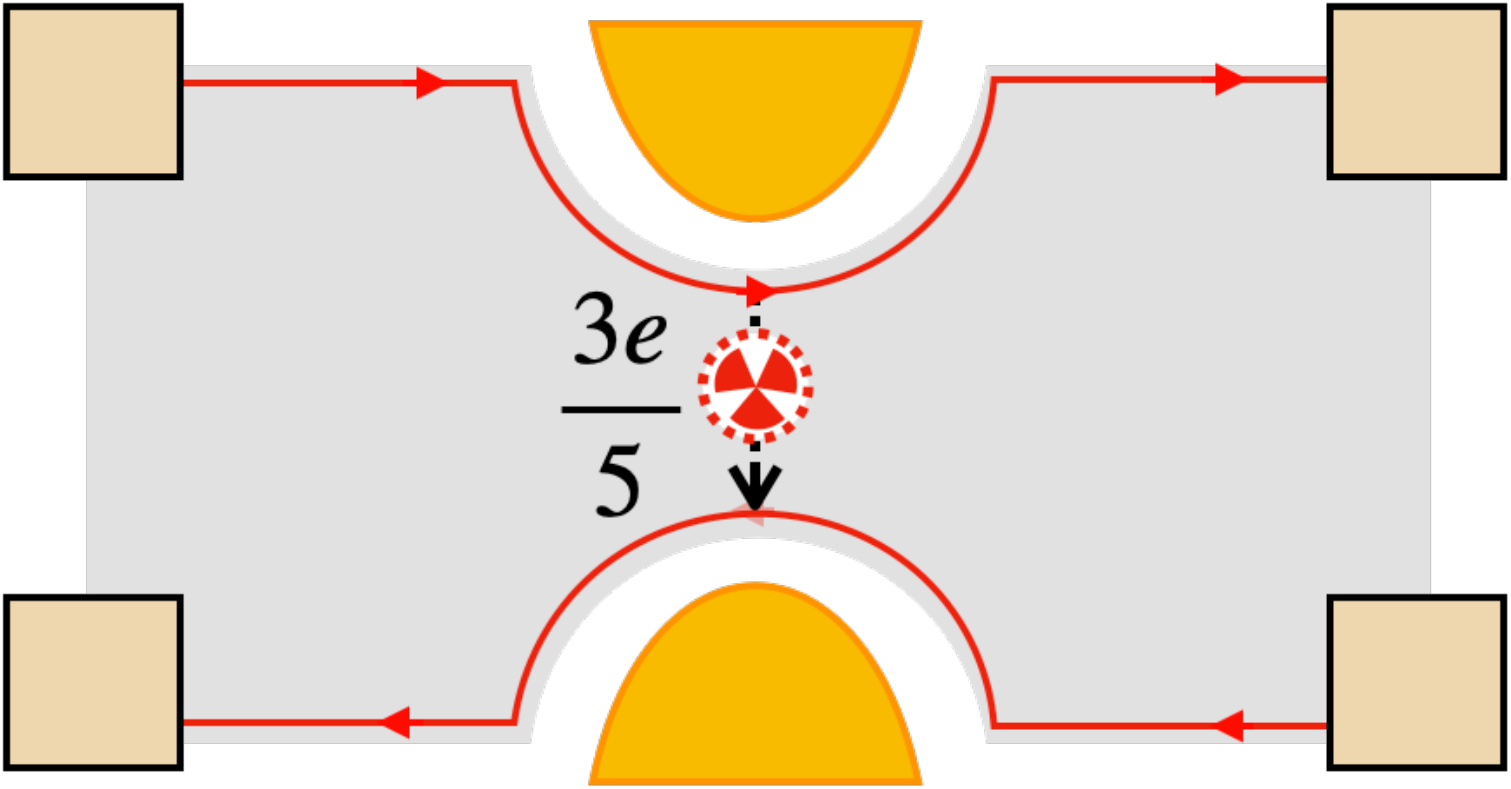}
\label{fig:QPC35}}
\subfloat[]{
\includegraphics[width =0.5\columnwidth]{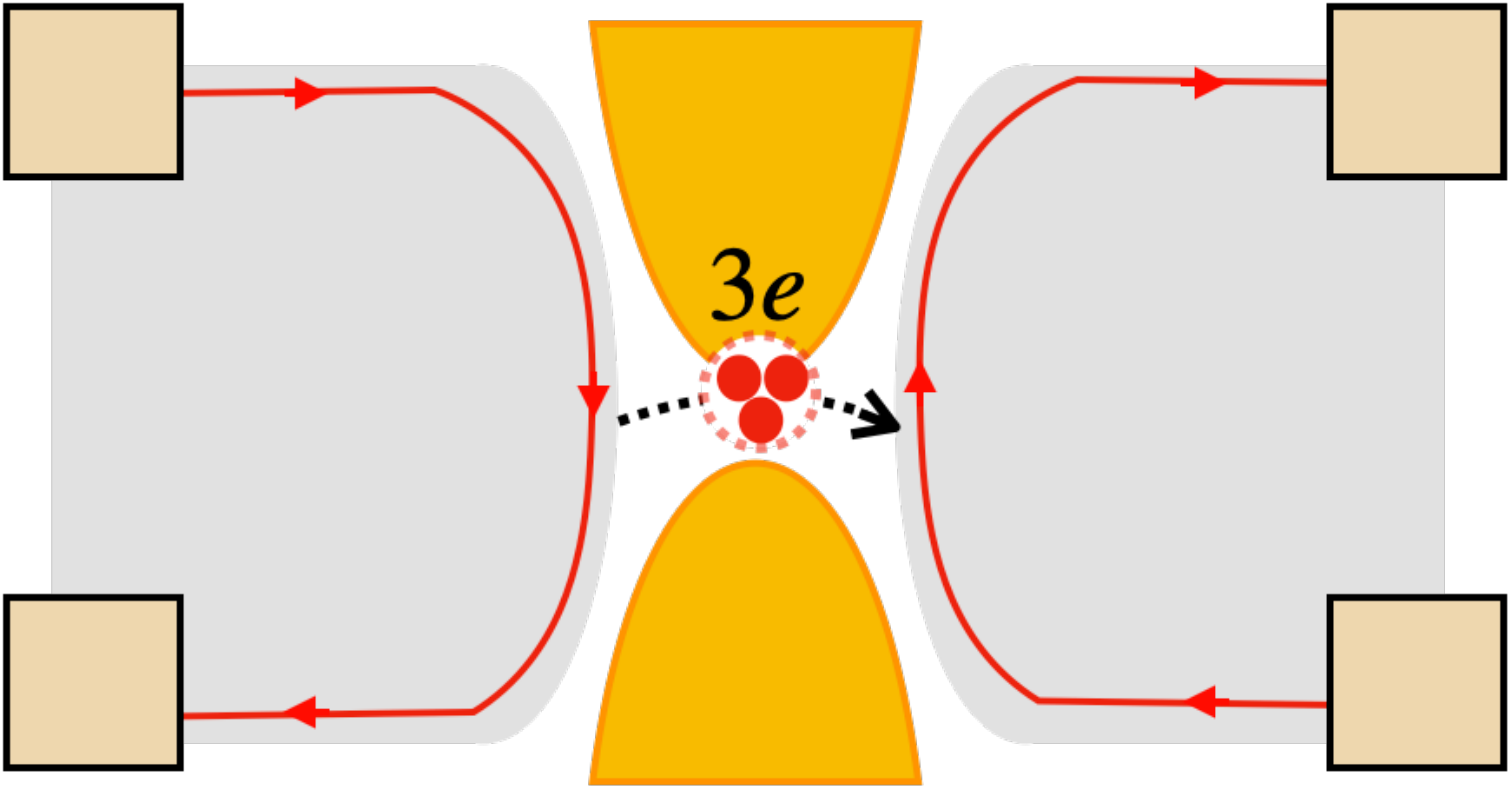}
\label{fig:QPC3}}
\caption{Schematics of a quantum point contact (QPC) device at filling $\nu=3/5$. (a)-(b): In the free phase, the weak back-scattering (WBS) regime favours tunneling of quasiparticles with charge $e/5$. For strong back-scattering (SBS), electrons with charge $e$ tunnel.  (c) In the bound phase, the WBS regime is dominated by tunneling of fractional charges $3e/5$. (d) In the SBS regime in the bound phase, single electron tunneling is strongly suppressed at low energies and three electron co-tunneling, i.e., charge packets of $3e$ is the most dominant process. All four types of charges can be detected in shot noise measurements.}
\end{figure}

As a complement to the transport characteristics in previous subsections, we here compute the electrical shot noise generated by current partitioning in a QPC device~\cite{KaneFisher1994Noise,Chamon1995,Fendley1995,Feldman2017}. We consider both the WBS regime (see Figs.~\ref{fig:QPC15} and \ref{fig:QPC35}), allowing tunneling of fractionally charged quasiparticles through the FQH bulk, and the SBS regime, which only allows electron tunneling across the non-topological vacuum (Figs.~\ref{fig:QPC1} and \ref{fig:QPC3}). 

We begin by deriving formulas for the current and shot noise in point tunneling between two counter-propagating edge channels. We then apply these formulas to obtain Fano factors for the bound and free phases, in both SBS and WBS regimes. We follow closely the perturbative Keldysh approach from Refs.~\cite{Spanslatt2021Contacts},~\cite{Nosiglia2018} and~\cite{Martin2005}. 

\subsubsection{Derivation of a general noise formula}
The edge channels in the QPC device is described by the effective Hamiltonian 
\begin{align}
\label{eq:BareHam}
	H_0 = \pi \int dx \Big[ v_1\frac{\rho_1^ 2 }{t_1^2}+v_2\frac{\rho_2^2}{t_2^2}\Big],
\end{align}
where $\rho_i=t_i\partial_x\phi_i/(2\pi)$ ($i=1,2$) are the neigbouring two counter-propagating charge densities (with velocities $v_i$) on each side of the constriction. For simplicity, we ignore the fully transmitted or reflected channels in the free phase. Their influence in the form of inter-channel interactions are discussed below Eqs.~\eqref{eq:Charge_Conservation},~\eqref{eq:F}, and~\eqref{eq:TunnelScaling}. We further ignore all interactions across the constriction. The bosons $\phi_i$ and their densities obey the following set of commutation relations
\begin{align}
\label{eq:BareComm}
&\left[\phi_i(x),\phi_j(x')\right] = i\pi\delta_{ij}\delta \nu_i \;\text{sgn}(x-x'),\\
	&\left[\rho_i(x),\rho_j(x')\right] = \frac{i}{2\pi}\delta_{ij}t_i^2\delta \nu_i  \partial_x \delta(x-x'),\\
	& \left[\phi_i(x),\rho_j(x')\right] = i\delta_{ij}t_i\delta \nu_i \delta(x-x').
\end{align}
Here, the filling factor discontinuities $\delta \nu_i$ and the charge vector entries $t_i$ are kept unspecified for full generality. To describe charge tunneling in the presence of a voltage bias, we introduce chemical potentials and point tunneling (at position $x=0$) with
\begin{align}
	&H_V = -\int dx \Big[\mu_1 \rho_1+\mu_2 \rho_2\Big],\\
	&H_{\tau} = \int dx \delta(x) \Big[ \Gamma_0 \mathcal{T}_{\mathbf{l}}(x) + h.c. \Big].
\end{align}
Here, $\Gamma_0$ is the tunneling amplitude and
\begin{align}
	\mathcal{T}_{\mathbf{l}}(x) = \frac{e^{il_1 \phi_1(x)+il_2 \phi_2(x)}}{2\pi a},
\end{align}
is a tunneling operator parametrized by $\mathbf{l}=(l_1,l_2)$ [cf. Eq.~\eqref{eq:tunnelingOperator}]. The length $a$ is our UV distance cutoff, e.g., the magnetic length. According to Eq.~\eqref{eq:Charge}, the charge created by $e^{i l_1 \phi_1}$ is
\begin{align}
\label{eq:q1}
	q_1 = t_1 \delta \nu_1 l_1.
\end{align}
Similarly, $e^{i l_2 \phi_2}$ creates charge
\begin{align}
\label{eq:q2}
	q_2=t_2 l_2\delta \nu_2.
\end{align}
Conservation of charge in the tunneling process restricts $l_1$ and $l_2$ such that
\begin{align}
\label{eq:Charge_Conservation}
 q_1=q_2\equiv q	
\end{align}
for a given set $\{t_1,t_2,\delta \nu_1 ,\delta \nu_2\}$. Note that since Eq.~\eqref{eq:Charge} is invariant under basis rotations, the charges $q_1$, $q_2$, and $q$ are all independent of possible interactions with fully transmitted and reflected channels in the free phase.

 Next, the channel $i$ current operator is defined as
\begin{align}
	I_i  &= \frac{d}{dt}\left( \int dx \rho_i \right) = \frac{1}{i}\left[ \int dx \rho_i, H_{0}+H_V+H_\tau \right] \notag\\
	&= \frac{1}{i}\left[ \int dx \rho_i, H_\tau \right],
	\end{align}
where we used that $\rho_i$ commutes with the quadratic $H_{0}$ and $H_V$. The commutator with $H_{\tau}$ yields
\begin{align}
	I_1 =-I_2\equiv I= i q \Gamma_0  \mathcal{T}_{\mathbf{l}}(0)+h.c.
\end{align}
In the interaction picture with $H_\tau$ as the interaction Hamiltonian, the operator $A$ time-evolve as
\begin{equation}
	A(t) = e^{i(H_0+H_V)t}Ae^{-i(H_0+H_V)t}.
\end{equation}
Applying this formula to $I$ and $H_\tau$, we find the time-evolutions
\begin{subequations}
	\begin{align}
		& H_\tau(t) = e^{-i\omega_0 t} \Gamma_0 \mathcal{T}_{\mathbf{l}}(0)+ h.c., \\
		& I(t) = i q e^{-i\omega_0 t}\Gamma_0 \mathcal{T}_{\mathbf{l}}(0) + h.c.,	
	\end{align}
\end{subequations}
where the characteristic ``Josephson frequency'' of the tunneling process
\begin{align}
\omega_0\equiv(t_1 l_1\delta \nu_1\mu_1-t_2l_2 \delta \nu_2\mu_2)=q(\mu_1-\mu_2)	.
\end{align}
In the second equality here, we used the charge conservation condition~\eqref{eq:Charge_Conservation}. The relation between $\omega_0$ and the voltage across the constriction reads $\hbar \omega_0 = qe(V_1-V_2)$.

Next, we compute the expectation value of the current $\langle I(t) \rangle$ on the Keldysh contour
\begin{equation}
	\langle I(t) \rangle = \frac{1}{2} \sum_{\lambda=\pm 1} \langle T_C \left[I(t^\lambda) \exp\left(-i\int_C dt_0 H_\tau(t_0^{\lambda'}) \right) \right] \rangle.
\end{equation}
Here, the Keldysh time-ordering operator, $T_C$ orders along the Keldysh contour $C$: $-\infty\rightarrow +\infty\rightarrow -\infty$. We denote the "upper" branch $-\infty\rightarrow +\infty$ by $\lambda=+1$ and lower branch $+\infty\rightarrow -\infty$ with $\lambda=-1$. The ordering acts according to $t^->t_0^+$ for all $t$ and $t_0$; $t^+>t_0^+$ for $t>t_0$; and $t^->t_0^-$ for $t<t_0$. Since our $H_\tau$ only depends on a single time argument, we have used a symmetric combination on both branches, compensated for with the factor of $1/2$. 

To second order in $\Gamma_0$, we find
\begin{align}
\label{eq:CurrentAverage}
	& \langle I(t) \rangle = \frac{q|\Gamma_0|^2}{2}\sum_{\lambda, \lambda'=\pm 1}\lambda' \int_{-\infty}^{\infty}dt_0
	\sin\big[\omega_0(t^\lambda-t_0^{\lambda'})\big]\times \notag \\
	&\langle T_C \mathcal{T}_{\mathbf{l}}(0,t^\lambda)\mathcal{T}_{\mathbf{l}}^\dagger(0,t_0^{\lambda'}) \rangle,
\end{align}
where we used that the tunneling operators are normal ordered. Next, we use the finite temperature Keldysh Green's functions for the tunneling operators
\begin{align}
\label{eq:GTVertex}
	&\langle T_C \mathcal{T}(0,t^\lambda)\mathcal{T}_{\mathbf{l}}^\dagger(0,t_0^{\lambda'}) \rangle = \frac{1}{(2\pi a)^{2}} \times \notag \\
	&  \prod_{j=1,2}\left(\frac{\pi a/(\beta_jv_j)}{\sin\left[\frac{\pi \left(a+i\chi_{\lambda,\lambda'}(t-t_0)\left(v_{j}(t-t_0)\right)\right)}{v_{j}\beta_{j}}\right]}\right)^{2d_j},
	\end{align}
where $\chi_{\lambda,\lambda'}(t-t_0)=\text{sgn}(t-t_0)(\lambda+\lambda')/2-(\lambda-\lambda')/2$, $\beta_j=1/T_j$ is the inverse temperature of channel $j$, and $d_j=l_j|\delta \nu_j|/2$ is the scaling dimension of $\exp(il_i \phi_i)$. The total scaling dimension is then $\Delta(\mathbf{l})=\sum_j d_j$. Inserting the Green's functions~\eqref{eq:GTVertex} into Eq.~\eqref{eq:CurrentAverage}, we obtain
\begin{align}
	&\langle I(t) \rangle =\frac{q |\Gamma_0|^2}{(2\pi a)^{2}} \prod_k \left(\frac{\pi a}{v_k \beta_k}\right)^{2d_k}\sum_{\lambda, \lambda'=\pm 1}\lambda' \times \notag \\ &\int_{-\infty}^{\infty} dt_0
	\frac{\sin\big[\omega_0(t^\lambda-t_0^{\lambda'})\big]}{\prod_j\sin\left[\frac{\pi}{v_{j}\beta_{j}}\left(a+i\chi_{\lambda,\lambda'}(t-t_0)v_{j}(t-t_0\right)\right]^{2d_j}}\notag \\
	& =\frac{q |\Gamma_0|^2}{(2\pi a)^{2}} \prod_k \left(\frac{\pi a}{v_k \beta_k}\right)^{2d_k}\sum_{\lambda=\pm 1}\lambda \times \notag\\
	&\int_{-\infty}^{\infty} dt_0
	\frac{\sin\big[\omega_0(t-t_0)\big]}{\prod_j\sin\left[\frac{\pi}{v_{j}\beta_{j}}\left(a-i\lambda v_{j}(t-t_0\right)\right]^{2d_j}},
\end{align}
In the second equality, we used that when $\lambda=\lambda'$, the integrand becomes odd in $t-t_0$ and hence this contribution vanishes. We next assume that the two channels are at the same temperature: $T_j=T$ and also that $aT/v_j<1$ for $j=1,2$. We then change variables $t-t_0\rightarrow \lambda (t_0+i/(2T))$. To lowest order in $b$, the integral becomes 
\begin{align}
	&\langle I(t) \rangle = \frac{q |\Gamma_0|^2}{(2\pi a)^{2}} \prod_k \left(\frac{\pi a}{v_k \beta_k}\right)^{2d_k}\sum_{\lambda=\pm 1}\lambda \times \notag \\  &\int_{-\infty}^{\infty} dt_0 \frac{\sin\big[-\lambda\omega_0(t_0+i/(2T))\big]}{\prod_j \cosh (\pi Tt_0)^{2d_j}} \notag \\
	& = \frac{2q |\Gamma_0|^2}{(2\pi a)^{2}} \prod_k \left(\frac{\pi a T}{v_k }\right)^{2d_k}\sinh\left(\frac{\omega_0}{2T}\right)\times \notag \\
	&\int_{-\infty}^{\infty} dt_0 \frac{\cos\big[\omega_0t_0\big]}{\prod_j \cosh (\pi Tt_0)^{2d_j}}\notag \\
	& =\frac{2q |\Gamma_0|^2}{(2\pi a)^{2}} \left(\pi a T\right)^{2\Delta(\mathbf{l})} \prod_k \left(\frac{1}{v_k }\right)^{2d_k}\sinh\left(\frac{\omega_0}{2T}\right)\times \notag \\
	&\int_{-\infty}^{\infty} dt_0 \frac{\cos\big[\omega_0t_0\big]}{\cosh (\pi Tt_0)^{2\Delta(\mathbf{l})}}.
\end{align}
The final integral can now be performed with the identity $\int_{-\infty}^{\infty}dt\cosh(2yt)/\cosh^{2x}t=2^{2x-1}\Gamma(x+y)\Gamma(x-y)/\Gamma(2x)$, for $\textrm{Re}\,x>|\textrm{Re}\,y|$ and $\textrm{Re}\,x>0$, with $\Gamma(z)$ the Gamma-function. We then find the time-independent tunneling current
 \begin{align}
 \label{eq:CurrentFull}
	&\langle I \rangle\equiv \langle I(t) \rangle  =2q |\Gamma_0|^2 \left(2\pi a \right)^{2\Delta(\mathbf{l})-2} T^{2\Delta(\mathbf{l})-1} \times \notag \\
	& \prod_k \left(\frac{1}{v_k }\right)^{2d_k}\sinh\left(\frac{\omega_0}{2T}\right) \frac{|\Gamma(\Delta(\mathbf{l})+i\frac{\omega_0}{2\pi T})|^{2}}{\Gamma(2\Delta(\mathbf{l}))}.	
\end{align}
In the high temperature regime $\omega_0 \ll T$, we find the Ohmic behaviour
\begin{align}
	\langle I \rangle = g_\tau\frac{e^2}{h} (V_1-V_2),
\end{align}
with the tunneling conductance
\begin{align}
\label{eq:gtauGen}
	g_\tau \equiv q^2\frac{|\Gamma_0|^2 \left(2\pi a T \right)^{2\Delta(\mathbf{l})-2}}{\prod_k v_k^{2d_k}}  \frac{|\Gamma(\Delta(\mathbf{l}))|^{2}}{\Gamma(2\Delta(\mathbf{l}))}. 
\end{align}
We next consider the symmetrized noise
\begin{align}
	& S(t,t') = \langle I(t) I(t')\rangle + \langle I(t') I(t)\rangle - 2\langle I(t) \rangle \langle I(t') \rangle.
\end{align}
To leading order in $\Gamma_0$, $S(t,t')$ is given on the Keldysh contour as
\begin{align}
	& S(t,t') \equiv S(t-t') = \sum_{\lambda=\pm 1} \langle T_C \left[I(t^\lambda)I(t'^{-\lambda})\right] \rangle \notag \\
	& = 2q^2|\Gamma_0|^2\sum_{\lambda=\pm 1} \cos\big[\omega_0(t^\lambda-t_0^{\lambda'})\big]\times  \notag \\
	&\langle T_C \mathcal{T}_{\mathbf{l}}(0,t^\lambda)\mathcal{T}_{\mathbf{l}}^\dagger(0,t_0^{\lambda'}) \rangle.
\end{align}
We are interested in the zero frequency (dc) noise $S(\omega=0)\equiv\int d(t-t_0) S(t-t_0)$. 
The calculations proceed in perfect analogy to those for $\langle I \rangle$, and using Eq.~\eqref{eq:CurrentFull}, we arrive at
\begin{align}
	S(\omega=0) = 2 q \cosh\left( \frac{\omega_0}{2T}\right) \frac{\langle I \rangle}{\sinh\left( \frac{\omega_0}{2T}\right)} = 2q\langle I \rangle \coth\left( \frac{\omega_0}{2T}\right).
\end{align}
In the limit $\omega_0 \ll T$, the noise approaches the equilibrium Nyqvist-Johnson noise 
\begin{align}
	S(\omega=0) = 4g_\tau\frac{e^2}{h} k_{\rm B} T.
\end{align}
In the shot-noise limit $\omega_0 \gg T$, the Fano factor $F$ becomes
\begin{align}
\label{eq:F}
	F\equiv \frac{S(\omega=0)}{2\langle I\rangle} = q \coth\left( \frac{\omega_0}{2T}\right) \xrightarrow[\omega_0 \gg T]{} q.
\end{align}
Eq.~\eqref{eq:F} manifest the well-known result that weak tunneling reveals the charges of the transferred particles. By use of Eqs.~\eqref{eq:q1},~\eqref{eq:q2}, the formula~\eqref{eq:F} manifests how the tunneling charge is affected by generic charge vector entries $t_1$ and $t_2$. We emphasize again that $F$ is independent of interactions in the free phase. 

Since the tunneling operator $\mathcal{T}_{\mathbf{l}}$ so far has been treated as general, we must now determine the most relevant tunneling processes in the SBS and WBS regimes. 

For point tunneling, the tree level RG equation for the tunneling amplitude $\Gamma_0$ can be read of directly from $H_0+H_\tau$. The equation reads
\begin{align}
\label{eq:TunnRenom}
	\frac{d \Gamma_0}{d \ell} = \left[1-\Delta(\mathbf{l})\right]\Gamma_0,
\end{align}
where $\Delta(\mathbf{l})$ is the scaling dimension of $\mathcal{T}_{\mathbf{l}}$ and $\ell$ is the running length scale. Under the assumption of no inter-channel interactions across the constriction, the scaling dimension is obtained from Eq.~\eqref{eq:GenScalingDim} as
\begin{align}
\label{eq:TunnelScaling}
	\Delta(\mathbf{l}) = \frac{1}{2} (d_1+d_2)=\frac{1}{2} (|\delta\nu_1| l_1^2+|\delta\nu_2| l_2^2).
\end{align}
This relation holds strictly only in the bound phase, where there is only a single channel propagating. In the free phase, with three edge channels, the scaling dimensions will be affected by inter-channel interactions. However, since the free phase is characterized by weak interactions, we expect that the difference between Eq.~\eqref{eq:TunnelScaling} and the true, ``renormalized'' scaling dimensions is small and can thus be ignored.

For the case of identical edge channels $|\delta \nu_1|=|\delta \nu_2|\equiv|\delta \nu|$, charge conservation requires $l_1=l_2\equiv l$. In this case Eq.~\eqref{eq:TunnRenom} becomes
\begin{align}
\label{eq:ScalingSimp}
	\frac{d \Gamma_0}{d \ell} = \left(1-l^2|\delta \nu|\right)\Gamma_0.
\end{align} 
Hence, the most relevant tunneling process is obtained for $l=1$. Physically, this means that single particle tunneling events dominate over multi particle events. In the next sections, we apply Eqs.~\eqref{eq:F}, and \eqref{eq:ScalingSimp} to determine the most dominant Fano factors for the two edge phases Eq.~\eqref{eq:KFree} and Eq.~\eqref{eq:Kbound}.

\subsubsection{Fano factors for the free phase}
For a $\nu=9/5$ device in the free phase, we have for the innermost channels $t_1=t_2=1$ and $\delta \nu_1=-\delta \nu_2 = 1/5\equiv \delta \nu$ [see Eq.~\eqref{eq:KFree}]. In the WBS regime, fractional charges may tunnel across the constriction through the Hall fluid, and, as stated above, the most relevant tunneling operator $\mathcal{T}_{\mathbf{l}}$ is obtained for $l=1$. The operator then describe the transfer of particles with charge 
\begin{align}
	q = t \delta \nu l = 1\times \frac{1}{5}\times 1 = \frac{1}{5}.
\end{align}

Using Eqs.~\eqref{eq:TunnelScaling} in Eq.~\eqref{eq:gtauGen}, we see that the tunneling conductance for this process scales with temperature in the Ohmic limit as $g_\tau\sim T^{-8/5}$. Note that this scaling law strictly holds in the absence of interactions. However, interactions are expected to only lead to small deviations from this value of the scaling exponent. The divergence at zero temperature signals the instability of the system towards quasiparticle tunneling. Thus, this type of tunneling is only visible at high temperatures/voltages. 

In the SBS regime, the left and right parts of the constriction are bridged by a region fully depleted of Hall fluid. No FQH quasiparticles may exist in this region, and hence only tunneling processes of electrons (integer charges) couple the edges. The most dominant such process is obtained for $l=5$, which gives a smallest possible integer charge of
\begin{align}
	q = t \delta \nu l = 1 \times \frac{1}{5} \times 5=1,
\end{align} 
i.e., single electron tunneling. For this process, Eqs.~\eqref{eq:TunnelScaling} and~\eqref{eq:gtauGen} shows that the tunneling conductance scaling with temperature becomes in the Ohmic limit $g_\tau\sim T^{8}$. Also this scaling exponent may be modified by interactions but we expect this effect to be  small. The SBS regime is a stable RG fixed point.

Direct application of Eq.~\eqref{eq:F} for these two types of processes give the Fano factors for the free phase
\begin{align}
	&F_{WBS} = \frac{1}{5},\\
	&F_{SBS} = 1.
\end{align}
Bases on these results, we predict that in the free phase, shot noise measurements in the WBS and SBS regimes should reveal tunneling of fractional charges $q=1/5$ and single electrons $q=1$ respectively.

\subsubsection{Fano factors for the bound phase}
In the bound phase, we have $t_1=t_2=3$ and $\delta \nu_1=-\delta \nu_2 = 1/5$ [see Eq.~\eqref{eq:Kbound}]. In the WBS regime, the most relevant tunneling operator is again obtained for $l=1$, which amounts to transfer of particles with charge
\begin{align}
	q = t \delta \nu l = 3\times \frac{1}{5}\times 1=\frac{3}{5}.
\end{align}
In the SBS regime, we seek the most relevant operator transferring an integer number of charges. From Eq.~\eqref{eq:q1}, we see that, once more, this operator is found for $l=5$, which amounts to the tunneling of charge
\begin{align}
	q = t \delta \nu l = 3\times \frac{1}{5}\times 5 = 3.
\end{align}
The conductance scaling laws in the WBS and SBS regimes are $g_\tau\sim T^{-8/5}$ and $g_\tau\sim T^{8}$, respectively, i.e., the same as those obtained for the bound phase.
Application of Eq.~\eqref{eq:F} for the WBS and SBS regimes gives
\begin{align}
	&F_{WBS} = \frac{3}{5},\\
	&F_{SBS} = 3 \label{eq:F3e}.
\end{align}
Eq.~\eqref{eq:F3e} is a key result of this section. It signals three-electron co-tunneling as the most relevant process in the SBS regime. Since, $t_1=t_2=t=3$, a process yielding $F_{SBS}=1$ is impossible at low energies in the bound phase. Similarly, for WBS it not possible to observe particles with charge $1/5$ in the bound phase. Measuring $F_{SBS}=3$ and $F_{WBS}=3/5$ in the SBS respectively WBS regimes, is therefore a striking manifestation of the $9/5$ edge in the bound phase. 

\subsubsection{Tunneling between the bound edge and a metal}
As a consistency check of our equations, we consider finally also the previously studied setup~\cite{Kao1999} of electron tunneling from an ordinary metal (or, equivalently, an integer quantum Hall edge) into the bound $9/5$ edge. As previously stated, only tunneling of electrons in bunches of three is possible. We therefore have $t_1=1$, $\delta \nu_1=1$, and $l_1=3$, respectively $t_2=3$, $\delta \nu_2=-1/5$, and $l_2=5$. Since $q_1=t_1|\delta \nu_1|l_1=q_2=t_2|\delta \nu_2|l_2=3$, we immediately find from Eq.~\eqref{eq:F} that $F=3$ for this tunneling. Furthermore, Eq~\eqref{eq:TunnelScaling} gives the scaling dimension
\begin{align}
	\Delta = \frac{1}{2}(1\times 3^2+\frac{1}{5}\times 5^2)=7,
\end{align}
for the most relevant tunneling operator. It follows that in the high-temperature limit, Eq.~\eqref{eq:gtauGen} gives that the tunneling conductance scales as $g_\tau \sim T^{2\Delta-2}=T^{12}$. In the low temperature limit we have from dimensional analysis $\langle I\rangle \sim V^{2\Delta-1}=V^{13}$. These scaling laws are precisely those derived in Ref.~\cite{Kao1999}.

We end Sec.~\ref{sec:Signatures} by summarizing the key results in Tab.~\ref{tab:TransportTab}. 

\section{\label{sec:Discussion}Discussion}
Our analysis at $\nu=9/5$ can straightforwardly be adapted to other T-unstable edge structures (several examples were given in Ref.~\cite{Kao1999}). Consider for example an interface between bulk fillings $\nu=2/5$ and $\nu=1/7$. Here, we assume that a FQH state and not a Wigner crystal forms at filling $\nu=1/7$ of the electron gas.   

The resulting edge structure is described by
\begin{align}
\label{eq:KFree3}
	K = \begin{pmatrix}
		3 & 0 & 0 \\
		0 & 15 & 0 \\
		0 & 0 & -7
	\end{pmatrix},\quad \mathbf{t}^T = (1,1,1).
\end{align}
According to Eq.~\eqref{eq:nu}, the ``effective filling'' of this structure is $\nu=9/35$. The two null vectors are $\mathbf{m}_1=(1,10,7)^T$ and $\mathbf{m}_2=(4,-5,7)^T$. Crossing over from the corresponding regimes $\mathcal{III}$ to $\mathcal{II}$, we therefore expect conductance transitions $G=9/35\rightarrow 19/35$ and $G^Q=1\rightarrow 3$ with decreasing temperature.

A basis transformation of Eq.~\eqref{eq:KFree3} on the form~\eqref{eq:rotations} leads to an alternative representation
\begin{align}
\label{eq:KFree3Alt}
	K' = \begin{pmatrix}
		35 & 0 & 0 \\
		0 & 1 & 0 \\
		0 & 0 & -1
	\end{pmatrix},\quad \mathbf{t'}^T = (3,1,1).
\end{align}
The binding transition amounts to the pair of counter-propagating integer channels localizing. The remaining  channel has a filling factor discontinuity $\delta\nu=1/35$ and it is made out of $3e$ composites. Hence, the bound phase is characterized by
\begin{align}
\label{eq:Kbound2}
	K' = (35),\quad \mathbf{t'}^T = (3).
\end{align}
In the bound phase, we therefore expect conductances $G=9/35$ and $G^Q=1$ in regime $\mathcal{I}$.

Tunneling in a QPC bridging two interfaced edges is geometrically complicated and we do not analyse it here. However, electron tunneling into the interface edge from a metal (e.g., an STM tip or a $\nu=1$ QH state) is conceptually simpler and we use Eq.~\eqref{eq:TunnelScaling} to compute the scaling dimension of the tunneling operator~\eqref{eq:tunnelingOperator} transferring three electrons. The result is
\begin{align}
	\Delta = \frac{1}{2}(1\times 3^2+\frac{1}{35}\times 35^2)=22,
\end{align}
where we used $t_1=1$, $\delta \nu_1=1$, and $l_1=3$, respectively $t_2=3$, $\delta \nu_2=-1/35$, and $l_2=35$. Then from Eqs.~\eqref{eq:q1}, \eqref{eq:q2} , and~\eqref{eq:Charge_Conservation}, we have $q_1=t_1|\delta \nu_1|l_1=q_2=t_2|\delta \nu_2|l_2=3$. With these values, Eq.~\eqref{eq:F} produces $F=3$ as expected for this tunneling.  From Eq.~\eqref{eq:gtauGen}, we have that the tunneling conductance scales with temperature as $g_\tau\sim T^{2\Delta-2}=T^{42}$. In the low temperature limit, the voltage scaling of the current reads $\langle I \rangle\sim V^{2\Delta-1}=V^{41}$. 

We now move on to discuss practical aspects of experimentally detecting the binding transition. The transition at filling $\nu=9/5$ can be expected to require tunneling between Landau levels with different spin-polarization. Breaking spin-rotation symmetry, for example by spin-orbit coupling, is therefore a necessary condition. A bi-layered device could be used to facilitate the binding transition~\cite{Kao1999}. Alternatively, carefully designed devices in the spirit of Ref.~\cite{Cohen2019} where all channels have the same spin may be another possibility.

In our view, the most standard probe of the binding transition should be a measurement of the shot noise in a QPC device. From Sec.~\ref{sec:QPC}, we anticipate that, with decreasing temperature, a binding transition at $\nu=9/5$ amounts to the crossover from $F_{SBS}=1$ to $F_{SBS}=3$ in the strong back-scattering regime of the QPC. Similarly, for weak back-scattering, one expects a crossover from $F_{WBS}=1/5$ to $F_{WBS}=3/5$. Changes in Fano factors with decreasing temperature were recently measured in Ref.~\cite{Biswas2022}. It would be interesting to investigate whether these changes could arise due to a combination of edge reconstruction and binding transitions.

\section{\label{sec:Summary}Summary and Outlook}
We proposed quantum transport signatures for the FQH edge binding transition, with focus on filling $\nu=9/5$. For this edge, we showed that interactions and disorder conspire to generate a rich phase diagram (Fig.~\ref{fig:phasediagram}) with distinct charge and heat transport regimes (see Tab.~\ref{tab:TransportTab}). The three regimes, labelled $\mathcal{I}$, $\mathcal{II}$, and $\mathcal{III}$ displays localized, non-equilibrated, and fully equilibrated characteristics. Probing the distinct transport behaviour, in terms of charge and heat conductances, of these regimes should be possible with present technology. As a complement to the conductance, we also estimated the shot noise produced of a single current biased edge segment. We demonstrated that such noise is only expected for the thermally non-equilibrated edge (regime $\mathcal{II}$) under the strong interaction condition~\eqref{eq:noisecondition} associated with the bound phase. 

We also studied a QPC device in the strong and weak backscattering regimes and derived shot noise Fano factors for tunneling processes across the constriction. The bound phase does not allow single electron tunneling in the strong back-scattering (SBS) regime. Instead of the typical Fano factor $F_{SBS}=1$ corresponding to single electron tunneling, we therefore found that the smallest Fano factor compatible with transferring an integer number of charges is $F_{SBS}=3$. This corresponds to three-electron co-tunneling. In addition, all higher order (but less relevant) tunneling processes of electrons are necessarily integer multiples of 3. In the weak back-scattering (WBS) regime, we found that the most relevant tunneling of quasiparticles yields $F_{SBS}=3/5$. These SBS and WBS Fano factors are in stark contrast to those for the free phase: $F_{SBS}=1$ respectively $F_{WBS}=1/5$. These contrasting values serve as a clear signature for the binding transition.

We hope that our predictions and proposals will stimulate further theoretical and experimental investigations of FQH binding transitions. While our present analysis focused on Abelian FQH edges, binding transitions for non-Abelian candidate edge theories for the state filling $\nu=5/2$ were studied in Ref.~\cite{Overbosch2008}. An experimentally oriented analysis similar to the present work could be useful for pin-pointing that state's underlying topological order.
 
\begin{acknowledgments}
C.S. thanks Jinhong Park, Gu Zhang, and Kyrylo Snizhko for discussions. C.S. and A.D.M acknowledge support from the DFG grant  No. MI 658/10-2 and the German-Israeli Foundation grant I-1505-303.10/2019. C.S. further acknowledge funding from the EI Nano Excellence Initiative at Chalmers University of Technology. This project has received funding from the European Union's Horizon 2020 research and innovation programme under grant agreement No 101031655 (TEAPOT). This work was also supported by the European Union's Horizon 2020 research and innovation programme (Grant Agreement LEGOTOP No. 788715), the DFG (CRC/Transregio 183, EI 519/7-1), ISF Quantum Science and Technology (2074/19).
\end{acknowledgments}

\appendix
\section{Dimensions of the disorder strength $D$}
\label{sec:DimAppendix}
A generic tunneling operator $\mathcal{T}_\mathbf{l}$ has length ($L$) dimensions 
\begin{align}
	L^{-m/2},
\end{align}
where $m$ is the number of involved vertex operators ($m=3$ for the $\nu=9/5$ edge). The action~\eqref{eq:Spert} is dimensionless, so the random tunneling amplitude $\xi(x)$ has dimensions
\begin{align}
	L^{m/2-1}s^{-1}.
\end{align}
where $s$ is the unit of time. From Eq.~\eqref{eq:RandTunn}, we then have that the units of $D$ are
\begin{align}
	L^{m-1}s^{-2}.
\end{align} 
Hence, to obtain a dimensionless $D$ we let
\begin{align}
	D \rightarrow D\times \frac{a^{3-m}}{v^2},
\end{align}
with some characteristic velocity  $v$ (a combination of all $v_i$ to various powers such that the total power is $2$).  This result is consistent with Ref.~\cite{Giamarchi1988} in which $m=2$. 


\begin{thebibliography}{62}%
\makeatletter
\providecommand \@ifxundefined [1]{%
 \@ifx{#1\undefined}
}%
\providecommand \@ifnum [1]{%
 \ifnum #1\expandafter \@firstoftwo
 \else \expandafter \@secondoftwo
 \fi
}%
\providecommand \@ifx [1]{%
 \ifx #1\expandafter \@firstoftwo
 \else \expandafter \@secondoftwo
 \fi
}%
\providecommand \natexlab [1]{#1}%
\providecommand \enquote  [1]{``#1''}%
\providecommand \bibnamefont  [1]{#1}%
\providecommand \bibfnamefont [1]{#1}%
\providecommand \citenamefont [1]{#1}%
\providecommand \href@noop [0]{\@secondoftwo}%
\providecommand \href [0]{\begingroup \@sanitize@url \@href}%
\providecommand \@href[1]{\@@startlink{#1}\@@href}%
\providecommand \@@href[1]{\endgroup#1\@@endlink}%
\providecommand \@sanitize@url [0]{\catcode `\\12\catcode `\$12\catcode
  `\&12\catcode `\#12\catcode `\^12\catcode `\_12\catcode `\%12\relax}%
\providecommand \@@startlink[1]{}%
\providecommand \@@endlink[0]{}%
\providecommand \url  [0]{\begingroup\@sanitize@url \@url }%
\providecommand \@url [1]{\endgroup\@href {#1}{\urlprefix }}%
\providecommand \urlprefix  [0]{URL }%
\providecommand \Eprint [0]{\href }%
\providecommand \doibase [0]{https://doi.org/}%
\providecommand \selectlanguage [0]{\@gobble}%
\providecommand \bibinfo  [0]{\@secondoftwo}%
\providecommand \bibfield  [0]{\@secondoftwo}%
\providecommand \translation [1]{[#1]}%
\providecommand \BibitemOpen [0]{}%
\providecommand \bibitemStop [0]{}%
\providecommand \bibitemNoStop [0]{.\EOS\space}%
\providecommand \EOS [0]{\spacefactor3000\relax}%
\providecommand \BibitemShut  [1]{\csname bibitem#1\endcsname}%
\let\auto@bib@innerbib\@empty
\bibitem [{\citenamefont {Tsui}\ \emph {et~al.}(1982)\citenamefont {Tsui},
  \citenamefont {Stormer},\ and\ \citenamefont {Gossard}}]{Stormer1982}%
  \BibitemOpen
  \bibfield  {author} {\bibinfo {author} {\bibfnamefont {D.~C.}\ \bibnamefont
  {Tsui}}, \bibinfo {author} {\bibfnamefont {H.~L.}\ \bibnamefont {Stormer}},\
  and\ \bibinfo {author} {\bibfnamefont {A.~C.}\ \bibnamefont {Gossard}},\
  }\bibfield  {title} {\bibinfo {title} {Two-dimensional magnetotransport in
  the extreme quantum limit},\ }\href
  {https://doi.org/10.1103/PhysRevLett.48.1559} {\bibfield  {journal} {\bibinfo
   {journal} {Phys. Rev. Lett.}\ }\textbf {\bibinfo {volume} {48}},\ \bibinfo
  {pages} {1559} (\bibinfo {year} {1982})}\BibitemShut {NoStop}%
\bibitem [{\citenamefont {Laughlin}(1983)}]{Laughlin1983}%
  \BibitemOpen
  \bibfield  {author} {\bibinfo {author} {\bibfnamefont {R.~B.}\ \bibnamefont
  {Laughlin}},\ }\bibfield  {title} {\bibinfo {title} {Anomalous quantum
  {{H}all} effect: An incompressible quantum fluid with fractionally charged
  excitations},\ }\href {https://doi.org/10.1103/PhysRevLett.50.1395}
  {\bibfield  {journal} {\bibinfo  {journal} {Phys. Rev. Lett.}\ }\textbf
  {\bibinfo {volume} {50}},\ \bibinfo {pages} {1395} (\bibinfo {year}
  {1983})}\BibitemShut {NoStop}%
\bibitem [{\citenamefont {Wen}(1990{\natexlab{a}})}]{Wen1990a}%
  \BibitemOpen
  \bibfield  {author} {\bibinfo {author} {\bibfnamefont {X.~G.}\ \bibnamefont
  {Wen}},\ }\bibfield  {title} {\bibinfo {title} {Topological orders in rigid
  states},\ }\href {https://doi.org/10.1142/S0217979290000139} {\bibfield
  {journal} {\bibinfo  {journal} {Int. J. Mod. Phys. B}\ }\textbf {\bibinfo
  {volume} {04}},\ \bibinfo {pages} {239} (\bibinfo {year}
  {1990}{\natexlab{a}})}\BibitemShut {NoStop}%
\bibitem [{\citenamefont {Wen}(1990{\natexlab{b}})}]{Wen1990b}%
  \BibitemOpen
  \bibfield  {author} {\bibinfo {author} {\bibfnamefont {X.~G.}\ \bibnamefont
  {Wen}},\ }\bibfield  {title} {\bibinfo {title} {Chiral {L}uttinger liquid and
  the edge excitations in the fractional quantum {{H}all} states},\ }\href
  {https://doi.org/10.1103/PhysRevB.41.12838} {\bibfield  {journal} {\bibinfo
  {journal} {Phys. Rev. B}\ }\textbf {\bibinfo {volume} {41}},\ \bibinfo
  {pages} {12838} (\bibinfo {year} {1990}{\natexlab{b}})}\BibitemShut {NoStop}%
\bibitem [{\citenamefont {Wen}(1992)}]{Wen1992}%
  \BibitemOpen
  \bibfield  {author} {\bibinfo {author} {\bibfnamefont {X.~G.}\ \bibnamefont
  {Wen}},\ }\bibfield  {title} {\bibinfo {title} {Theory of the edge states in
  fractional quantum {{H}all} effects},\ }\href
  {https://doi.org/10.1142/S0217979292000840} {\bibfield  {journal} {\bibinfo
  {journal} {Int. J. Mod. Phys. B}\ }\textbf {\bibinfo {volume} {06}},\
  \bibinfo {pages} {1711} (\bibinfo {year} {1992})}\BibitemShut {NoStop}%
\bibitem [{\citenamefont {Wen}(1994)}]{Wen1994}%
  \BibitemOpen
  \bibfield  {author} {\bibinfo {author} {\bibfnamefont {X.-G.}\ \bibnamefont
  {Wen}},\ }\bibfield  {title} {\bibinfo {title} {Impurity effects on chiral
  one-dimensional electron systems},\ }\href
  {https://doi.org/10.1103/PhysRevB.50.5420} {\bibfield  {journal} {\bibinfo
  {journal} {Phys. Rev. B}\ }\textbf {\bibinfo {volume} {50}},\ \bibinfo
  {pages} {5420} (\bibinfo {year} {1994})}\BibitemShut {NoStop}%
\bibitem [{\citenamefont {Wen}(1995)}]{Wen1995}%
  \BibitemOpen
  \bibfield  {author} {\bibinfo {author} {\bibfnamefont {X.-G.}\ \bibnamefont
  {Wen}},\ }\bibfield  {title} {\bibinfo {title} {Topological orders and edge
  excitations in fractional quantum {{H}all} states},\ }\href
  {https://doi.org/10.1080/00018739500101566} {\bibfield  {journal} {\bibinfo
  {journal} {Advances in Physics}\ }\textbf {\bibinfo {volume} {44}},\ \bibinfo
  {pages} {405} (\bibinfo {year} {1995})}\BibitemShut {NoStop}%
\bibitem [{\citenamefont {Chang}(2003)}]{Chang2003}%
  \BibitemOpen
  \bibfield  {author} {\bibinfo {author} {\bibfnamefont {A.~M.}\ \bibnamefont
  {Chang}},\ }\bibfield  {title} {\bibinfo {title} {Chiral {L}uttinger liquids
  at the fractional quantum {{H}all} edge},\ }\href
  {https://doi.org/10.1103/RevModPhys.75.1449} {\bibfield  {journal} {\bibinfo
  {journal} {Rev. Mod. Phys.}\ }\textbf {\bibinfo {volume} {75}},\ \bibinfo
  {pages} {1449} (\bibinfo {year} {2003})}\BibitemShut {NoStop}%
\bibitem [{\citenamefont {Saminadayar}\ \emph {et~al.}(1997)\citenamefont
  {Saminadayar}, \citenamefont {Glattli}, \citenamefont {Jin},\ and\
  \citenamefont {Etienne}}]{Saminadayar1997}%
  \BibitemOpen
  \bibfield  {author} {\bibinfo {author} {\bibfnamefont {L.}~\bibnamefont
  {Saminadayar}}, \bibinfo {author} {\bibfnamefont {D.~C.}\ \bibnamefont
  {Glattli}}, \bibinfo {author} {\bibfnamefont {Y.}~\bibnamefont {Jin}},\ and\
  \bibinfo {author} {\bibfnamefont {B.}~\bibnamefont {Etienne}},\ }\bibfield
  {title} {\bibinfo {title} {Observation of the $\mathit{e}\mathit{/}3$
  fractionally charged {L}aughlin quasiparticle},\ }\href
  {https://doi.org/10.1103/PhysRevLett.79.2526} {\bibfield  {journal} {\bibinfo
   {journal} {Phys. Rev. Lett.}\ }\textbf {\bibinfo {volume} {79}},\ \bibinfo
  {pages} {2526} (\bibinfo {year} {1997})}\BibitemShut {NoStop}%
\bibitem [{\citenamefont {de~Picciotto}\ \emph {et~al.}(1997)\citenamefont
  {de~Picciotto}, \citenamefont {Reznikov}, \citenamefont {Heiblum},
  \citenamefont {Umansky}, \citenamefont {Bunin},\ and\ \citenamefont
  {Mahalu}}]{DePicciotto1997}%
  \BibitemOpen
  \bibfield  {author} {\bibinfo {author} {\bibfnamefont {R.}~\bibnamefont
  {de~Picciotto}}, \bibinfo {author} {\bibfnamefont {M.}~\bibnamefont
  {Reznikov}}, \bibinfo {author} {\bibfnamefont {M.}~\bibnamefont {Heiblum}},
  \bibinfo {author} {\bibfnamefont {V.}~\bibnamefont {Umansky}}, \bibinfo
  {author} {\bibfnamefont {G.}~\bibnamefont {Bunin}},\ and\ \bibinfo {author}
  {\bibfnamefont {D.}~\bibnamefont {Mahalu}},\ }\bibfield  {title} {\bibinfo
  {title} {Direct observation of a fractional charge},\ }\href
  {https://doi.org/10.1038/38241} {\bibfield  {journal} {\bibinfo  {journal}
  {Nature}\ }\textbf {\bibinfo {volume} {389}},\ \bibinfo {pages} {162}
  (\bibinfo {year} {1997})}\BibitemShut {NoStop}%
\bibitem [{\citenamefont {Lin}\ \emph {et~al.}(2021)\citenamefont {Lin},
  \citenamefont {Hashisaka}, \citenamefont {Akiho}, \citenamefont {Muraki},\
  and\ \citenamefont {Fujisawa}}]{Lin2021}%
  \BibitemOpen
  \bibfield  {author} {\bibinfo {author} {\bibfnamefont {C.}~\bibnamefont
  {Lin}}, \bibinfo {author} {\bibfnamefont {M.}~\bibnamefont {Hashisaka}},
  \bibinfo {author} {\bibfnamefont {T.}~\bibnamefont {Akiho}}, \bibinfo
  {author} {\bibfnamefont {K.}~\bibnamefont {Muraki}},\ and\ \bibinfo {author}
  {\bibfnamefont {T.}~\bibnamefont {Fujisawa}},\ }\bibfield  {title} {\bibinfo
  {title} {Quantized charge fractionalization at quantum {{H}all} y junctions
  in the disorder dominated regime},\ }\href
  {https://doi.org/10.1038/s41467-020-20395-7} {\bibfield  {journal} {\bibinfo
  {journal} {Nature Communications}\ }\textbf {\bibinfo {volume} {12}},\
  \bibinfo {pages} {131} (\bibinfo {year} {2021})}\BibitemShut {NoStop}%
\bibitem [{\citenamefont {Nakamura}\ \emph {et~al.}(2020)\citenamefont
  {Nakamura}, \citenamefont {Liang}, \citenamefont {Gardner},\ and\
  \citenamefont {Manfra}}]{Nakamura2020}%
  \BibitemOpen
  \bibfield  {author} {\bibinfo {author} {\bibfnamefont {J.}~\bibnamefont
  {Nakamura}}, \bibinfo {author} {\bibfnamefont {S.}~\bibnamefont {Liang}},
  \bibinfo {author} {\bibfnamefont {G.~C.}\ \bibnamefont {Gardner}},\ and\
  \bibinfo {author} {\bibfnamefont {M.~J.}\ \bibnamefont {Manfra}},\ }\bibfield
   {title} {\bibinfo {title} {Direct observation of anyonic braiding
  statistics},\ }\href
  {https://doi.org/https://doi.org/10.1038/s41567-020-1019-1} {\bibfield
  {journal} {\bibinfo  {journal} {Nature Physics}\ }\textbf {\bibinfo {volume}
  {16}},\ \bibinfo {pages} {931} (\bibinfo {year} {2020})}\BibitemShut
  {NoStop}%
\bibitem [{\citenamefont {Bartolomei}\ \emph {et~al.}(2020)\citenamefont
  {Bartolomei}, \citenamefont {Kumar}, \citenamefont {Bisognin}, \citenamefont
  {Marguerite}, \citenamefont {Berroir}, \citenamefont {Bocquillon},
  \citenamefont {Placais}, \citenamefont {Cavanna}, \citenamefont {Dong},
  \citenamefont {Gennser} \emph {et~al.}}]{Bartolomei2020}%
  \BibitemOpen
  \bibfield  {author} {\bibinfo {author} {\bibfnamefont {H.}~\bibnamefont
  {Bartolomei}}, \bibinfo {author} {\bibfnamefont {M.}~\bibnamefont {Kumar}},
  \bibinfo {author} {\bibfnamefont {R.}~\bibnamefont {Bisognin}}, \bibinfo
  {author} {\bibfnamefont {A.}~\bibnamefont {Marguerite}}, \bibinfo {author}
  {\bibfnamefont {J.-M.}\ \bibnamefont {Berroir}}, \bibinfo {author}
  {\bibfnamefont {E.}~\bibnamefont {Bocquillon}}, \bibinfo {author}
  {\bibfnamefont {B.}~\bibnamefont {Placais}}, \bibinfo {author} {\bibfnamefont
  {A.}~\bibnamefont {Cavanna}}, \bibinfo {author} {\bibfnamefont
  {Q.}~\bibnamefont {Dong}}, \bibinfo {author} {\bibfnamefont {U.}~\bibnamefont
  {Gennser}}, \emph {et~al.},\ }\bibfield  {title} {\bibinfo {title}
  {Fractional statistics in anyon collisions},\ }\href
  {https://doi.org/https://doi.org/10.1126/science.aaz5601} {\bibfield
  {journal} {\bibinfo  {journal} {Science}\ }\textbf {\bibinfo {volume}
  {368}},\ \bibinfo {pages} {173} (\bibinfo {year} {2020})}\BibitemShut
  {NoStop}%
\bibitem [{\citenamefont {Nayak}\ \emph {et~al.}(2008)\citenamefont {Nayak},
  \citenamefont {Simon}, \citenamefont {Stern}, \citenamefont {Freedman},\ and\
  \citenamefont {Das~Sarma}}]{Nayak2008}%
  \BibitemOpen
  \bibfield  {author} {\bibinfo {author} {\bibfnamefont {C.}~\bibnamefont
  {Nayak}}, \bibinfo {author} {\bibfnamefont {S.~H.}\ \bibnamefont {Simon}},
  \bibinfo {author} {\bibfnamefont {A.}~\bibnamefont {Stern}}, \bibinfo
  {author} {\bibfnamefont {M.}~\bibnamefont {Freedman}},\ and\ \bibinfo
  {author} {\bibfnamefont {S.}~\bibnamefont {Das~Sarma}},\ }\bibfield  {title}
  {\bibinfo {title} {Non-{A}belian anyons and topological quantum
  computation},\ }\href {https://doi.org/10.1103/RevModPhys.80.1083} {\bibfield
   {journal} {\bibinfo  {journal} {Rev. Mod. Phys.}\ }\textbf {\bibinfo
  {volume} {80}},\ \bibinfo {pages} {1083} (\bibinfo {year}
  {2008})}\BibitemShut {NoStop}%
\bibitem [{\citenamefont {Kane}\ \emph {et~al.}(1994)\citenamefont {Kane},
  \citenamefont {Fisher},\ and\ \citenamefont {Polchinski}}]{Kane1994}%
  \BibitemOpen
  \bibfield  {author} {\bibinfo {author} {\bibfnamefont {C.~L.}\ \bibnamefont
  {Kane}}, \bibinfo {author} {\bibfnamefont {M.~P.~A.}\ \bibnamefont
  {Fisher}},\ and\ \bibinfo {author} {\bibfnamefont {J.}~\bibnamefont
  {Polchinski}},\ }\bibfield  {title} {\bibinfo {title} {Randomness at the
  edge: Theory of quantum {{H}all} transport at filling \ensuremath{\nu}=2/3},\
  }\href {https://doi.org/10.1103/PhysRevLett.72.4129} {\bibfield  {journal}
  {\bibinfo  {journal} {Phys. Rev. Lett.}\ }\textbf {\bibinfo {volume} {72}},\
  \bibinfo {pages} {4129} (\bibinfo {year} {1994})}\BibitemShut {NoStop}%
\bibitem [{\citenamefont {Kane}\ and\ \citenamefont
  {Fisher}(1995{\natexlab{a}})}]{Kane1995b}%
  \BibitemOpen
  \bibfield  {author} {\bibinfo {author} {\bibfnamefont {C.~L.}\ \bibnamefont
  {Kane}}\ and\ \bibinfo {author} {\bibfnamefont {M.~P.~A.}\ \bibnamefont
  {Fisher}},\ }\bibfield  {title} {\bibinfo {title} {Impurity scattering and
  transport of fractional quantum {{H}all} edge states},\ }\href
  {https://doi.org/10.1103/PhysRevB.51.13449} {\bibfield  {journal} {\bibinfo
  {journal} {Phys. Rev. B}\ }\textbf {\bibinfo {volume} {51}},\ \bibinfo
  {pages} {13449} (\bibinfo {year} {1995}{\natexlab{a}})}\BibitemShut {NoStop}%
\bibitem [{\citenamefont {Moore}\ and\ \citenamefont {Wen}(1998)}]{Moore1998}%
  \BibitemOpen
  \bibfield  {author} {\bibinfo {author} {\bibfnamefont {J.~E.}\ \bibnamefont
  {Moore}}\ and\ \bibinfo {author} {\bibfnamefont {X.-G.}\ \bibnamefont
  {Wen}},\ }\bibfield  {title} {\bibinfo {title} {Classification of disordered
  phases of quantum {{H}all} edge states},\ }\href
  {https://doi.org/10.1103/PhysRevB.57.10138} {\bibfield  {journal} {\bibinfo
  {journal} {Phys. Rev. B}\ }\textbf {\bibinfo {volume} {57}},\ \bibinfo
  {pages} {10138} (\bibinfo {year} {1998})}\BibitemShut {NoStop}%
\bibitem [{\citenamefont {Kao}\ \emph {et~al.}(1999)\citenamefont {Kao},
  \citenamefont {Chang},\ and\ \citenamefont {Wen}}]{Kao1999}%
  \BibitemOpen
  \bibfield  {author} {\bibinfo {author} {\bibfnamefont {H.-c.}\ \bibnamefont
  {Kao}}, \bibinfo {author} {\bibfnamefont {C.-H.}\ \bibnamefont {Chang}},\
  and\ \bibinfo {author} {\bibfnamefont {X.-G.}\ \bibnamefont {Wen}},\
  }\bibfield  {title} {\bibinfo {title} {Binding transition in quantum {{H}all}
  edge states},\ }\href {https://doi.org/10.1103/PhysRevLett.83.5563}
  {\bibfield  {journal} {\bibinfo  {journal} {Phys. Rev. Lett.}\ }\textbf
  {\bibinfo {volume} {83}},\ \bibinfo {pages} {5563} (\bibinfo {year}
  {1999})}\BibitemShut {NoStop}%
\bibitem [{\citenamefont {Haldane}(1995)}]{Haldane1995}%
  \BibitemOpen
  \bibfield  {author} {\bibinfo {author} {\bibfnamefont {F.~D.~M.}\
  \bibnamefont {Haldane}},\ }\bibfield  {title} {\bibinfo {title} {Stability of
  chiral {L}uttinger liquids and {A}belian quantum {{H}all} states},\ }\href
  {https://doi.org/10.1103/PhysRevLett.74.2090} {\bibfield  {journal} {\bibinfo
   {journal} {Phys. Rev. Lett.}\ }\textbf {\bibinfo {volume} {74}},\ \bibinfo
  {pages} {2090} (\bibinfo {year} {1995})}\BibitemShut {NoStop}%
\bibitem [{\citenamefont {Heiblum}\ and\ \citenamefont
  {Feldman}(2020)}]{Heiblum2020}%
  \BibitemOpen
  \bibfield  {author} {\bibinfo {author} {\bibfnamefont {M.}~\bibnamefont
  {Heiblum}}\ and\ \bibinfo {author} {\bibfnamefont {D.~E.}\ \bibnamefont
  {Feldman}},\ }\bibfield  {title} {\bibinfo {title} {Edge probes of
  topological order},\ }\href {https://doi.org/10.1142/S0217751X20300094}
  {\bibfield  {journal} {\bibinfo  {journal} {International Journal of Modern
  Physics A}\ }\textbf {\bibinfo {volume} {35}},\ \bibinfo {pages} {2030009}
  (\bibinfo {year} {2020})}\BibitemShut {NoStop}%
\bibitem [{\citenamefont {Banerjee}\ \emph {et~al.}(2017)\citenamefont
  {Banerjee}, \citenamefont {Heiblum}, \citenamefont {Rosenblatt},
  \citenamefont {Oreg}, \citenamefont {Feldman}, \citenamefont {Stern},\ and\
  \citenamefont {Umansky}}]{Banerjee2017}%
  \BibitemOpen
  \bibfield  {author} {\bibinfo {author} {\bibfnamefont {M.}~\bibnamefont
  {Banerjee}}, \bibinfo {author} {\bibfnamefont {M.}~\bibnamefont {Heiblum}},
  \bibinfo {author} {\bibfnamefont {A.}~\bibnamefont {Rosenblatt}}, \bibinfo
  {author} {\bibfnamefont {Y.}~\bibnamefont {Oreg}}, \bibinfo {author}
  {\bibfnamefont {D.~E.}\ \bibnamefont {Feldman}}, \bibinfo {author}
  {\bibfnamefont {A.}~\bibnamefont {Stern}},\ and\ \bibinfo {author}
  {\bibfnamefont {V.}~\bibnamefont {Umansky}},\ }\bibfield  {title} {\bibinfo
  {title} {Observed quantization of anyonic heat flow},\ }\href
  {https://doi.org/10.1038/nature22052} {\bibfield  {journal} {\bibinfo
  {journal} {Nature}\ }\textbf {\bibinfo {volume} {545}},\ \bibinfo {pages} {75
  EP } (\bibinfo {year} {2017})}\BibitemShut {NoStop}%
\bibitem [{\citenamefont {Banerjee}\ \emph {et~al.}(2018)\citenamefont
  {Banerjee}, \citenamefont {Heiblum}, \citenamefont {Umansky}, \citenamefont
  {Feldman}, \citenamefont {Oreg},\ and\ \citenamefont {Stern}}]{Banerjee2018}%
  \BibitemOpen
  \bibfield  {author} {\bibinfo {author} {\bibfnamefont {M.}~\bibnamefont
  {Banerjee}}, \bibinfo {author} {\bibfnamefont {M.}~\bibnamefont {Heiblum}},
  \bibinfo {author} {\bibfnamefont {V.}~\bibnamefont {Umansky}}, \bibinfo
  {author} {\bibfnamefont {D.~E.}\ \bibnamefont {Feldman}}, \bibinfo {author}
  {\bibfnamefont {Y.}~\bibnamefont {Oreg}},\ and\ \bibinfo {author}
  {\bibfnamefont {A.}~\bibnamefont {Stern}},\ }\bibfield  {title} {\bibinfo
  {title} {Observation of half-integer thermal {{H}all} conductance},\ }\href
  {https://doi.org/10.1038/s41586-018-0184-1} {\bibfield  {journal} {\bibinfo
  {journal} {Nature}\ }\textbf {\bibinfo {volume} {559}},\ \bibinfo {pages}
  {205} (\bibinfo {year} {2018})}\BibitemShut {NoStop}%
\bibitem [{\citenamefont {Srivastav}\ \emph {et~al.}(2019)\citenamefont
  {Srivastav}, \citenamefont {Sahu}, \citenamefont {Watanabe}, \citenamefont
  {Taniguchi}, \citenamefont {Banerjee},\ and\ \citenamefont
  {Das}}]{Srivastav2019}%
  \BibitemOpen
  \bibfield  {author} {\bibinfo {author} {\bibfnamefont {S.~K.}\ \bibnamefont
  {Srivastav}}, \bibinfo {author} {\bibfnamefont {M.~R.}\ \bibnamefont {Sahu}},
  \bibinfo {author} {\bibfnamefont {K.}~\bibnamefont {Watanabe}}, \bibinfo
  {author} {\bibfnamefont {T.}~\bibnamefont {Taniguchi}}, \bibinfo {author}
  {\bibfnamefont {S.}~\bibnamefont {Banerjee}},\ and\ \bibinfo {author}
  {\bibfnamefont {A.}~\bibnamefont {Das}},\ }\bibfield  {title} {\bibinfo
  {title} {Universal quantized thermal conductance in graphene},\ }\href
  {https://advances.sciencemag.org/content/5/7/eaaw5798} {\bibfield  {journal}
  {\bibinfo  {journal} {Science Advances}\ }\textbf {\bibinfo {volume} {5}}
  (\bibinfo {year} {2019})}\BibitemShut {NoStop}%
\bibitem [{\citenamefont {Srivastav}\ \emph {et~al.}(2021)\citenamefont
  {Srivastav}, \citenamefont {Kumar}, \citenamefont {Sp\aa{}nsl\"att},
  \citenamefont {Watanabe}, \citenamefont {Taniguchi}, \citenamefont {Mirlin},
  \citenamefont {Gefen},\ and\ \citenamefont {Das}}]{Srivastav2021}%
  \BibitemOpen
  \bibfield  {author} {\bibinfo {author} {\bibfnamefont {S.~K.}\ \bibnamefont
  {Srivastav}}, \bibinfo {author} {\bibfnamefont {R.}~\bibnamefont {Kumar}},
  \bibinfo {author} {\bibfnamefont {C.}~\bibnamefont {Sp\aa{}nsl\"att}},
  \bibinfo {author} {\bibfnamefont {K.}~\bibnamefont {Watanabe}}, \bibinfo
  {author} {\bibfnamefont {T.}~\bibnamefont {Taniguchi}}, \bibinfo {author}
  {\bibfnamefont {A.~D.}\ \bibnamefont {Mirlin}}, \bibinfo {author}
  {\bibfnamefont {Y.}~\bibnamefont {Gefen}},\ and\ \bibinfo {author}
  {\bibfnamefont {A.}~\bibnamefont {Das}},\ }\bibfield  {title} {\bibinfo
  {title} {Vanishing thermal equilibration for hole-conjugate fractional
  quantum {{H}all} states in graphene},\ }\href
  {https://doi.org/10.1103/PhysRevLett.126.216803} {\bibfield  {journal}
  {\bibinfo  {journal} {Phys. Rev. Lett.}\ }\textbf {\bibinfo {volume} {126}},\
  \bibinfo {pages} {216803} (\bibinfo {year} {2021})}\BibitemShut {NoStop}%
\bibitem [{\citenamefont {Melcer}\ \emph {et~al.}(2022)\citenamefont {Melcer},
  \citenamefont {Dutta}, \citenamefont {Sp{\aa}nsl{\"a}tt}, \citenamefont
  {Park}, \citenamefont {Mirlin},\ and\ \citenamefont {Umansky}}]{Melcer2022}%
  \BibitemOpen
  \bibfield  {author} {\bibinfo {author} {\bibfnamefont {R.~A.}\ \bibnamefont
  {Melcer}}, \bibinfo {author} {\bibfnamefont {B.}~\bibnamefont {Dutta}},
  \bibinfo {author} {\bibfnamefont {C.}~\bibnamefont {Sp{\aa}nsl{\"a}tt}},
  \bibinfo {author} {\bibfnamefont {J.}~\bibnamefont {Park}}, \bibinfo {author}
  {\bibfnamefont {A.~D.}\ \bibnamefont {Mirlin}},\ and\ \bibinfo {author}
  {\bibfnamefont {V.}~\bibnamefont {Umansky}},\ }\bibfield  {title} {\bibinfo
  {title} {Absent thermal equilibration on fractional quantum {{H}all} edges
  over macroscopic scale},\ }\href {https://doi.org/10.1038/s41467-022-28009-0}
  {\bibfield  {journal} {\bibinfo  {journal} {Nature Communications}\ }\textbf
  {\bibinfo {volume} {13}},\ \bibinfo {pages} {376} (\bibinfo {year}
  {2022})}\BibitemShut {NoStop}%
\bibitem [{\citenamefont {Kumar}\ \emph {et~al.}(2022)\citenamefont {Kumar},
  \citenamefont {Srivastav}, \citenamefont {Sp{\aa}nsl{\"a}tt}, \citenamefont
  {Watanabe}, \citenamefont {Taniguchi}, \citenamefont {Gefen}, \citenamefont
  {Mirlin},\ and\ \citenamefont {Das}}]{Kumar2022}%
  \BibitemOpen
  \bibfield  {author} {\bibinfo {author} {\bibfnamefont {R.}~\bibnamefont
  {Kumar}}, \bibinfo {author} {\bibfnamefont {S.~K.}\ \bibnamefont
  {Srivastav}}, \bibinfo {author} {\bibfnamefont {C.}~\bibnamefont
  {Sp{\aa}nsl{\"a}tt}}, \bibinfo {author} {\bibfnamefont {K.}~\bibnamefont
  {Watanabe}}, \bibinfo {author} {\bibfnamefont {T.}~\bibnamefont {Taniguchi}},
  \bibinfo {author} {\bibfnamefont {Y.}~\bibnamefont {Gefen}}, \bibinfo
  {author} {\bibfnamefont {A.~D.}\ \bibnamefont {Mirlin}},\ and\ \bibinfo
  {author} {\bibfnamefont {A.}~\bibnamefont {Das}},\ }\bibfield  {title}
  {\bibinfo {title} {Observation of ballistic upstream modes at fractional
  quantum {{H}all} edges of graphene},\ }\href
  {https://doi.org/10.1038/s41467-021-27805-4} {\bibfield  {journal} {\bibinfo
  {journal} {Nature Communications}\ }\textbf {\bibinfo {volume} {13}},\
  \bibinfo {pages} {213} (\bibinfo {year} {2022})}\BibitemShut {NoStop}%
\bibitem [{\citenamefont {Srivastav}\ \emph {et~al.}(2022)\citenamefont
  {Srivastav}, \citenamefont {Kumar}, \citenamefont
  {Sp{\aa}nsl{\ifmmode\ddot{a}\else\"{a}\fi}tt}, \citenamefont {Watanabe},
  \citenamefont {Taniguchi}, \citenamefont {Mirlin}, \citenamefont {Gefen},\
  and\ \citenamefont {Das}}]{Srivastav2022}%
  \BibitemOpen
  \bibfield  {author} {\bibinfo {author} {\bibfnamefont {S.~K.}\ \bibnamefont
  {Srivastav}}, \bibinfo {author} {\bibfnamefont {R.}~\bibnamefont {Kumar}},
  \bibinfo {author} {\bibfnamefont {C.}~\bibnamefont
  {Sp{\aa}nsl{\ifmmode\ddot{a}\else\"{a}\fi}tt}}, \bibinfo {author}
  {\bibfnamefont {K.}~\bibnamefont {Watanabe}}, \bibinfo {author}
  {\bibfnamefont {T.}~\bibnamefont {Taniguchi}}, \bibinfo {author}
  {\bibfnamefont {A.~D.}\ \bibnamefont {Mirlin}}, \bibinfo {author}
  {\bibfnamefont {Y.}~\bibnamefont {Gefen}},\ and\ \bibinfo {author}
  {\bibfnamefont {A.}~\bibnamefont {Das}},\ }\bibfield  {title} {\bibinfo
  {title} {{Determination of topological edge quantum numbers of fractional
  quantum {H}all phases by thermal conductance measurements}},\ }\href
  {https://doi.org/10.1038/s41467-022-32956-z} {\bibfield  {journal} {\bibinfo
  {journal} {Nat. Commun.}\ }\textbf {\bibinfo {volume} {13}},\ \bibinfo
  {pages} {1} (\bibinfo {year} {2022})}\BibitemShut {NoStop}%
\bibitem [{\citenamefont {Le~Breton}\ \emph {et~al.}(2022)\citenamefont
  {Le~Breton}, \citenamefont {Delagrange}, \citenamefont {Hong}, \citenamefont
  {Garg}, \citenamefont {Watanabe}, \citenamefont {Taniguchi}, \citenamefont
  {Ribeiro-Palau}, \citenamefont {Roulleau}, \citenamefont {Roche},\ and\
  \citenamefont {Parmentier}}]{LeBreton2022Sep}%
  \BibitemOpen
  \bibfield  {author} {\bibinfo {author} {\bibfnamefont {G.}~\bibnamefont
  {Le~Breton}}, \bibinfo {author} {\bibfnamefont {R.}~\bibnamefont
  {Delagrange}}, \bibinfo {author} {\bibfnamefont {Y.}~\bibnamefont {Hong}},
  \bibinfo {author} {\bibfnamefont {M.}~\bibnamefont {Garg}}, \bibinfo {author}
  {\bibfnamefont {K.}~\bibnamefont {Watanabe}}, \bibinfo {author}
  {\bibfnamefont {T.}~\bibnamefont {Taniguchi}}, \bibinfo {author}
  {\bibfnamefont {R.}~\bibnamefont {Ribeiro-Palau}}, \bibinfo {author}
  {\bibfnamefont {P.}~\bibnamefont {Roulleau}}, \bibinfo {author}
  {\bibfnamefont {P.}~\bibnamefont {Roche}},\ and\ \bibinfo {author}
  {\bibfnamefont {F.~D.}\ \bibnamefont {Parmentier}},\ }\bibfield  {title}
  {\bibinfo {title} {{Heat Equilibration of Integer and Fractional Quantum
  {{H}all} Edge Modes in Graphene}},\ }\href
  {https://doi.org/10.1103/PhysRevLett.129.116803} {\bibfield  {journal}
  {\bibinfo  {journal} {Phys. Rev. Lett.}\ }\textbf {\bibinfo {volume} {129}},\
  \bibinfo {pages} {116803} (\bibinfo {year} {2022})}\BibitemShut {NoStop}%
\bibitem [{\citenamefont {Cohen}\ \emph {et~al.}(2019)\citenamefont {Cohen},
  \citenamefont {Ronen}, \citenamefont {Yang}, \citenamefont {Banitt},
  \citenamefont {Park}, \citenamefont {Heiblum}, \citenamefont {Mirlin},
  \citenamefont {Gefen},\ and\ \citenamefont {Umansky}}]{Cohen2019}%
  \BibitemOpen
  \bibfield  {author} {\bibinfo {author} {\bibfnamefont {Y.}~\bibnamefont
  {Cohen}}, \bibinfo {author} {\bibfnamefont {Y.}~\bibnamefont {Ronen}},
  \bibinfo {author} {\bibfnamefont {W.}~\bibnamefont {Yang}}, \bibinfo {author}
  {\bibfnamefont {D.}~\bibnamefont {Banitt}}, \bibinfo {author} {\bibfnamefont
  {J.}~\bibnamefont {Park}}, \bibinfo {author} {\bibfnamefont {M.}~\bibnamefont
  {Heiblum}}, \bibinfo {author} {\bibfnamefont {A.~D.}\ \bibnamefont {Mirlin}},
  \bibinfo {author} {\bibfnamefont {Y.}~\bibnamefont {Gefen}},\ and\ \bibinfo
  {author} {\bibfnamefont {V.}~\bibnamefont {Umansky}},\ }\bibfield  {title}
  {\bibinfo {title} {Synthesizing a \ensuremath{\nu}=2/3 fractional quantum
  {{H}all} effect edge state from counter-propagating \ensuremath{\nu}=1 and
  \ensuremath{\nu}=1/3 states},\ }\href
  {https://doi.org/10.1038/s41467-019-09920-5} {\bibfield  {journal} {\bibinfo
  {journal} {Nature Communications}\ }\textbf {\bibinfo {volume} {10}},\
  \bibinfo {pages} {1920} (\bibinfo {year} {2019})}\BibitemShut {NoStop}%
\bibitem [{\citenamefont {Lafont}\ \emph {et~al.}(2019)\citenamefont {Lafont},
  \citenamefont {Rosenblatt}, \citenamefont {Heiblum},\ and\ \citenamefont
  {Umansky}}]{Lafont2019}%
  \BibitemOpen
  \bibfield  {author} {\bibinfo {author} {\bibfnamefont {F.}~\bibnamefont
  {Lafont}}, \bibinfo {author} {\bibfnamefont {A.}~\bibnamefont {Rosenblatt}},
  \bibinfo {author} {\bibfnamefont {M.}~\bibnamefont {Heiblum}},\ and\ \bibinfo
  {author} {\bibfnamefont {V.}~\bibnamefont {Umansky}},\ }\bibfield  {title}
  {\bibinfo {title} {Counter-propagating charge transport in the quantum
  {{H}all} effect regime},\ }\href {https://doi.org/10.1126/science.aar3766}
  {\bibfield  {journal} {\bibinfo  {journal} {Science}\ }\textbf {\bibinfo
  {volume} {363}},\ \bibinfo {pages} {54} (\bibinfo {year} {2019})}\BibitemShut
  {NoStop}%
\bibitem [{\citenamefont {Dutta}\ \emph
  {et~al.}(2022{\natexlab{a}})\citenamefont {Dutta}, \citenamefont {Umansky},
  \citenamefont {Banerjee},\ and\ \citenamefont {Heiblum}}]{Dutta2022Sep}%
  \BibitemOpen
  \bibfield  {author} {\bibinfo {author} {\bibfnamefont {B.}~\bibnamefont
  {Dutta}}, \bibinfo {author} {\bibfnamefont {V.}~\bibnamefont {Umansky}},
  \bibinfo {author} {\bibfnamefont {M.}~\bibnamefont {Banerjee}},\ and\
  \bibinfo {author} {\bibfnamefont {M.}~\bibnamefont {Heiblum}},\ }\bibfield
  {title} {\bibinfo {title} {{Isolated ballistic non-{A}belian interface
  channel}},\ }\href {https://doi.org/10.1126/science.abm6571} {\bibfield
  {journal} {\bibinfo  {journal} {Science}\ }\textbf {\bibinfo {volume}
  {377}},\ \bibinfo {pages} {1198} (\bibinfo {year}
  {2022}{\natexlab{a}})}\BibitemShut {NoStop}%
\bibitem [{\citenamefont {Dutta}\ \emph
  {et~al.}(2022{\natexlab{b}})\citenamefont {Dutta}, \citenamefont {Yang},
  \citenamefont {Melcer}, \citenamefont {Kundu}, \citenamefont {Heiblum},
  \citenamefont {Umansky}, \citenamefont {Oreg}, \citenamefont {Stern},\ and\
  \citenamefont {Mross}}]{Dutta2022}%
  \BibitemOpen
  \bibfield  {author} {\bibinfo {author} {\bibfnamefont {B.}~\bibnamefont
  {Dutta}}, \bibinfo {author} {\bibfnamefont {W.}~\bibnamefont {Yang}},
  \bibinfo {author} {\bibfnamefont {R.}~\bibnamefont {Melcer}}, \bibinfo
  {author} {\bibfnamefont {H.~K.}\ \bibnamefont {Kundu}}, \bibinfo {author}
  {\bibfnamefont {M.}~\bibnamefont {Heiblum}}, \bibinfo {author} {\bibfnamefont
  {V.}~\bibnamefont {Umansky}}, \bibinfo {author} {\bibfnamefont
  {Y.}~\bibnamefont {Oreg}}, \bibinfo {author} {\bibfnamefont {A.}~\bibnamefont
  {Stern}},\ and\ \bibinfo {author} {\bibfnamefont {D.}~\bibnamefont {Mross}},\
  }\bibfield  {title} {\bibinfo {title} {Distinguishing between non-{A}belian
  topological orders in a quantum {{H}all} system},\ }\href
  {https://doi.org/10.1126/science.abg6116} {\bibfield  {journal} {\bibinfo
  {journal} {Science}\ }\textbf {\bibinfo {volume} {375}},\ \bibinfo {pages}
  {193} (\bibinfo {year} {2022}{\natexlab{b}})}\BibitemShut {NoStop}%
\bibitem [{\citenamefont {Kane}\ and\ \citenamefont {Fisher}(1997)}]{Kane1997}%
  \BibitemOpen
  \bibfield  {author} {\bibinfo {author} {\bibfnamefont {C.~L.}\ \bibnamefont
  {Kane}}\ and\ \bibinfo {author} {\bibfnamefont {M.~P.~A.}\ \bibnamefont
  {Fisher}},\ }\bibfield  {title} {\bibinfo {title} {Quantized thermal
  transport in the fractional quantum {{H}all} effect},\ }\href
  {https://doi.org/10.1103/PhysRevB.55.15832} {\bibfield  {journal} {\bibinfo
  {journal} {Phys. Rev. B}\ }\textbf {\bibinfo {volume} {55}},\ \bibinfo
  {pages} {15832} (\bibinfo {year} {1997})}\BibitemShut {NoStop}%
\bibitem [{\citenamefont {Cappelli}\ \emph {et~al.}(2002)\citenamefont
  {Cappelli}, \citenamefont {Huerta},\ and\ \citenamefont
  {Zemba}}]{Capelli2002}%
  \BibitemOpen
  \bibfield  {author} {\bibinfo {author} {\bibfnamefont {A.}~\bibnamefont
  {Cappelli}}, \bibinfo {author} {\bibfnamefont {M.}~\bibnamefont {Huerta}},\
  and\ \bibinfo {author} {\bibfnamefont {G.~R.}\ \bibnamefont {Zemba}},\
  }\bibfield  {title} {\bibinfo {title} {Thermal transport in chiral conformal
  theories and hierarchical quantum {{H}all} states},\ }\href
  {https://doi.org/https://doi.org/10.1016/S0550-3213(02)00340-1} {\bibfield
  {journal} {\bibinfo  {journal} {Nuclear Physics B}\ }\textbf {\bibinfo
  {volume} {636}},\ \bibinfo {pages} {568 } (\bibinfo {year}
  {2002})}\BibitemShut {NoStop}%
\bibitem [{\citenamefont {Giamarchi}\ and\ \citenamefont
  {Schulz}(1988)}]{Giamarchi1988}%
  \BibitemOpen
  \bibfield  {author} {\bibinfo {author} {\bibfnamefont {T.}~\bibnamefont
  {Giamarchi}}\ and\ \bibinfo {author} {\bibfnamefont {H.~J.}\ \bibnamefont
  {Schulz}},\ }\bibfield  {title} {\bibinfo {title} {Anderson localization and
  interactions in one-dimensional metals},\ }\href
  {https://doi.org/10.1103/PhysRevB.37.325} {\bibfield  {journal} {\bibinfo
  {journal} {Phys. Rev. B}\ }\textbf {\bibinfo {volume} {37}},\ \bibinfo
  {pages} {325} (\bibinfo {year} {1988})}\BibitemShut {NoStop}%
\bibitem [{\citenamefont {Gornyi}\ \emph {et~al.}(2007)\citenamefont {Gornyi},
  \citenamefont {Mirlin},\ and\ \citenamefont {Polyakov}}]{Gornyi2007}%
  \BibitemOpen
  \bibfield  {author} {\bibinfo {author} {\bibfnamefont {I.~V.}\ \bibnamefont
  {Gornyi}}, \bibinfo {author} {\bibfnamefont {A.~D.}\ \bibnamefont {Mirlin}},\
  and\ \bibinfo {author} {\bibfnamefont {D.~G.}\ \bibnamefont {Polyakov}},\
  }\bibfield  {title} {\bibinfo {title} {Electron transport in a disordered
  {L}uttinger liquid},\ }\href {https://doi.org/10.1103/PhysRevB.75.085421}
  {\bibfield  {journal} {\bibinfo  {journal} {Phys. Rev. B}\ }\textbf {\bibinfo
  {volume} {75}},\ \bibinfo {pages} {085421} (\bibinfo {year}
  {2007})}\BibitemShut {NoStop}%
\bibitem [{\citenamefont {Murthy}\ and\ \citenamefont
  {Nayak}(2020)}]{Murthy2020}%
  \BibitemOpen
  \bibfield  {author} {\bibinfo {author} {\bibfnamefont {C.}~\bibnamefont
  {Murthy}}\ and\ \bibinfo {author} {\bibfnamefont {C.}~\bibnamefont {Nayak}},\
  }\bibfield  {title} {\bibinfo {title} {Almost perfect metals in one
  dimension},\ }\href {https://doi.org/10.1103/PhysRevLett.124.136801}
  {\bibfield  {journal} {\bibinfo  {journal} {Phys. Rev. Lett.}\ }\textbf
  {\bibinfo {volume} {124}},\ \bibinfo {pages} {136801} (\bibinfo {year}
  {2020})}\BibitemShut {NoStop}%
\bibitem [{\citenamefont {Protopopov}\ \emph {et~al.}(2017)\citenamefont
  {Protopopov}, \citenamefont {Gefen},\ and\ \citenamefont
  {Mirlin}}]{Protopopov2017}%
  \BibitemOpen
  \bibfield  {author} {\bibinfo {author} {\bibfnamefont {I.}~\bibnamefont
  {Protopopov}}, \bibinfo {author} {\bibfnamefont {Y.}~\bibnamefont {Gefen}},\
  and\ \bibinfo {author} {\bibfnamefont {A.}~\bibnamefont {Mirlin}},\
  }\bibfield  {title} {\bibinfo {title} {Transport in a disordered
  $\ensuremath{\nu}=2/3$ fractional quantum {{H}all} junction},\ }\href
  {https://doi.org/https://doi.org/10.1016/j.aop.2017.07.015} {\bibfield
  {journal} {\bibinfo  {journal} {Annals of Physics}\ }\textbf {\bibinfo
  {volume} {385}},\ \bibinfo {pages} {287 } (\bibinfo {year}
  {2017})}\BibitemShut {NoStop}%
\bibitem [{\citenamefont {Sp\aa{}nsl\"att}\ \emph {et~al.}(2019)\citenamefont
  {Sp\aa{}nsl\"att}, \citenamefont {Park}, \citenamefont {Gefen},\ and\
  \citenamefont {Mirlin}}]{Spanslatt2019}%
  \BibitemOpen
  \bibfield  {author} {\bibinfo {author} {\bibfnamefont {C.}~\bibnamefont
  {Sp\aa{}nsl\"att}}, \bibinfo {author} {\bibfnamefont {J.}~\bibnamefont
  {Park}}, \bibinfo {author} {\bibfnamefont {Y.}~\bibnamefont {Gefen}},\ and\
  \bibinfo {author} {\bibfnamefont {A.~D.}\ \bibnamefont {Mirlin}},\ }\bibfield
   {title} {\bibinfo {title} {Topological classification of shot noise on
  fractional quantum {{H}all} edges},\ }\href
  {https://doi.org/10.1103/PhysRevLett.123.137701} {\bibfield  {journal}
  {\bibinfo  {journal} {Phys. Rev. Lett.}\ }\textbf {\bibinfo {volume} {123}},\
  \bibinfo {pages} {137701} (\bibinfo {year} {2019})}\BibitemShut {NoStop}%
\bibitem [{\citenamefont {Asasi}\ and\ \citenamefont
  {Mulligan}(2020)}]{Asasi2020}%
  \BibitemOpen
  \bibfield  {author} {\bibinfo {author} {\bibfnamefont {H.}~\bibnamefont
  {Asasi}}\ and\ \bibinfo {author} {\bibfnamefont {M.}~\bibnamefont
  {Mulligan}},\ }\bibfield  {title} {\bibinfo {title} {Partial equilibration of
  anti-{P}faffian edge modes at $\ensuremath{\nu}=5/2$},\ }\href
  {https://doi.org/10.1103/PhysRevB.102.205104} {\bibfield  {journal} {\bibinfo
   {journal} {Phys. Rev. B}\ }\textbf {\bibinfo {volume} {102}},\ \bibinfo
  {pages} {205104} (\bibinfo {year} {2020})}\BibitemShut {NoStop}%
\bibitem [{\citenamefont {Kane}\ and\ \citenamefont
  {Fisher}(1995{\natexlab{b}})}]{Kane1995}%
  \BibitemOpen
  \bibfield  {author} {\bibinfo {author} {\bibfnamefont {C.~L.}\ \bibnamefont
  {Kane}}\ and\ \bibinfo {author} {\bibfnamefont {M.~P.~A.}\ \bibnamefont
  {Fisher}},\ }\bibfield  {title} {\bibinfo {title} {Contacts and edge-state
  equilibration in the fractional quantum {{H}all} effect},\ }\href
  {https://doi.org/10.1103/PhysRevB.52.17393} {\bibfield  {journal} {\bibinfo
  {journal} {Phys. Rev. B}\ }\textbf {\bibinfo {volume} {52}},\ \bibinfo
  {pages} {17393} (\bibinfo {year} {1995}{\natexlab{b}})}\BibitemShut {NoStop}%
\bibitem [{\citenamefont {Nosiglia}\ \emph {et~al.}(2018)\citenamefont
  {Nosiglia}, \citenamefont {Park}, \citenamefont {Rosenow},\ and\
  \citenamefont {Gefen}}]{Nosiglia2018}%
  \BibitemOpen
  \bibfield  {author} {\bibinfo {author} {\bibfnamefont {C.}~\bibnamefont
  {Nosiglia}}, \bibinfo {author} {\bibfnamefont {J.}~\bibnamefont {Park}},
  \bibinfo {author} {\bibfnamefont {B.}~\bibnamefont {Rosenow}},\ and\ \bibinfo
  {author} {\bibfnamefont {Y.}~\bibnamefont {Gefen}},\ }\bibfield  {title}
  {\bibinfo {title} {Incoherent transport on the $\ensuremath{\nu}=2/3$ quantum
  {{H}all} edge},\ }\href {https://doi.org/10.1103/PhysRevB.98.115408}
  {\bibfield  {journal} {\bibinfo  {journal} {Phys. Rev. B}\ }\textbf {\bibinfo
  {volume} {98}},\ \bibinfo {pages} {115408} (\bibinfo {year}
  {2018})}\BibitemShut {NoStop}%
\bibitem [{\citenamefont {Park}\ \emph {et~al.}(2019)\citenamefont {Park},
  \citenamefont {Mirlin}, \citenamefont {Rosenow},\ and\ \citenamefont
  {Gefen}}]{Park2019}%
  \BibitemOpen
  \bibfield  {author} {\bibinfo {author} {\bibfnamefont {J.}~\bibnamefont
  {Park}}, \bibinfo {author} {\bibfnamefont {A.~D.}\ \bibnamefont {Mirlin}},
  \bibinfo {author} {\bibfnamefont {B.}~\bibnamefont {Rosenow}},\ and\ \bibinfo
  {author} {\bibfnamefont {Y.}~\bibnamefont {Gefen}},\ }\bibfield  {title}
  {\bibinfo {title} {Noise on complex quantum {{H}all} edges: Chiral anomaly
  and heat diffusion},\ }\href {https://doi.org/10.1103/PhysRevB.99.161302}
  {\bibfield  {journal} {\bibinfo  {journal} {Phys. Rev. B}\ }\textbf {\bibinfo
  {volume} {99}},\ \bibinfo {pages} {161302} (\bibinfo {year}
  {2019})}\BibitemShut {NoStop}%
\bibitem [{\citenamefont {Sp\aa{}nsl\"att}\ \emph {et~al.}(2020)\citenamefont
  {Sp\aa{}nsl\"att}, \citenamefont {Park}, \citenamefont {Gefen},\ and\
  \citenamefont {Mirlin}}]{Spanslatt2020}%
  \BibitemOpen
  \bibfield  {author} {\bibinfo {author} {\bibfnamefont {C.}~\bibnamefont
  {Sp\aa{}nsl\"att}}, \bibinfo {author} {\bibfnamefont {J.}~\bibnamefont
  {Park}}, \bibinfo {author} {\bibfnamefont {Y.}~\bibnamefont {Gefen}},\ and\
  \bibinfo {author} {\bibfnamefont {A.~D.}\ \bibnamefont {Mirlin}},\ }\bibfield
   {title} {\bibinfo {title} {Conductance plateaus and shot noise in fractional
  quantum {{H}all} point contacts},\ }\href
  {https://doi.org/10.1103/PhysRevB.101.075308} {\bibfield  {journal} {\bibinfo
   {journal} {Phys. Rev. B}\ }\textbf {\bibinfo {volume} {101}},\ \bibinfo
  {pages} {075308} (\bibinfo {year} {2020})}\BibitemShut {NoStop}%
\bibitem [{\citenamefont {Park}\ \emph {et~al.}(2020)\citenamefont {Park},
  \citenamefont {Sp\aa{}nsl\"att}, \citenamefont {Gefen},\ and\ \citenamefont
  {Mirlin}}]{Park2020NAB}%
  \BibitemOpen
  \bibfield  {author} {\bibinfo {author} {\bibfnamefont {J.}~\bibnamefont
  {Park}}, \bibinfo {author} {\bibfnamefont {C.}~\bibnamefont
  {Sp\aa{}nsl\"att}}, \bibinfo {author} {\bibfnamefont {Y.}~\bibnamefont
  {Gefen}},\ and\ \bibinfo {author} {\bibfnamefont {A.~D.}\ \bibnamefont
  {Mirlin}},\ }\bibfield  {title} {\bibinfo {title} {Noise on the non-{A}belian
  $\ensuremath{\nu}=5/2$ fractional quantum {{H}all} edge},\ }\href
  {https://doi.org/10.1103/PhysRevLett.125.157702} {\bibfield  {journal}
  {\bibinfo  {journal} {Phys. Rev. Lett.}\ }\textbf {\bibinfo {volume} {125}},\
  \bibinfo {pages} {157702} (\bibinfo {year} {2020})}\BibitemShut {NoStop}%
\bibitem [{\citenamefont {Hein}\ and\ \citenamefont
  {Sp{\aa}nsl{\ifmmode\ddot{a}\else\"{a}\fi}tt}(2022)}]{Hein2022Nov}%
  \BibitemOpen
  \bibfield  {author} {\bibinfo {author} {\bibfnamefont {M.}~\bibnamefont
  {Hein}}\ and\ \bibinfo {author} {\bibfnamefont {C.}~\bibnamefont
  {Sp{\aa}nsl{\ifmmode\ddot{a}\else\"{a}\fi}tt}},\ }\bibfield  {title}
  {\bibinfo {title} {{Thermal conductance and noise of Majorana modes along
  interfaced $\nu=5/2$ fractional quantum Hall states}},\ }\href
  {https://arxiv.org/abs/2211.08000} {\bibfield  {journal} {\bibinfo  {journal}
  {arXiv preprint arXiv:2211.08000}\ } (\bibinfo {year} {2022})}\BibitemShut
  {NoStop}%
\bibitem [{\citenamefont {Fidkowski}\ \emph {et~al.}(2013)\citenamefont
  {Fidkowski}, \citenamefont {Chen},\ and\ \citenamefont
  {Vishwanath}}]{Fidkowski2013}%
  \BibitemOpen
  \bibfield  {author} {\bibinfo {author} {\bibfnamefont {L.}~\bibnamefont
  {Fidkowski}}, \bibinfo {author} {\bibfnamefont {X.}~\bibnamefont {Chen}},\
  and\ \bibinfo {author} {\bibfnamefont {A.}~\bibnamefont {Vishwanath}},\
  }\bibfield  {title} {\bibinfo {title} {Non-{A}belian topological order on the
  surface of a 3d topological superconductor from an exactly solved model},\
  }\href {https://doi.org/10.1103/PhysRevX.3.041016} {\bibfield  {journal}
  {\bibinfo  {journal} {Phys. Rev. X}\ }\textbf {\bibinfo {volume} {3}},\
  \bibinfo {pages} {041016} (\bibinfo {year} {2013})}\BibitemShut {NoStop}%
\bibitem [{\citenamefont {Son}(2015)}]{Son2015}%
  \BibitemOpen
  \bibfield  {author} {\bibinfo {author} {\bibfnamefont {D.~T.}\ \bibnamefont
  {Son}},\ }\bibfield  {title} {\bibinfo {title} {Is the composite fermion a
  {D}irac particle?},\ }\href {https://doi.org/10.1103/PhysRevX.5.031027}
  {\bibfield  {journal} {\bibinfo  {journal} {Phys. Rev. X}\ }\textbf {\bibinfo
  {volume} {5}},\ \bibinfo {pages} {031027} (\bibinfo {year}
  {2015})}\BibitemShut {NoStop}%
\bibitem [{\citenamefont {Zucker}\ and\ \citenamefont
  {Feldman}(2016)}]{Zucker2016}%
  \BibitemOpen
  \bibfield  {author} {\bibinfo {author} {\bibfnamefont {P.~T.}\ \bibnamefont
  {Zucker}}\ and\ \bibinfo {author} {\bibfnamefont {D.~E.}\ \bibnamefont
  {Feldman}},\ }\bibfield  {title} {\bibinfo {title} {Stabilization of the
  particle-hole {P}faffian order by landau-level mixing and impurities that
  break particle-hole symmetry},\ }\href
  {https://doi.org/10.1103/PhysRevLett.117.096802} {\bibfield  {journal}
  {\bibinfo  {journal} {Phys. Rev. Lett.}\ }\textbf {\bibinfo {volume} {117}},\
  \bibinfo {pages} {096802} (\bibinfo {year} {2016})}\BibitemShut {NoStop}%
\bibitem [{\citenamefont {Antoni\ifmmode~\acute{c}\else \'{c}\fi{}}\ \emph
  {et~al.}(2018)\citenamefont {Antoni\ifmmode~\acute{c}\else \'{c}\fi{}},
  \citenamefont {Vu\ifmmode \check{c}\else \v{c}\fi{}i\ifmmode \check{c}\else
  \v{c}\fi{}evi\ifmmode~\acute{c}\else \'{c}\fi{}},\ and\ \citenamefont
  {Milovanovi\ifmmode~\acute{c}\else \'{c}\fi{}}}]{Antonic2018}%
  \BibitemOpen
  \bibfield  {author} {\bibinfo {author} {\bibfnamefont {L.}~\bibnamefont
  {Antoni\ifmmode~\acute{c}\else \'{c}\fi{}}}, \bibinfo {author} {\bibfnamefont
  {J.}~\bibnamefont {Vu\ifmmode \check{c}\else \v{c}\fi{}i\ifmmode
  \check{c}\else \v{c}\fi{}evi\ifmmode~\acute{c}\else \'{c}\fi{}}},\ and\
  \bibinfo {author} {\bibfnamefont {M.~V.}\ \bibnamefont
  {Milovanovi\ifmmode~\acute{c}\else \'{c}\fi{}}},\ }\bibfield  {title}
  {\bibinfo {title} {Paired states at 5/2: Particle-hole {P}faffian and
  particle-hole symmetry breaking},\ }\href
  {https://doi.org/10.1103/PhysRevB.98.115107} {\bibfield  {journal} {\bibinfo
  {journal} {Phys. Rev. B}\ }\textbf {\bibinfo {volume} {98}},\ \bibinfo
  {pages} {115107} (\bibinfo {year} {2018})}\BibitemShut {NoStop}%
\bibitem [{\citenamefont {Willett}\ \emph {et~al.}(1987)\citenamefont
  {Willett}, \citenamefont {Eisenstein}, \citenamefont {St\"ormer},
  \citenamefont {Tsui}, \citenamefont {Gossard},\ and\ \citenamefont
  {English}}]{Willet1987}%
  \BibitemOpen
  \bibfield  {author} {\bibinfo {author} {\bibfnamefont {R.}~\bibnamefont
  {Willett}}, \bibinfo {author} {\bibfnamefont {J.~P.}\ \bibnamefont
  {Eisenstein}}, \bibinfo {author} {\bibfnamefont {H.~L.}\ \bibnamefont
  {St\"ormer}}, \bibinfo {author} {\bibfnamefont {D.~C.}\ \bibnamefont {Tsui}},
  \bibinfo {author} {\bibfnamefont {A.~C.}\ \bibnamefont {Gossard}},\ and\
  \bibinfo {author} {\bibfnamefont {J.~H.}\ \bibnamefont {English}},\
  }\bibfield  {title} {\bibinfo {title} {Observation of an even-denominator
  quantum number in the fractional quantum {{H}all} effect},\ }\href
  {https://doi.org/10.1103/PhysRevLett.59.1776} {\bibfield  {journal} {\bibinfo
   {journal} {Phys. Rev. Lett.}\ }\textbf {\bibinfo {volume} {59}},\ \bibinfo
  {pages} {1776} (\bibinfo {year} {1987})}\BibitemShut {NoStop}%
\bibitem [{\citenamefont {Moore}\ and\ \citenamefont {Read}(1991)}]{Moore1991}%
  \BibitemOpen
  \bibfield  {author} {\bibinfo {author} {\bibfnamefont {G.}~\bibnamefont
  {Moore}}\ and\ \bibinfo {author} {\bibfnamefont {N.}~\bibnamefont {Read}},\
  }\bibfield  {title} {\bibinfo {title} {Nonabelions in the fractional quantum
  {{H}all} effect},\ }\href
  {https://doi.org/https://doi.org/10.1016/0550-3213(91)90407-O} {\bibfield
  {journal} {\bibinfo  {journal} {Nuclear Physics B}\ }\textbf {\bibinfo
  {volume} {360}},\ \bibinfo {pages} {362 } (\bibinfo {year}
  {1991})}\BibitemShut {NoStop}%
\bibitem [{\citenamefont {Sp\aa{}nsl\"att}\ \emph {et~al.}(2021)\citenamefont
  {Sp\aa{}nsl\"att}, \citenamefont {Gefen}, \citenamefont {Gornyi},\ and\
  \citenamefont {Polyakov}}]{Spanslatt2021Contacts}%
  \BibitemOpen
  \bibfield  {author} {\bibinfo {author} {\bibfnamefont {C.}~\bibnamefont
  {Sp\aa{}nsl\"att}}, \bibinfo {author} {\bibfnamefont {Y.}~\bibnamefont
  {Gefen}}, \bibinfo {author} {\bibfnamefont {I.~V.}\ \bibnamefont {Gornyi}},\
  and\ \bibinfo {author} {\bibfnamefont {D.~G.}\ \bibnamefont {Polyakov}},\
  }\bibfield  {title} {\bibinfo {title} {Contacts, equilibration, and
  interactions in fractional quantum {{H}all} edge transport},\ }\href
  {https://doi.org/10.1103/PhysRevB.104.115416} {\bibfield  {journal} {\bibinfo
   {journal} {Phys. Rev. B}\ }\textbf {\bibinfo {volume} {104}},\ \bibinfo
  {pages} {115416} (\bibinfo {year} {2021})}\BibitemShut {NoStop}%
\bibitem [{\citenamefont {Jezouin}\ \emph {et~al.}(2013)\citenamefont
  {Jezouin}, \citenamefont {Parmentier}, \citenamefont {Anthore}, \citenamefont
  {Gennser}, \citenamefont {Cavanna}, \citenamefont {Jin},\ and\ \citenamefont
  {Pierre}}]{Jezouin2013}%
  \BibitemOpen
  \bibfield  {author} {\bibinfo {author} {\bibfnamefont {S.}~\bibnamefont
  {Jezouin}}, \bibinfo {author} {\bibfnamefont {F.~D.}\ \bibnamefont
  {Parmentier}}, \bibinfo {author} {\bibfnamefont {A.}~\bibnamefont {Anthore}},
  \bibinfo {author} {\bibfnamefont {U.}~\bibnamefont {Gennser}}, \bibinfo
  {author} {\bibfnamefont {A.}~\bibnamefont {Cavanna}}, \bibinfo {author}
  {\bibfnamefont {Y.}~\bibnamefont {Jin}},\ and\ \bibinfo {author}
  {\bibfnamefont {F.}~\bibnamefont {Pierre}},\ }\bibfield  {title} {\bibinfo
  {title} {Quantum limit of heat flow across a single electronic channel},\
  }\href {https://doi.org/10.1126/science.1241912} {\bibfield  {journal}
  {\bibinfo  {journal} {Science}\ }\textbf {\bibinfo {volume} {342}},\ \bibinfo
  {pages} {601} (\bibinfo {year} {2013})}\BibitemShut {NoStop}%
\bibitem [{\citenamefont {Krive}(1998)}]{Krive1998}%
  \BibitemOpen
  \bibfield  {author} {\bibinfo {author} {\bibfnamefont {I.~V.}\ \bibnamefont
  {Krive}},\ }\bibfield  {title} {\bibinfo {title} {Thermal transport through
  {L}uttinger liquid constriction},\ }\href {https://doi.org/10.1063/1.593605}
  {\bibfield  {journal} {\bibinfo  {journal} {Low Temperature Physics}\
  }\textbf {\bibinfo {volume} {24}},\ \bibinfo {pages} {377} (\bibinfo {year}
  {1998})}\BibitemShut {NoStop}%
\bibitem [{\citenamefont {Kane}\ and\ \citenamefont
  {Fisher}(1994)}]{KaneFisher1994Noise}%
  \BibitemOpen
  \bibfield  {author} {\bibinfo {author} {\bibfnamefont {C.~L.}\ \bibnamefont
  {Kane}}\ and\ \bibinfo {author} {\bibfnamefont {M.~P.~A.}\ \bibnamefont
  {Fisher}},\ }\bibfield  {title} {\bibinfo {title} {Nonequilibrium noise and
  fractional charge in the quantum {{H}all} effect},\ }\href
  {https://doi.org/10.1103/PhysRevLett.72.724} {\bibfield  {journal} {\bibinfo
  {journal} {Phys. Rev. Lett.}\ }\textbf {\bibinfo {volume} {72}},\ \bibinfo
  {pages} {724} (\bibinfo {year} {1994})}\BibitemShut {NoStop}%
\bibitem [{\citenamefont {Chamon}\ \emph {et~al.}(1995)\citenamefont {Chamon},
  \citenamefont {Freed},\ and\ \citenamefont {Wen}}]{Chamon1995}%
  \BibitemOpen
  \bibfield  {author} {\bibinfo {author} {\bibfnamefont {C.~d.~C.}\
  \bibnamefont {Chamon}}, \bibinfo {author} {\bibfnamefont {D.~E.}\
  \bibnamefont {Freed}},\ and\ \bibinfo {author} {\bibfnamefont {X.~G.}\
  \bibnamefont {Wen}},\ }\bibfield  {title} {\bibinfo {title} {Tunneling and
  quantum noise in one-dimensional {L}uttinger liquids},\ }\href
  {https://doi.org/10.1103/PhysRevB.51.2363} {\bibfield  {journal} {\bibinfo
  {journal} {Phys. Rev. B}\ }\textbf {\bibinfo {volume} {51}},\ \bibinfo
  {pages} {2363} (\bibinfo {year} {1995})}\BibitemShut {NoStop}%
\bibitem [{\citenamefont {Fendley}\ \emph {et~al.}(1995)\citenamefont
  {Fendley}, \citenamefont {Ludwig},\ and\ \citenamefont
  {Saleur}}]{Fendley1995}%
  \BibitemOpen
  \bibfield  {author} {\bibinfo {author} {\bibfnamefont {P.}~\bibnamefont
  {Fendley}}, \bibinfo {author} {\bibfnamefont {A.~W.~W.}\ \bibnamefont
  {Ludwig}},\ and\ \bibinfo {author} {\bibfnamefont {H.}~\bibnamefont
  {Saleur}},\ }\bibfield  {title} {\bibinfo {title} {Exact nonequilibrium dc
  shot noise in {L}uttinger liquids and fractional quantum {{H}all} devices},\
  }\href {https://doi.org/10.1103/PhysRevLett.75.2196} {\bibfield  {journal}
  {\bibinfo  {journal} {Phys. Rev. Lett.}\ }\textbf {\bibinfo {volume} {75}},\
  \bibinfo {pages} {2196} (\bibinfo {year} {1995})}\BibitemShut {NoStop}%
\bibitem [{\citenamefont {Feldman}\ and\ \citenamefont
  {Heiblum}(2017)}]{Feldman2017}%
  \BibitemOpen
  \bibfield  {author} {\bibinfo {author} {\bibfnamefont {D.~E.}\ \bibnamefont
  {Feldman}}\ and\ \bibinfo {author} {\bibfnamefont {M.}~\bibnamefont
  {Heiblum}},\ }\bibfield  {title} {\bibinfo {title} {Why a noninteracting
  model works for shot noise in fractional charge experiments},\ }\href
  {https://doi.org/10.1103/PhysRevB.95.115308} {\bibfield  {journal} {\bibinfo
  {journal} {Phys. Rev. B}\ }\textbf {\bibinfo {volume} {95}},\ \bibinfo
  {pages} {115308} (\bibinfo {year} {2017})}\BibitemShut {NoStop}%
\bibitem [{\citenamefont {Martin}(2005)}]{Martin2005}%
  \BibitemOpen
  \bibfield  {author} {\bibinfo {author} {\bibfnamefont {T.}~\bibnamefont
  {Martin}},\ }\bibfield  {title} {\bibinfo {title} {Noise in mesoscopic
  physics},\ }in\ \href@noop {} {\emph {\bibinfo {booktitle} {Proceedings of
  the Les Houches Summer School, Session LXXXI}}},\ \bibinfo {editor} {edited
  by\ \bibinfo {editor} {\bibfnamefont {H.~B.}\ \bibnamefont {\textit{et
  al.}}}}\ (\bibinfo  {publisher} {Elsevier},\ \bibinfo {address} {New York},\
  \bibinfo {year} {2005})\BibitemShut {NoStop}%
\bibitem [{\citenamefont {Biswas}\ \emph {et~al.}(2022)\citenamefont {Biswas},
  \citenamefont {Bhattacharyya}, \citenamefont {Kundu}, \citenamefont {Das},
  \citenamefont {Heiblum}, \citenamefont {Umansky}, \citenamefont {Goldstein},\
  and\ \citenamefont {Gefen}}]{Biswas2022}%
  \BibitemOpen
  \bibfield  {author} {\bibinfo {author} {\bibfnamefont {S.}~\bibnamefont
  {Biswas}}, \bibinfo {author} {\bibfnamefont {R.}~\bibnamefont
  {Bhattacharyya}}, \bibinfo {author} {\bibfnamefont {H.~K.}\ \bibnamefont
  {Kundu}}, \bibinfo {author} {\bibfnamefont {A.}~\bibnamefont {Das}}, \bibinfo
  {author} {\bibfnamefont {M.}~\bibnamefont {Heiblum}}, \bibinfo {author}
  {\bibfnamefont {V.}~\bibnamefont {Umansky}}, \bibinfo {author} {\bibfnamefont
  {M.}~\bibnamefont {Goldstein}},\ and\ \bibinfo {author} {\bibfnamefont
  {Y.}~\bibnamefont {Gefen}},\ }\bibfield  {title} {\bibinfo {title} {Shot
  noise does not always provide the quasiparticle charge},\ }\href
  {https://doi.org/10.1038/s41567-022-01758-x} {\bibfield  {journal} {\bibinfo
  {journal} {Nature Physics}\ }\textbf {\bibinfo {volume} {18}},\ \bibinfo
  {pages} {1476} (\bibinfo {year} {2022})}\BibitemShut {NoStop}%
\bibitem [{\citenamefont {Overbosch}\ and\ \citenamefont
  {Wen}(2008)}]{Overbosch2008}%
  \BibitemOpen
  \bibfield  {author} {\bibinfo {author} {\bibfnamefont {B.}~\bibnamefont
  {Overbosch}}\ and\ \bibinfo {author} {\bibfnamefont {X.-G.}\ \bibnamefont
  {Wen}},\ }\bibfield  {title} {\bibinfo {title} {Phase transitions on the edge
  of the $\nu= 5/2$ {P}faffian and anti-{P}faffian quantum {{H}all} state},\
  }\href {https://arxiv.org/abs/0804.2087} {\bibfield  {journal} {\bibinfo
  {journal} {arXiv preprint arXiv:0804.2087}\ } (\bibinfo {year}
  {2008})}\BibitemShut {NoStop}%
\end{thebibliography}
%

\end{document}